\documentclass[final,3p,times]{elsarticle}

\usepackage{graphicx}

\usepackage{amssymb}

\usepackage[english,francais]{babel}

\def\oge{``}
\def\fge{"}

\usepackage[applemac]{inputenc}
\usepackage[T1]{fontenc}
\usepackage{amsmath}
\usepackage{color}
\usepackage{hyperref}
\hypersetup{breaklinks=true}
\hypersetup{colorlinks=true}
\hypersetup{urlcolor=blue}
\hypersetup{citecolor=black}
\hypersetup{linkcolor=black}
\usepackage{cleveref}
\newcommand{\bea}{\begin{eqnarray}}
\newcommand{\eea}{\end{eqnarray}}
\newcommand{\be}{\begin{equation}}
\newcommand{\ee}{\end{equation}}
\newcommand{\rr}{\mathbf{r}}

\newcommand{\pp}{\mathbf{p}}
\newcommand{\qq}{\mathbf{q}}

\newcommand{\epsc}{\check{\epsilon}}

\newcommand{\ii}{\textrm{i}}
\newcommand{\eee}{\textrm{e}}
\newcommand{\dd}{\mathrm{d}}

\DeclareMathOperator\acos{acos}
\DeclareMathOperator\atan{atan}

\DeclareMathOperator\argsh{argsh}

\DeclareMathOperator\re{Re}

\newcommand{\yvan}{}
\usepackage{float}

\begin{document}

\begin{frontmatter}


\selectlanguage{english}
\title{Coherence time of a Bose-Einstein condensate in an isolated harmonically trapped gas}


\author{Yvan Castin}
\author{Alice Sinatra}

\address{Laboratoire Kastler Brossel, ENS-PSL, CNRS, Sorbonne Universit\'e and Coll\`ege de France, Paris, France}



\begin{abstract}
We study the condensate phase dynamics in a low-temperature equilibrium gas of weakly interacting bosons, harmonically trapped and isolated from the environment. 
We find that at long times, much longer than the collision time between Bogoliubov quasiparticles, 
the variance of the phase accumulated by the condensate grows with a ballistic term quadratic in time and a diffusive term affine in time.
We give the corresponding analytical expressions in the limit of a large system, in the collisionless regime and 
in the ergodic approximation for the quasiparticle motion. When properly rescaled, they are described
by universal functions of the temperature divided by the Thomas-Fermi chemical potential. The same conclusion holds for the mode damping rates. 
Such universality class differs from the previously studied one of the homogeneous gas.  

\noindent 
\vskip 0.5\baselineskip

\end{abstract} 
\end{frontmatter}



\section{Introduction and overview} 

We consider here an unsolved problem of the theory of quantum gases: the coherence time of a 
spinless boson gas in the weakly interacting regime, in a harmonic trap. The gas is prepared at thermal equilibrium at a temperature $ T $ much lower than the critical temperature $ T_c $, 
that is in the strongly condensed regime, and it is perfectly isolated in its subsequent evolution. 
The coherence time of the bosonic field is then intrinsic and dominated by that of the condensate. In view of recent technical developments 
\cite{Treutlein, Oberthaler, Schmiedmayer},
this question could soon receive an experimental response in cold gases of atoms confined in 
non-dissipative magnetic potentials \cite{Cornell, Ketterle, Stringari_revue} and, unlike other 
 solid state systems \cite{JBloch, LeSiDang, Bramati, Dubin}, 
well decoupled from their environment and showing only low particle losses. Our 
theoretical study is also important for future applications in atom optics and matter wave interferometry. 

Following the pioneering work of references \cite{Zoller, Grahamdiff, Kuklov}, 
our previous studies \cite{Superdiff, Genuine, PRAlong, chapitre}
performed in a spatially homogeneous boson gas, rely on the Bogoliubov method, 
which reduces the system to a weakly interacting quasiparticle gas. They have identified two mechanisms limiting the 
coherence time, both involving the dynamics of the phase operator $\hat{\theta} (t) $ of the condensate: 
\begin{itemize} 
\item{\sl phase blurring}: when the conserved quantities (the energy $ E $ of the gas and its number of particles $ N $) fluctuate 
from one experimental realization to another, the average rate of evolution of the phase $ [\hat{\theta} (t) - \hat{\theta} (0)] / t $ in one realization, as a 
function of these conserved quantities, fluctuates too. After averaging over realizations, this induces a ballistic spread of the phase shift $ \hat {\theta} (t) - \hat {\theta} (0) $, that is a quadratic divergence of its variance, with a ballistic coefficient 
$ A $ \cite{Superdiff}: 
\be
\mbox{Var}[\hat{\theta}(t)-\hat{\theta}(0)] \sim A t^2
\label{eq:bali}
\ee
this holds at long times with respect to $ \gamma _ {\rm coll} ^ {- 1} $, where $ \gamma _ {\rm coll} $ is the typical collision rate between thermal Bogoliubov quasiparticles; 
\item{\sl phase diffusion}: even if the system is prepared in the microcanonical ensemble, where $ E $ and $ N $ are fixed, 
the interactions between quasiparticles cause their occupation numbers, and therefore the instantaneous speed $ \frac{\dd} {\dd t} \hat {\theta} $ 
of the phase, which depends on them, to fluctuate. This induces a diffusive spread of $ \hat {\theta} (t) - \hat {\theta} (0) $ at the times 
$ t \gg \gamma _ {\rm coll}^{- 1} $, with a coefficient $ D $ \cite {Genuine, PRAlong}: 
\be
\mbox{Var}_{\rm mc} [\hat{\theta}(t)-\hat{\theta}(0)] \sim 2 D t
\label{eq:diff} 
\ee
\end{itemize}
In the general case, both mechanisms are present and the variance of the phase shift admits (\ref{eq:bali}) as the dominant term, and (\ref{eq:diff}) as a subdominant term. The condensate phase spreading directly affects its 
first-order temporal coherence function 
\be
g_1(t) = \langle \hat{a}_0^\dagger(t) \hat{a}_0(0)\rangle
\ee
where  $\hat{a}_0$  is the annihilation operator of a boson in the condensate mode, thanks to the the approximate relation 
\be
g_1(t) \simeq \eee^{-\ii \langle\hat{\theta}(t)-\hat{\theta}(0)\rangle} \eee^{-\mathrm{Var}[\hat{\theta}(t)-\hat{\theta}(0)]/2}
\label{eq:g1}
\ee
admitted in reference {\yvan \cite {chapitre} under the hypothesis of a Gaussian distribution of $\hat{\theta}(t)-\hat{\theta}(0)$},
then justified in reference \cite {CohFer} at sufficiently low temperature {\yvan under fairly general assumptions.
\footnote{{\yvan 
Let us recall the assumptions used in reference \cite{CohFer} to establish equation (\ref{eq:g1}). $(i)$ The relative fluctuations of the modulus of $\hat{a}_0$ are small, the system being strongly Bose condensed. $(ii)$ The system is close enough to the thermodynamic limit, with normal fluctuations and asymptotically Gaussian laws for the energy and the number of particles. This is used in particular to put the ballistic contribution of the phase shift to $g_1(t)$ in the form 
(\ref{eq:g1}). $(iii)$ The diffusion coefficient of the phase (of order $1/N$) must be much smaller than the typical collision rate $\gamma_{\rm coll}$ 
between Bogolioubov quasi-particles (of order $N^0$) but much larger than the spacing of the energy levels (of order $N^{-2}$) of quasiparticle pairs created 
or annihilated during Beliaev collision processes. This is used, for a system prepared in the microcanonical ensemble, to show that $g_1(t)$ is of the form 
(\ref{eq:g1}) on time intervals $t=O(N^0)$ and $t=O(N^1)$, with the same diffusion coefficient. $(iv)$ The correlation function of $\frac{\dd}{\dd t}\hat{\theta}$ 
is real, as predicted by kinetic equations. $(v)$ We ignore the commutator of $\hat{\theta}(t)$ with $\hat{\theta}(0)$, 
which introduces a $O(t/N)$ phase error into the factor $\exp[-\ii\langle\hat{\theta}(t)-\hat{\theta}(0)\rangle]$. This is an error of order unity 
at times $t\approx N$ but $g_1(t)$ then began to decrease under the effect of phase diffusion in the microcanonical ensemble (and is otherwise 
already very strongly damped under the effect of ballistic phase blurring after a time $t\approx N^{1/2}$).
}}}
We propose here to generalize these first studies to the experimentally more usual case of a harmonically 
trapped system (see however reference \cite {ZoranBoite}). 
As the dependence of the damping rates of the Bogoliubov modes on the energy or the temperature 
are already very different from those of the homogeneous case, as it was shown in reference \cite {PRLGora}, it will certainly be the same for the spreading of the condensate phase. The trapped case is non-trivial, since the Bogoliubov modes 
are not known analytically, and there is no  local density approximation applicable {\yvan to the phase evolution} (as verified by 
reference \cite {EPLspin}). Fortunately we have the possibility to consider: 
\begin{itemize} 
\item the classical limit for the Bogoliubov quasiparticles motion in the trapped gas. Indeed, at the thermodynamic limit 
($ N \to + \infty $ with constant Gross-Pitaevskii's chemical potential $ \mu _ {\rm GP} $ and constant temperature), the trapping angular frequencies 
$ \omega_ \alpha $, $ \alpha \in \{x, y, z \} $, tend to zero {\yvan as $1/N^{1/3}$}:
\be
\hbar\omega_\alpha \ll \mu_{\rm GP}, k_B T
\label{eq:condval1}
\ee
so that we can cleverly reinterpret the thermodynamic limit as a classical limit $ \hbar \to 0 $; 
\item the limit of very weak interactions between Bogoliubov quasiparticles: 
\be
\gamma_{\rm coll} \ll \omega_\alpha
\label{eq:condval2}
\ee
This implies that all the modes, even those of weaker angular frequency $ \approx \omega_ \alpha $, are in the 
collisionless regime (by opposition to hydrodynamics), and makes it possible to make a secular approximation on the kinetic equations 
describing the collisions between the quasiparticles; 
\item ergodicity in a completely anisotropic trap: as shown by references \cite {Graham, GrahamD}, the 
classical motion of quasiparticles in a non isotropic harmonic trap with cylindrical symmetry 
is highly chaotic at energies $ \epsilon \approx \mu _ {\rm GP} $ but 
almost integrable when $ \epsilon \to 0 $ or $ \epsilon \to + \infty $. In a completely anisotropic trap, 
at temperatures neither too small nor too large with respect to $ \mu _ {\rm GP} / k_B $,
we can hope to complete the secular approximation by the hypothesis of ergodicity, which we will endeavor to show. 
\end{itemize} 

\medskip
Our article is articulated as follows. In section \ref{sec:formalisme_resultats}, after a few reminders about 
Bogoliubov's theory in a trap, we specify the state of the system and introduce the quantities to 
formally describe the phase spread, namely the derivative of the condensate phase operator and its time correlation function. 
In section \ref{sec:balistique}, we give an expression of the ballistic coefficient $ A $ in the thermodynamic limit in 
any harmonic trap (including isotropic), first in the most general state of system considered here, then in the simpler case of a 
statistical mixture of canonical ensembles of the same temperature $ T $. In the long section \ref{sec:micro}, we tackle the heart of the problem, 
calculating the correlation function $ C _ {\rm mc} (\tau) $ of $ \dd \hat {\theta} / \dd t $ in the microcanonical ensemble, 
which gives access in general 
terms to the sub-ballistic spreading terms of the phase, since they are independent of the state of the system in the thermodynamic limit at fixed average energy and fixed average number of particles. 
We first introduce the semiclassical limit in subsection \ref{subsec:formesemiclassique}, 
the Bogoliubov quasiparticles motion being treated 
classically but the bosonic field of quasiparticles remaining quantum~; the semiclassical form of $ \dd \hat {\theta} / \dd t $ is deduced from a 
correspondence principle. We then write, in subsection \ref{subsec:equacin}, 
kinetic equations on the quasiparticle occupation numbers in the classical phase space $ (\rr, \pp) $ 
and we show how, 
once linearized they formally lead to $C_ {\rm mc} (\tau) $. The problem remains formidable, since the occupation numbers 
depend on the six variables ($ \rr, \pp) $ and time. In the secular limit $ \gamma _ {\rm coll} \ll \omega_ \alpha $ and in the ergodic approximation 
for the quasiparticles motion (which excludes isotropic or cylindrically-symmetric traps), we 
reduce in subsection \ref{subsec:solution} to occupation numbers that are functions only of the energy of the classical motion $\epsilon $ and the time, which leads to 
explicit results on $ C _ {\rm mc} (\tau) $, on phase diffusion and, an interesting by-product, on the damping rate 
of Bogoliubov's modes in the trap, {\yvan in subsection \ref{subsec:resetdiscus} where we also evaluate the condensate phase shift due to particle losses}.
Finally, we make a critical discussion of the ergodic approximation in subsection 
\ref{subsec:discuss_ergo}, estimating in particular the error that it introduces on the quantities controlling the phase diffusion of the condensate. 
We conclude in section \ref{sec:conclusion}.

\section{Summary of the formalism and results} 
\label{sec:formalisme_resultats}

{\sl The derivative of the phase} - 
As we recalled in the introduction, the coherence time of a condensate is controlled by the dynamics of its 
phase operator $ \hat {\theta} (t) $ at times long with respect to the typical collision time $ \gamma _ {\rm coll} ^ {- 1} $ of quasiparticles.
The starting point of our study is therefore the expression of the temporal derivative of $ \hat {\theta} (t) $, smoothed temporally 
(that is, coarse-grained over a short time with respect to $ \gamma _ {\rm coll} ^ {- 1} $ but long with respect to the typical inverse frequency $ \epsilon_k ^ {\rm th} / \hbar $ of thermal quasiparticles).
As it has been established in all generality 
in reference \cite {CohFer}, to order one in the non-condensed fraction: 
\be
-\hbar \frac{\dd\hat{\theta}}{\dd t} = \mu_0(\hat{N}) + \sum_{k\in {\cal F}_+} \frac{\dd\epsilon_k}{\dd N} \:  \hat{n}_k \equiv \hat{\mu}
\label{eq:thetadot}
\ee
Here $ \mu_0 (N) $ is the chemical potential of the gas in the ground state and $ \hat {N} $ is the total number of particles operator. 
The sum over the generic quantum number $ k $ (it's not a wavenumber) 
deals with the Bogoliubov modes of eigenenergy $ \epsilon_k $, and $ \hat {n} _k $ is the operator number of quasiparticles 
in the $ k $ mode. The expression (\ref{eq:thetadot}) is a quantum version of the second Josephson relation:
its right-hand side is a chemical potential operator $ \hat {\mu} $ of the gas, 
since it is the adiabatic derivative (with the occupation numbers $ \hat {n} _k $ fixed) with respect to $ N $ 
of the Bogoliubov Hamiltonian 
\be
\hat{H}_{\rm Bog} = E_0(\hat{N}) + \sum_{k\in {\cal F}_+} \epsilon_k \hat{n}_k
\label{eq:H_Bog}
\ee
The Bogoliubov modes are of the are family $ \mathcal {F} _ + $, according to the terminology of reference \cite {CastinDum} in the sense 
that their modal functions $ (u_k (\rr) v_k (\rr)) $ are solutions of the eigenvalue equation 
\be
\epsilon_k \binom{|u_k\rangle}{|v_k\rangle} =  \left(\begin{array}{cc}
 H_{\rm GP}+Q g\rho_0(\hat{\rr}) Q &  Q g\rho_0(\hat{\rr}) Q \\
 -Q g\rho_0(\hat{\rr}) Q & -[ H_{\rm GP}+Q g\rho_0(\hat{\rr}) Q ]
 \end{array} \right)  \binom{|u_k\rangle}{|v_k\rangle} \equiv {\cal L}(\hat{\rr},\hat{\pp}) \binom{|u_k\rangle}{|v_k\rangle}
\label{eq:calL}
\ee
with the normalization condition $ \int \dd ^ 3r \, (| u_k (\rr) | ^ 2- | v_k (\rr) | ^ 2) = 1> 0 $. We took the wave function $ \phi_0 (\rr) $ 
of the condensate real, normalized to one ($ \int \dd ^ 3r \, \phi_0 ^ 2 (\rr) = 1 $), 
and written to the order zero in the non-condensed fraction, that is to the Gross-Pitaevskii approximation: 
\be
H_{\rm GP} |\phi_0\rangle = 0 \quad \mbox{with} \quad 
H_{\rm GP}=\frac{\hat{\pp}^2}{2m}+U(\hat{\rr})+g\rho_0(\hat{\rr})-\mu_{\rm GP}
\ee
so that at this order, the condensed density is $ \rho_0 (\rr) = N \phi ^ 2_0 (\rr) $. Here, $ g = 4 \pi \hbar ^ 2 a / m $ is the coupling constant, 
proportional to the $s$-wave scattering length $ a $ between bosons of mass $ m $, and $ U (\rr) = \sum_ \alpha m \omega_ \alpha ^ 2 r_ \alpha ^ 2/2 $ 
is the trapping potential. The projector $ Q $ projects orthogonally to $ | \phi_0 \rangle $ 
and ensures that $ | \phi_0 \rangle \bot | u_k \rangle $ and $ | \phi_0 \rangle \bot | v_k \rangle $ as it 
should be \cite{CastinDum}. Since the condensate is in its ground mode ($ \phi_0 $ minimizes the Gross-Pitaevskii energy functional), the $ \epsilon_k $ are positive. 

{\sl The state of the system} - Evaporative cooling in cold atom gases does not lead {\sl a priori} to any of the 
usual ensembles of statistical physics. To cover all reasonable cases, we therefore suppose that the gas is prepared at 
time $ 0 $ in a generalized ensemble, statistical mixture of eigenstates $ | \psi_ \lambda \rangle $ 
of the complete Hamiltonian $ \hat {H} $, with $ N_ \lambda $ particles and energy $ E_ \lambda $, hence of density operator \be
\hat{\sigma} = \sum_\lambda \Pi_\lambda |\psi_\lambda\rangle \langle \psi_\lambda|
\label{eq:OpDens}
\ee
with, as the only restriction, the existence of narrow laws on $ E_ \lambda $ and $ N_ \lambda $, of variances and covariance not growing faster 
than the averages $ \bar {E} $ and $ \bar {N} $ in the thermodynamic limit. 

{\sl Average phase shift} - Let's average expression (\ref{eq:thetadot}) in the steady state $ | \psi_ \lambda \rangle $. At the right-hand side appears 
the expectation  in $ | \psi_ \lambda \rangle $ of the chemical potential operator. 
Because of the interactions between Bogoliubov quasiparticles, 
the $ N $ body system is expected to be ergodic in the quantum sense of the term, that is, to obey to the,
so called in the Anglo-American literature, ``Eigenstate Thermalisation Hypothesis" 
(see references \cite {Deutsch1991, Srednicki1994, Olshanii2008}), 
\be
\langle\psi_\lambda| \hat{\mu} | \psi_\lambda\rangle = \mu_{\rm mc} (E_\lambda,N_\lambda) 
\ee
where $ \mu _ {\rm mc} (E, N) $ is the chemical potential in the microcanonical ensemble of energy $ E $ with $ N $ particles. 
For a large system, it suffices to expand to first order in the fluctuations, to obtain: 
\be 
\mu_{\rm mc}(E_\lambda,N_\lambda) = \mu_{\rm mc}(\bar{E},\bar{N}) + (E_\lambda-\bar{E}) \partial_E \mu_{\rm mc}(\bar{E},\bar{N})
+(N_\lambda-\bar{N}) \partial_N \mu_{\rm mc}(\bar{E},\bar{N}) + O(1/\bar{N})
\ee
It remains to average on the states $ | \psi_ \lambda \rangle $ with the weights $ \Pi_ \lambda $ as in equation 
(\ref{eq:OpDens}) to 
get the first brick of the time coherence function (\ref{eq:g1}), that is the average phase shift: 
\be
\langle\hat{\theta}(t)-\hat{\theta}(0)\rangle = -\mu_{\rm mc}(\bar{E},\bar{N}) t/\hbar
\label{eq:dephmoy}
\ee
with an error $ O (1 / \bar {N}) $ on the coefficient of $ t $. 

{\sl Average quadratic phase shift} - Proceeding in the same way for the second moment of the phase shift of the condensate, we 
find as it is written implicitly in \cite {PRAlong, CohFer} that 
\be
\mbox{Var}\,[\hat{\theta}(t)-\hat{\theta}(0)] = A t^2 + 2\int_0^t \dd\tau\, (t-\tau)\,{\yvan \re}\, C_{\rm mc}(\tau)
\label{eq:vartot}
\ee
with the ballistic coefficient 
\be
A = \mbox{Var}[(N_\lambda-\bar{N})\partial_N\mu_{\rm mc}(\bar{E},\bar{N}) + (E_\lambda-\bar{E}) \partial_E \mu_{\rm mc}(\bar{E},\bar{N})]
/\hbar^2
\label{eq:grandA}
\ee
and the correlation function of the phase derivative in the 
microcanonical ensemble of energy  $ \bar {E} $ and $ \bar {N} $ particles:
\be
C_{\rm mc}(\tau) = \left\langle \frac{\dd\hat{\theta}}{\dd t}(\tau) \frac{\dd\hat{\theta}}{\dd t}(0)\right\rangle_{\rm mc} - 
\left\langle \frac{\dd\hat{\theta}}{\dd t}\right\rangle_{\rm mc}^2
\label{eq:defCmc}
\ee
 This completes our formal knowledge of $ g_1 (t) $. 

In view of future experimental observations, however, it remains to calculate {\sl explicitly} $ A $ and $ C _ {\rm mc} (\tau) $ 
for a harmonically trapped system. It will be necessary in particular to verify that $ C _ {\rm mc} (\tau) $ in the trapped case 
decreases fast enough so that one finds a diffusive law (\ref{eq:diff}) as in the spatially homogeneous case.

\section {Calculation of the ballistic coefficient in the phase shift variance} 
\label{sec:balistique}

{\sl In the generalized statistical ensemble} - 
To calculate the average phase shift (\ref{eq:dephmoy}) and the ballistic coefficient (\ref{eq:grandA}) in the general case, 
we must know the 
microcanonical chemical potential $ \mu _ {\rm mc} (\bar {E}, \bar {N}) $ and its derivatives in the harmonic trap. 
At the thermodynamic limit, $ \mu _ {\rm mc} $ coincides with the chemical potential 
$ \mu _ {\rm can} $ in the canonical ensemble of temperature $ T $ and number of particles $ \bar {N} $, more convenient to calculate,
provided that the temperature $ T $ is adjusted so that there is equality of mean energies $ E _ {\rm can} (T, \bar {N}) $ and $ \bar {E} $. 
In other words, 
\be
\mu_{\rm mc}(E_{\rm can}(T,\bar{N}),\bar{N}) \sim \mu_{\rm can}(T,\bar{N})
\ee
one just takes the derivative of this relation with respect to $ T $ or $ \bar {N} $ to get the useful derivatives of $ \mu _ {\rm mc} $, 
then one replaces $ E _ {\rm can} $ by $ \bar {E} $, to obtain: 
\bea
\label{eq:acalc1}
\partial_E \mu_{\rm mc}(\bar{E},\bar{N}) &\sim &  \frac{\partial_T \mu_{\rm can}(T,\bar{N})}{\partial_T E_{\rm can}(T,\bar{N})} \\
\partial_N \mu_{\rm mc}(\bar{E},\bar{N}) &\sim & \partial_N \mu_{\rm can}(T,\bar{N}) 
-\frac{\partial_N E_{\rm can}(T,\bar{N})}{\partial_T E_{\rm can}(T,\bar{N})} \partial_T \mu_{\rm can}(T,\bar{N})
\label{eq:acalc2}
\eea

At the first order in the non-condensed fraction, the canonical chemical potential is deduced from the free energy $ F $ of the ideal gas 
of Bogoliubov quasiparticles of Hamiltonian (\ref{eq:H_Bog}) by the usual thermodynamic relation 
$ \mu _ {\rm can} = \partial_N F $. The free energy is a simple functional of the density of states $ \rho(\epsilon)$ of quasiparticles,
\be
F(T,\bar{N})=E_0(\bar{N}) + k_B T \int_0^{+\infty} \dd\epsilon\,\rho(\epsilon) \ln\left(1-e^{-\beta\epsilon}\right)
\ee
with $ \beta = 1 / k_B T $.  
At the thermodynamic limit, the ground state energy {\yvan $E_0$} {\yvan of the gas} in the harmonic trap is deduced from that of the homogeneous system 
\cite {LY} by a local density approximation, and the density of states $ \rho (\epsilon) $ is obtained by taking 
the classical limit $ \hbar \to 0 $, thanks to {\yvan inequality} (\ref{eq:condval1}) \cite {Stringari_revue}: 
\be
\rho(\epsilon)= \int \frac{\dd^3r\dd^3p}{(2\pi\hbar)^3} \delta(\epsilon-\epsilon(\rr,\pp))
\label{eq:rhoeps}
\ee
The classical Hamiltonian $ \epsilon (\rr, \pp) $ is the positive eigenvalue of the Bogoliubov's $ 2 \times 2 $ matrix of equation 
(\ref{eq:calL}) with the position $ \rr $ and the momentum $ \pp $ treated classically \footnote{The projector $ Q $, 
projecting on a space of codimension one, can be omitted at the thermodynamic limit.} and the condensed density $ \rho_0 (\rr) $ written 
at the classical limit that is in the Thomas-Fermi approximation: 
\be
g \rho_0^{\rm TF}(\rr) = \left\{ 
\begin{array}{lll} 
\mu_{\rm TF}-U(\rr) \equiv \mu_{\rm loc}(\rr) \quad \quad \quad \quad & \mbox{if} & \quad U(\rr) <  \mu_{\rm TF} \\
0 \quad \quad & \mbox{elsewhere} &
\end{array}  \right.
\ee
Here, the Thomas-Fermi chemical potential, classical limit of $ \mu _ {\rm GP} $ of Gross-Pitaevskii, is 
\be
\mu_{\rm TF}=\frac{1}{2}\hbar\bar{\omega}[15\bar{N}a(m\bar{\omega}/\hbar)^{1/2}]^{2/5}
\label{eq:mutf}
\ee
and $ \bar {\omega} = (\omega_x \omega_y \omega_z) ^ {1/3} $ is the geometric average of the trapping angular frequencies. 
We deduce that 
\be
\epsilon(\rr,\pp) =\left\{ 
\begin{array}{lll} 
\left\{ {\displaystyle\frac{p^2}{2m}}\left[{\displaystyle\frac{p^2}{2m}}+2 \mu_{\rm loc}(\rr)\right] \right\}^{1/2} \quad \quad & \mbox{if} & \quad U(\rr) <  \mu_{\rm TF} \\
{\displaystyle\frac{p^2}{2m}}+ U(\rr)-\mu_{\rm TF} \quad \quad & \mbox{elsewhere} &
\end{array} \right.
\label{eq:eps_rp}
\ee
The six-fold integral (\ref{eq:rhoeps}) has been calculated in reference \cite {PRLLincoln}.\footnote{The case of an anisotropic harmonic trap comes down to the 
isotropic case treated in \cite {PRLLincoln} by performing the change of variable (with unit Jacobian) $ r_ \alpha = \lambda_ \alpha r '_ \alpha $, with 
$ \omega_ \alpha \lambda_ \alpha = \bar {\omega} $, such that $ U (\rr) = \frac{1} {2} m \bar {\omega} ^ 2 r '^ {2} $. 
\label{note:isotropisation}} 
Here we give the result in a somewhat more compact form: 
\bea
\rho(\epsilon) &=& \frac{\mu_{\rm TF}^2}{(\hbar\bar{\omega})^3} f(\check{\epsilon}\equiv \epsilon/\mu_{\rm TF}) \\
f(\check{\epsilon}) &=& \frac{1}{\pi}\left[-2\sqrt{2}\epsc^2 \acos\frac{\epsc-1}{(1+\epsc^2)^{1/2}} 
              +2\sqrt{2}\epsc\ln\frac{1+\sqrt{2\epsc}+\epsc}{(1+\epsc^2)^{1/2}}
              +\sqrt{\epsc}(5\epsc-1)+(1+\epsc)^2\acos\frac{1}{(1+\epsc)^{1/2}}\right]
\label{eq:deff}
\eea
We finally obtain the canonical chemical potential 
\be
\mu_{\rm can}(T,\bar{N})=\mu_0(\bar{N})+\frac{6k_B T}{5\bar{N}} \left(\frac{\mu_{\rm TF}}{\hbar\bar{\omega}}\right)^3
\int_0^{+\infty} \dd\epsc f(\epsc) \ln\left(1-\eee^{-\check{\beta}\epsc}\right)
+\frac{2\mu_{\rm TF}}{5\bar{N}} \left(\frac{\mu_{\rm TF}}{\hbar\bar{\omega}}\right)^3 
\int_0^{+\infty} \dd\epsc \frac{f(\epsc)\epsc}{\eee^{\check{\beta}\epsc}-1} 
\label{eq:mucanfin}
\ee
with the contribution of the ground state \cite {Stringari_revue} 
\be
\mu_0(\bar{N})= \mu_{\rm TF} \left[1+\pi^{1/2} \left({\mu_{\rm TF}a^3}/{g}\right)^{1/2}\right]
\ee
When one takes the derivative of (\ref{eq:mucanfin}) with respect to $ T $ and $ \bar {N} $ to evaluate expressions (\ref{eq:acalc1}) and (\ref{eq:acalc2}), 
one will remember that $ \check {\beta} = \mu_ { \rm TF} / k_B T $ depends on $ \bar {N} $ through $ \mu _ {\rm TF} $. For brevity, we do 
not give the result here. 

\begin{figure} [t] 
\centerline {\includegraphics [width = 0.4 \textwidth,clip=] {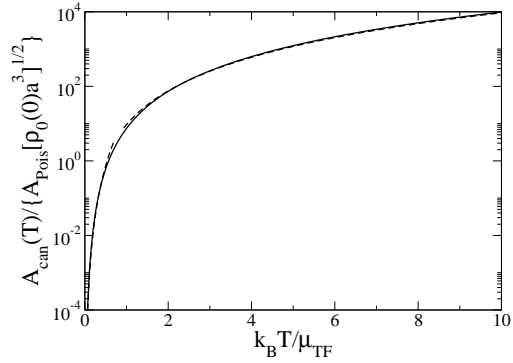}} 
\caption {Coefficient of ballistic spreading (\ref{eq:bali}) of the condensate phase in the long time limit with respect to the collision time 
$ \gamma _ {\rm coll} ^ {- 1} $ of quasiparticles, for a gas of $ \bar {N} $ bosons prepared in the canonical ensemble in an 
harmonic trap (isotropic or not), depending on the temperature. The result is at the thermodynamic limit where 
the trapping angular frequencies $ \omega_ \alpha $ are negligible compared to Thomas-Fermi's chemical potential $ \mu _ {\rm TF} $ (\ref{eq:mutf}). 
Full line: second term of equation 
(\ref{eq:Apart}), deduced from the canonical chemical potential (\ref{eq:mucanfin}) 
to the Bogoliubov approximation (weak interactions, $ T \ll T_c $). Dashes: 
equivalents at low and high temperature (dominant terms of equations (\ref{eq:equivbT}, \ref{eq:equivhT})). 
The division of $ A _ {\rm can} (T) $ 
by the small parameter (\ref{eq:rhoa3}) of Bogoliubov's theory and 
by the value (\ref{eq:APois}) of the ballistic coefficient for Poissonian fluctuations of $ N $ 
leads to a universal function of $ k_B T / \mu _ {\rm TF} $. } 
\label{fig:Acan} 
\end{figure}

{\sl In a slightly less general ensemble} - A simpler expression 
\footnote{The general expression (\ref{eq:grandA}) of $ A $ is a little tricky to grasp. Since 
the energy of the ground state depends on $ N $, fluctuations of $ N $ mechanically cause 
energy fluctuations. For example, if $ N $ fluctuates at $ T = 0 $ (in each subspace of fixed $ N $, the system is in the 
ground state), we can, to find $ A (T = 0) $ of equation (\ref{eq:Apart}) from equation (\ref{eq:grandA}), 
use the fact that $ E_ \lambda- \bar {E} = (N_ \lambda- \bar {N}) \mu_0 (\bar {N}) + O (\bar {N} ^ 0) $ 
and that $ \partial_E \mu _ {\rm mc} (\bar {E}, \bar {N}) \underset {T \to 0} {\sim} -2 / (25 \bar {N}) $, whose report in (\ref{eq:acalc2}) 
gives $ \partial_N \mu _ {\rm mc} (\bar {E}, \bar {N}) \underset {T \to 0} {\sim} \partial_N \mu_0 (\bar {N}) + 2 \mu_0 (\bar {N}) / (25 \bar {N}) $. 
} 
of the ballistic coefficient $ A $ can be obtained when 
the state of the system is a statistical mixture of canonical ensembles of the same temperature $ T $ but of 
variable number of particles. By expressing the various coefficients in (\ref{eq:grandA}, \ref{eq:acalc1}, \ref{eq:acalc2}) 
as derivatives of the free energy $ F (T, \bar {N}) $ with respect to $ \bar {N} $ and $ T $, and remembering the expression 
$ \mbox {Var} _ {\rm can} E = k_B T ^ 2 \partial_T E _ {\rm can} $ of the variance of the energy in the canonical ensemble, we find 
to the dominant order $ 1 / \bar {N} $ that 
\be
A(T) = (\mbox{Var}\,N) \left(\frac{\partial_N\mu_{\rm can}(T,\bar{N})}{\hbar}\right)^2
+\frac{k_B T^2 \left[\partial_T\mu_{\rm can}(T,\bar{N})\right]^2}{\hbar^2 \partial_T E_{\rm can}(T,\bar{N})}
\label{eq:Apart}
\ee
At zero temperature, only the first term contributes, and we find the prediction of references \cite {Walls, DalibardCastin} pushed 
to order one in the non-condensed fraction $ f _ {\rm nc} $. 
A $ T \neq $ 0 but in the absence of fluctuations of $ N $, only the second term contributes; it is none other than the ballistic coefficient 
$ A _ {\rm can} (T) $ in the canonical ensemble. 
In the validity regime of the Bogoliubov approximation, $ f _ {\rm nc} \ll 1 $, the 
chemical potential $ \mu_ {\rm can} (T, \bar {N}) $ of the gas remains close to that of a Thomas-Fermi pure condensate, so that 
\be
\partial_N \mu_{\rm can}(T,\bar{N}) = \partial_N \mu_{\rm TF} + O\left(\frac{f_{\rm nc}}{\bar{N}}\right)
\ee
$ \partial_T \mu _ {\rm can} (T, \bar {N}) $ is immediately first-order in $ f _ {\rm nc} $, and the same goes for
the second term in equation (\ref{eq:Apart}). It is therefore only for strongly subpoissonian fluctuations of $ N $ 
($ \mbox {Var} \, N \ll \mbox {Var} _ {\rm Pois} N \equiv \bar {N} $) that the second term of (\ref{eq:Apart}), that is the effect 
of thermal fluctuations, is not dominated by the first one. Assuming this condition satisfied in the experiment, 
we represent in figure \ref{fig:Acan} the canonical coefficient $ A _ {\rm can} (T) $ scaled by the $ A _ {\rm Pois} $ value 
of $ A $ in a pure condensate with  Poissonian fluctuations of $ N $,
\be
A_{\rm Pois} = \bar{N} \left(\frac{\partial_N \mu_{\rm TF}}{\hbar}\right)^2
\label{eq:APois}
\ee
all divided by the small parameter of the Bogoliubov theory at zero temperature,\footnote{One sometimes prefers to take 
as a small parameter $ 1 / [\rho_0 (\mathbf {0 }) \xi ^ 3] $, where the healing length $ \xi $ of the condensate at the center of the trap 
is such that $ \hbar ^ 2 / (m \xi ^ 2) = \mu _ {\rm TF} $. One can easily go from one small parameter to the other using the 
relation $ [\rho_0 (\mathbf {0}) a ^ 3] ^ {1/2} \rho_0 (\mathbf {0}) \xi ^ 3 = 1 / (8 \pi ^ {3/2}) $.} 
proportional to $ f _ {\rm nc} (T = 0) $: 
\be
[\rho_0(\mathbf{0})a^3]^{1/2} = \frac{2\sqrt{2}}{15\pi^{1/2}\bar{N}} \left(\frac{\mu_{\rm TF}}{\hbar\bar{\omega}}\right)^3
\label{eq:rhoa3}
\ee
The ratio thus formed is a universal function of $ k_B T / \mu _ {\rm TF} $. 
From the low and high energy expansions of the quasiparticle density of states, 
\bea
\label{eq:fbasse}
f(\epsc) &\underset{\epsc\to 0}{=}& \frac{32}{3\pi} \epsc^{3/2} -2\sqrt{2}\,\epsc^2 + O(\epsc^{5/2}) \\
f(\epsc) &\underset{\epsc\to+\infty}{=}& \frac{1}{2} \epsc^2 + \epsc + \frac{1}{2} +O(\epsc^{-1/2})
\label{eq:fhaute}
\eea
we obtain low and high temperature expansions ($ \check {T} = k_B T / \mu _ {\rm TF} = 1 / \check {\beta} $) 
\bea
\label{eq:equivbT}
\frac{A_{\rm can}(T)}{A_{\rm Pois}[\rho_0(\mathbf{0})a^3]^{1/2}} &\underset{\check{T}\to 0}{=}& \frac{21\zeta(7/2)}{\sqrt{2}}
\check{T}^{9/2} \left[1+\frac{4\sqrt{2}\pi^{9/2}}{525\zeta(7/2)}\check{T}^{1/2}+O(\check{T})\right] \\
&\underset{\check{T}\to+\infty}{=}&\frac{15\pi^{1/2}}{2\sqrt{2}}\frac{3\zeta(3)^2}{4\zeta(4)}\check{T}^3 
\left[1+\check{\beta}\left(\frac{4\zeta(2)}{3\zeta(3)}-\frac{\zeta(3)}{2\zeta(4)}\right)+O(\check{\beta}^{3/2})\right]
\label{eq:equivhT}
\eea
whose dominant terms \footnote{In the value window of figure \ref{fig:Acan}, in practice $ 1/10 \leq \check {T} \leq $ 10, 
the inclusion of subdominant terms does not usefully approximate the exact result.} 
are shown as dashed lines on figure \ref{fig:Acan}. Let us note a particularly simple and beautiful reworking of the 
high temperature equivalent, accidentally already operational at $ k_B T / \mu _ {\rm TF} \geq 2 $: 
\be
\frac{A_{\rm can}(T)}{A_{\rm Pois}}\underset{k_BT\gg\mu_{\rm TF}}{\sim}\frac{3\zeta(3)}{4\zeta(4)}\left(\frac{T}{T_c^{(0)}}\right)^3
\ee
where $ T_c ^ {(0)} $ is the critical temperature of an ideal gas of bosons in a harmonic trap at the thermodynamic limit, 
$ k_B T_c ^ {(0)} = \hbar \bar {\omega} [\bar {N} / \zeta (3)] ^ {1/3 } $. In this limit, $ A _ {\rm can} (T) $ is therefore lower than $ A _ {\rm Pois} $ by a factor 
proportional to the non-condensed fraction $ (T / T_c ^ {(0)}) ^ 3 \ll 1 $. 

\section{Variance of the condensate phase shift in the microcanonical ensemble} 
\label{sec:micro} 
Here we calculate the correlation function of $ \dd \hat {\theta} / \dd t $, namely $ C _ {\rm mc} (\tau) $, for a system prepared in the microcanonical ensemble, using at the thermodynamic limit $ \frac{\hbar \omega_ \alpha} {\mu _ {\rm TF}} \to0 $ a semiclassical description of the quasiparticles and taking into account the effect of their interaction by 
quantum Boltzmann kinetic equations on their classical phase space distribution $n (\rr, \pp) $. 

\subsection {Semi-classical form of Bogoliubov's Hamiltonian and $ {\dd \hat {\theta}} / {\dd t} $} 
\label{subsec:formesemiclassique}
In the semiclassical description, the motion of Bogoliubov quasiparticles is treated {\it classically}, that is that they have at each moment a  well defined position $ \rr $  and momentum $ \pp $ \cite {Stringari_revue}, whose evolution in phase space derives from the Hamiltonian $ \epsilon (\rr, \pp) $ given in equation (\ref{eq:eps_rp}) \cite{Graham}: 
\bea
\label{eq:drdt}
\frac{\dd\rr}{\dd t} &=& \partial_{\pp} \, \epsilon(\rr,\pp) \\
\frac{\dd\pp}{\dd t} &=& -\partial_{\rr} \, \epsilon(\rr,\pp) 
\label{eq:dpdt}
\eea
but we treat {\it in a quantum way} the bosonic field of quasiparticles by introducing their occupation numbers operators 
$ \hat {n} (\rr, \pp) $ in the phase space, {\yvan which allows us to take into account the discrete nature of the numbers of quasiparticles
and the quantum statistical effects (with the Bose law rather than the equipartition law of the classical field at equilibrium).}

In this semiclassical limit, the Bogoliubov Hamiltonian (\ref{eq:H_Bog}) (without interaction between quasiparticles) is written immediately
\be
H_{\rm Bog}^{\rm sc} = E_0(\hat{N}) + \int \frac{\dd^3r\, \dd^3p}{(2\pi \hbar)^3} \: \epsilon(\rr,\pp) \, \hat{n}(\rr,\pp)
\label{eq:H_Bog_sc}
\ee
One might think, given formula (\ref{eq:thetadot}), that $ {\dd \hat {\theta}} / {\dd t} $ admits a similar writing, with 
$ \epsilon (\rr , \pp) $ replaced by $ \frac {\dd} {\dd N} \epsilon (\rr, \pp) $. This is not so, the reason being that {\yvan the derivative} $ \frac {\dd} {\dd N} \epsilon (\rr, \pp) $ is not constant on the classical trajectory. The $ {\dd \hat {\theta}} / {\dd t} $ operator is part of a general class of so-called Fock quantum observables (diagonal in the Fock basis of quasiparticles thus - here linear - functionals of Bogoliubov's occupation numbers): 
\be
\hat{A}=\sum_{k\in {\cal F}_+} a_k \hat{n}_k \quad \quad \mbox{with} \quad 
a_k = (\langle u_k|, \langle v_k|) \, {\cal A}(\hat{\rr},\hat{\pp}) \, \binom{|u_k\rangle}{|v_k\rangle}
\label{eq:A}
\ee 
where $ {\cal A} (\hat {\rr}, \hat {\pp}) $ is a $ 2 \times 2 $ hermitian matrix operator and $ a_k$ its average in the Bogoliubov mode of eigenenergy $ \epsilon_k $. The observable $ {\dd \hat {\theta}} / {\dd t} $ corresponds to the choice 
$ {\cal A} _ {\dot {\theta}} = \sigma_z \frac{\dd}{\dd N} {\cal L} $ 
where $ \sigma_z $ is the third Pauli matrix and $ {\cal L} (\hat {\rr}, \hat {\pp}) $ is the operator appearing in equation (\ref{eq:calL}). By using Hellmann-Feynman's theorem \footnote{The theorem is here generalized to the case of a non-Hermitian operator $ {\cal L} $, $ (\langle u_k |, - \langle v_k |) $ being the dual vector of the eigenvector $ (| u_k \rangle, | v_k \rangle) $ of $ {\cal L} $.}, we have indeed 
\be
(\langle u_k|, -\langle v_k|) \left(\frac{\dd}{\dd N} \, {\cal L}\right)  \binom{|u_k\rangle}{|v_k\rangle} = \frac{\dd\epsilon_k}{\dd N}
\ee
For these Fock operators we use the semiclassical correspondence principle 
\be
\hat{A}^{\rm sc} = \int \frac{\dd^3r\, \dd^3p}{(2\pi \hbar)^3} \: \overline{a(\rr,\pp)} \, \hat{n}(\rr,\pp)
\label{eq:Asc}
\ee
where $ a (\rr, \pp) = (U (\rr, \pp), V (\rr, \pp)) \, {\cal A} (\rr, \pp) \, \binom{U (\rr, \pp)} {V (\rr, \pp)} $, $ {\cal A} (\rr, \pp) $ being the classical equivalent 
 of $ { \cal A} (\hat {\rr}, \hat {\pp}) $, and $ \overline {a (\rr, \pp)} $ represents the time average of $ a (\rr, \pp) $ on the only classical trajectory passing through $ (\rr, \pp) $ at time $ t = 0 $: 
\be
\overline{a(\rr,\pp)} \equiv \lim_{t \to {\yvan +}\infty} \frac{1}{t} \int_0^t \, \dd\tau \,  a(\rr(\tau),\pp(\tau))
\ee
The vector $ (U (\rr, \pp) , V (\rr, \pp)) $, normalized according to the condition $ U ^ 2-V ^ 2 = 1 $, is eigenvector of the classical equivalent 
$ {\cal L} (\rr, \pp) $ of  $ {\cal L} (\hat {\rr}, \hat {\pp}) $ with eigenvalue $ \epsilon (\rr, \pp) $:
\be
\binom{U(\rr,\pp)}{V(\rr,\pp)} = \left\{ 
\begin{array}{lll} 
\left(
\begin{array}{l}
\frac{1}{2} \left[ \left({\displaystyle\frac{p^2/2m}{\epsilon(\rr,\pp)}} \right)^{1/2} + 
\left({\displaystyle\frac{p^2/2m}{\epsilon(\rr,\pp)}} \right)^{-1/2}\right] \\\\
\frac{1}{2} \left[ \left({\displaystyle\frac{p^2/2m}{\epsilon(\rr,\pp)}} \right)^{1/2} -  \left({\displaystyle\frac{p^2/2m}{\epsilon(\rr,\pp)}} \right)^{-1/2}\right]
\end{array} 
\right)
\quad \quad \quad \quad & \mbox{if} & \quad U(\rr) <  \mu_{\rm TF} \\\\
{\displaystyle\binom{1}{0}} & \mbox{elsewhere} &
\end{array}
\right.
\ee
At the basis of this correspondence principle lies the idea that the equivalent of a stationary quantum mode 
$ (| u_k \rangle, | v_k \rangle) $ in the classical world is a classical trajectory of the same energy, itself also stationary as a whole by temporal evolution. To the quantum expectation $ a_k $ of the observable $ {\cal A} (\hat {\rr}, \hat {\pp}) $ 
in the mode $ (| u_k \rangle, | v_k \rangle) $ thus we must associate {\it an average} over a trajectory of the expectation 
$ a (\rr, \pp) $ of the classical equivalent $ {\cal A} (\rr, \pp) $ in the local mode $ (U (\rr, \pp), V (\rr, \pp)) $. We therefore retain for the semiclassical version of the derivative of the condensate phase operator: 
\be
-\hbar \frac{\dd\hat{\theta}^{\rm sc}}{\dd t} = \mu_0(\hat{N}) + \int \frac{\dd^3r\, \dd^3p}{(2\pi \hbar)^3} \: 
\overline{ \frac{\dd\epsilon(\rr,\pp)}{\dd N} }\:  \hat{n}(\rr,\pp)
\label{eq:thetadot_sc}
\ee
Here, let us repeat it, the expectation $ a (\rr, \pp) = \frac {\dd \epsilon (\rr , \pp)} {\dd N} $ is not a constant of motion, unlike 
$ \epsilon (\rr, \pp) $, so we can not omit as in (\ref{eq:H_Bog_sc}) the temporal average. 

\subsection{About the usefulness of kinetic equations in calculating the correlation function of $ {\dd \hat {\theta}} / {\dd t} $} 
\label{subsec:equacin} 

\begin{figure}[t] 
\begin{center}
\begin{tabular}{ccccccc} 
Beliaev direct & \quad \quad & Beliaev inverse & \quad \quad \quad \quad & Landau direct & \quad \quad & Landau inverse \\
\includegraphics [width = 0.11 \textwidth,clip=] {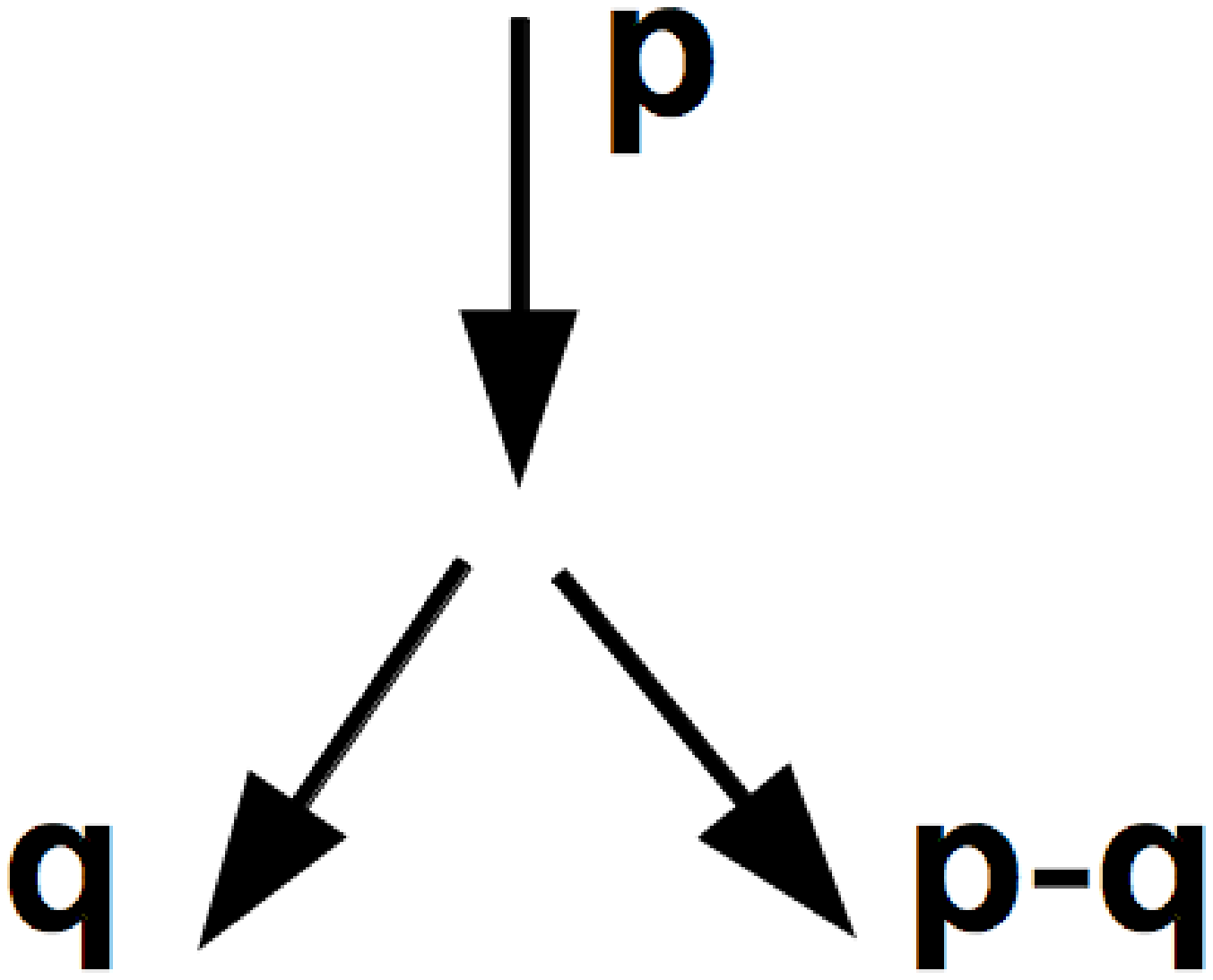} & & \includegraphics [width = 0.1 \textwidth,clip=] {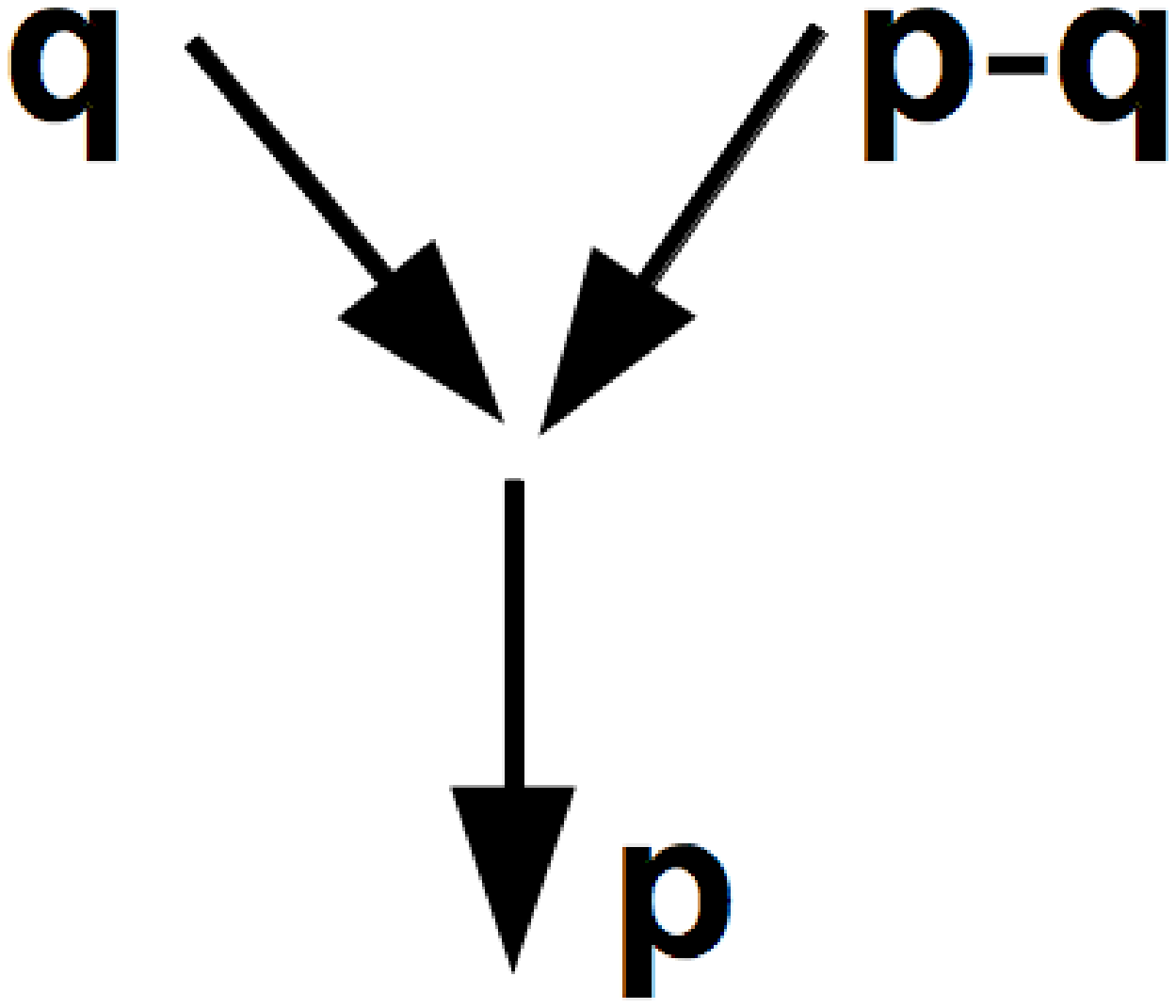} & & 
\includegraphics [width = 0.08 \textwidth,clip=] {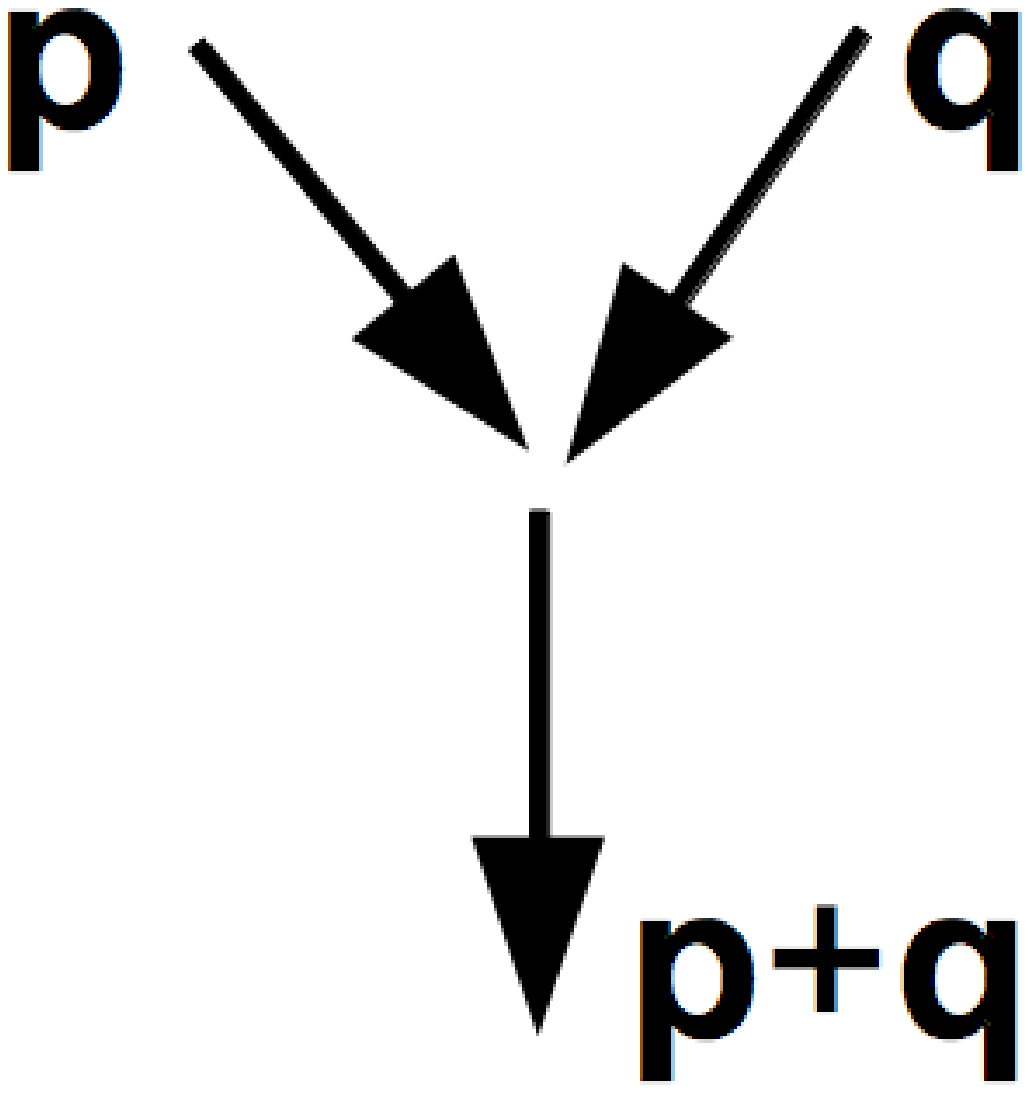} & & \includegraphics [width = 0.09 \textwidth,clip=] {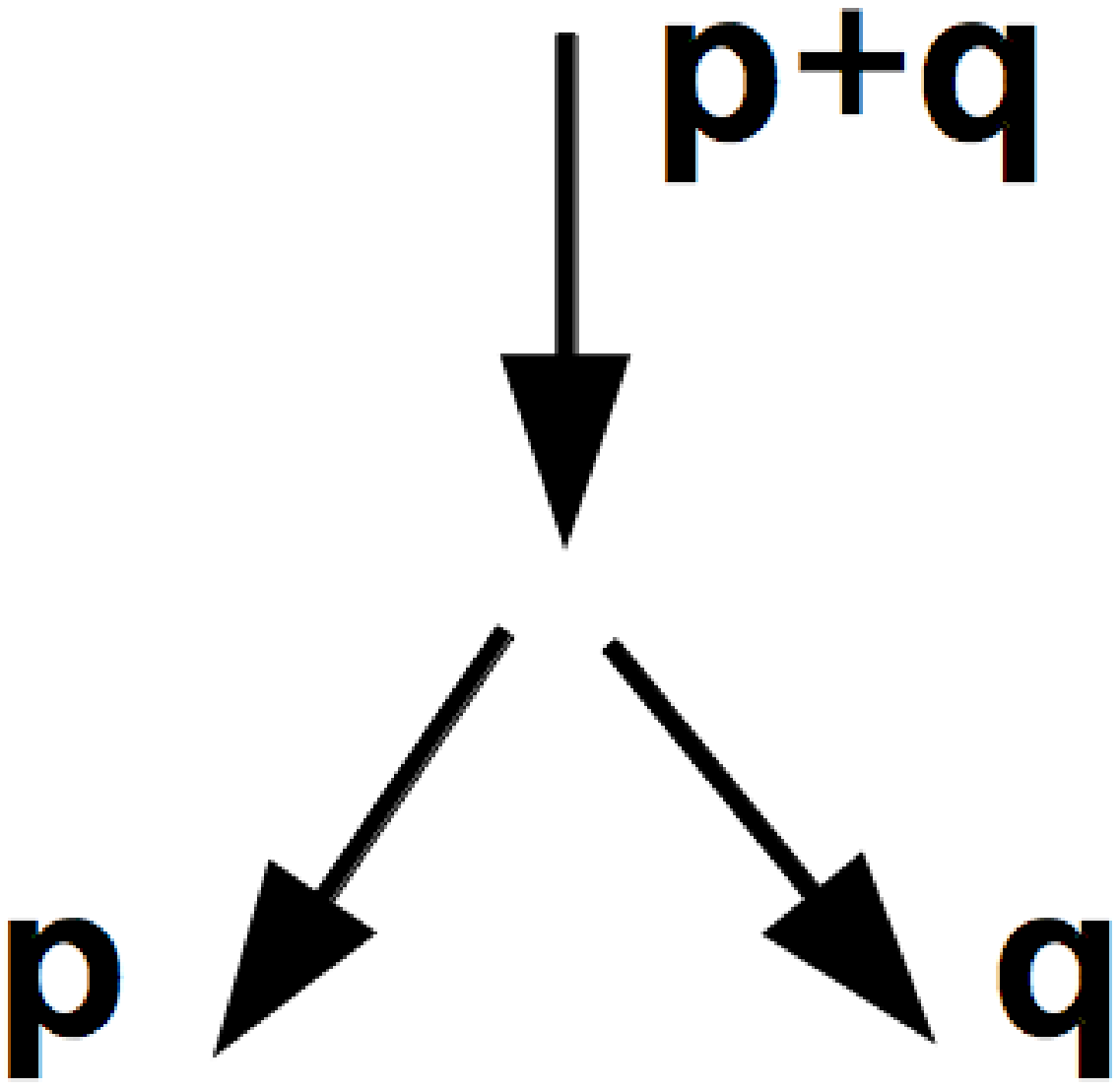} 
\end{tabular} 
\begin{tabular} {ccc} 
amplitude $ 2g \rho_0 ^ {1/2} (\rr) \, {\cal A} _ {\qq, | \pp- \qq |} ^ {\pp} (\rr) $ & \quad \quad \quad \quad \quad & amplitude $ 2g \rho_0 ^ {1/2} (\rr) \, {\cal A} _ {\pp, \qq} ^ {| \pp + \qq |} (\rr) $ 
\end{tabular} 
\end{center} 
\caption{Beliaev and Landau process involving three quasiparticles and the corresponding coupling amplitudes.} 
\label {fig:LK} 
\end{figure}

We must determine, in the semiclassical limit, the correlation function of $ {\dd \hat {\theta}} / {\dd t} $, for a system prepared in the microcanonical ensemble. Given {\yvan equations (\ref{eq:defCmc}) and (\ref{eq:thetadot_sc})} we must calculate 
\be
C_{\rm mc}^{\rm sc}(\tau)=\int \frac{\dd^3r\, \dd^3p}{(2\pi \hbar)^3} \int  \frac{\dd^3r'\, \dd^3p'}{(2\pi \hbar)^3} \: 
\overline{ \frac{\dd\epsilon(\rr,\pp)}{\hbar \dd N} }\:  \overline{ \frac{\dd\epsilon(\rr',\pp')}{\hbar \dd N} }\:  
\langle  \delta \hat{n}(\rr,\pp,\tau) \,  \delta \hat{n}(\rr',\pp',0)\rangle
\label{eq:C_sc}
\ee 
where $ \langle \ldots \rangle $ represents the mean in the state of the system and where we introduced the fluctuations of the occupation number operators in phase space at time $ \tau $, 
\be
 \delta \hat{n}(\rr,\pp,\tau) =  \hat{n}(\rr,\pp,\tau) -  \bar{n}(\rr,\pp)
\ee
The microcanonical ensemble can be seen in the semiclassical phase space as a constant-energy statistical mixture of Fock states 
$ | {\cal F} \rangle = | n (\rr '', \pp '') _ {(\rr '', \pp '') \in \mathbb {R} ^ 6} \rangle $, eigenstates of $ H_ {\rm Bog} ^ {\rm sc} $, 
where all $ n (\rr '', \pp '') $ are integers. It is assumed at first that the system is prepared in one of such Fock states 
$ | {\cal F} \rangle $ at the initial time $ t = 0 $, eigenstate of $ \delta \hat {n} (\rr ', \pp', 0) $ with the eigenvalue 
$ n(\rr',\pp') - \bar{n}(\rr',\pp')$;  it remains then to calculate in equation (\ref{eq:C_sc}) the quantity 
\be
\langle {\cal F} | \delta \hat{n}(\rr,\pp,\tau) |{\cal F}\rangle=n(\rr,\pp,\tau)-\bar{n}(\rr,\pp)\equiv  \delta n(\rr,\pp,\tau)
\ee
at $ \tau> 0 $, that is the evolution of the mean occupation numbers {\yvan $n(\rr,\pp,\tau)$} in phase space, their initial values being known, 
taking into account $ (i) $ the Hamiltonian 
quasiparticle transport and $ (ii) $ the effect of quasiparticle collisions by the Beliaev or Landau three-quasiparticles processes \footnote{Four-quasiparticle processes, of higher order in the non-condensed fraction, are assumed here negligible.} 
represented in figure~\ref{fig:LK}. 
This is exactly what {\yvan the usual Boltzmann-type quantum} kinetic equations can do, 
{\yvan with the difference that the semiclassical distribution function $n(\rr,\pp,\tau)$ does not correspond here to a local thermal equilibrium state
of the system, but to the mean occupation number at time $\tau$ knowing that the initial state of the system is a quasiparticle Fock state.}
The evolution equation of the average occupation numbers $ n (\rr, \pp, \tau) $ is of the form 
\be
\frac{\rm D}{{\rm D}\tau} n(\rr,\pp,\tau) + I_{\rm coll}(\rr,\pp,\tau) =0
\label{eq:eqcin}
\ee
The first term is the convective derivative resulting from the classical Hamilton equations: 
\be
\frac{\rm D}{{\rm D}\tau} = \partial_\tau + \partial_\pp \epsilon(\rr,\pp) \cdot \partial_\rr - \partial_\rr \epsilon(\rr,\pp) \cdot \partial_\pp
\ee
It preserves the density in the phase space along a classical trajectory (Liouville's theorem). The second term describes the effect of collisions between quasiparticles, local in position space, and which can only occur, at the order of Beliaev-Landau, 
at points where the Thomas-Fermi density of the condensate $ \rho_0 (\rr) $ is nonzero (see the diagrams in figure \ref{fig:LK}): 
\footnote{These diagrams imply a hidden process of absorption or stimulated emission in the condensate mode.} 
\bea
I_{\rm coll}(\rr,\pp,\tau) &=& \frac{1}{2} \int \frac{\dd^3q}{(2\pi \hbar)^3} \; \frac{2\pi}{\hbar} \left[ 2g\rho_0^{1/2}(\rr) \, {\cal A}_{\qq , |\pp-\qq|}^{\pp}(\rr) \right]^2 \delta\left(\epsilon(\rr,\qq)+\epsilon(\rr,\pp-\qq)-\epsilon(\rr,\pp)\right) \nonumber\\
&&\times \left\{ -n(\rr,\pp,\tau)[1+n(\rr,\qq,\tau)][1+n(\rr,\pp-\qq,\tau)]+ n(\rr,\qq,\tau) n(\rr,\pp-\qq,\tau)[1+n(\rr,\pp,\tau)]\right\} \nonumber\\
&+& \phantom{\frac{1}{2} } \int \frac{\dd^3q}{(2\pi \hbar)^3} \; \frac{2\pi}{\hbar} \left[ 2g\rho_0^{1/2}(\rr) \, {\cal A}_{\pp ,\qq}^{|\pp+\qq|}(\rr) \right]^2 \delta\left(\epsilon(\rr,\pp)+\epsilon(\rr,\qq)-\epsilon(\rr,\pp+\qq)\right) \nonumber\\
&&\times \left\{ -n(\rr,\pp,\tau)n(\rr,\qq,\tau)[1+n(\rr,\pp+\qq,\tau)]+ n(\rr,\pp+\qq,\tau)[1+n(\rr,\pp,\tau)][1+n(\rr,\qq,\tau)]\right\} 
\label{eq:Icoll}
\eea
In this process are involved, at point $ \rr $, a quasiparticle of momentum $ \pp $ (whose evolution of the average occupation number $ n (\rr, \pp \, \tau) $ has to be determined), a second outgoing or incoming quasiparticle of momentum $ \qq $ on which it is necessary to integrate, and a 
third quasiparticle whose momentum is fixed by momentum conservation. In equation (\ref{eq:Icoll}) 
the first integral takes into account the Beliaev processes; it shows a factor of $ 1/2 $ to avoid double counting of the final or initial two quasiparticle states $ (\qq, \pp- \qq) $ and $ (\pp- \qq, \qq) $; the second integral takes into account the Landau processes. Note in both cases: $ (i) $ the factor $ \frac {2 \pi} {\hbar} $, from Fermi's golden rule, 
$ (ii) $ the $ - $ sign taking direct processes into account (they depopulate the $ \pp $ mode at point $ \rr $) and the $ + $ sign for
inverse processes, with the bosonic amplification factors $ 1 + n $, $ (iii) $ the presence of an energy conservation Dirac function at $ \rr $. The reduced coupling amplitudes for three-quasiparticles processes at point $ \rr $ are given  by \cite {Superdiff, Giorgini} \be
{\cal A}_{\pp_2,\pp_3}^{\pp_1}(\rr)=\frac{s^2(\rr,\pp_2)+s^2(\rr,\pp_3)-s^2(\rr,\pp_1)}{4s(\rr,\pp_1)s(\rr,\pp_2)s(\rr,\pp_3)}
+\frac{3}{4}s(\rr,\pp_1)s(\rr,\pp_2)s(\rr,\pp_3)
\label{eq:ampred}
\ee
with $ s (\rr, \pp) = U (\rr, \pp) + V (\rr , \pp) $. As expected, the kinetic equations admit the average thermal equilibrium occupation numbers as a stationary solution \footnote{Strictly speaking, this stationary solution corresponds to the average occupation numbers in the canonical ensemble, rather than in the microcanonical one. The difference, computable as in Appendix C of reference \cite{PRAlong}, but out of reach of our kinetic equations, tends to zero at the thermodynamic limit and is negligible here. It should also be noted that the non-conservation of the total number of quasiparticles by the Beliaev-Landau processes requires the Bose's $ \bar {n} $ law to have {\yvan unit} fugacity.} 
\be
\bar{n}(\rr,\pp) = \frac{1}{\eee^{\beta \epsilon(\rr,\pp)}-1}
\label{eq:solstat}
\ee
The well-known property of the Bose law $ 1 + \bar {n} = \eee ^ {\beta \epsilon} \bar {n} $ allows to verify it easily: supplemented with energy conservation, it leads to the perfect compensation in all points of direct and inverse processes, that is to the cancellation of the quantities between curly brackets in equation (\ref{eq:Icoll}), following the principle of microreversibility; we also have 
$ \frac {\rm D} {{\rm D} \tau} \bar {n} = 0 $ since $ \bar {n} (\rr, \pp) $ is a function of $ \epsilon (\rr, \pp) $, a quantity that is conserved by the Hamiltonian transport. 

As our system fluctuates weakly around the equilibrium, we linearise the kinetic equations around $ n = \bar {n} $ 
as in reference \cite {PRAlong} to get 
\be
\frac{\rm D}{{\rm D}\tau} \delta n(\rr,\pp,\tau) =-\Gamma(\rr,\pp,\tau) \delta n(\rr,\pp,\tau) + \int \frac{\dd^3q}{(2\pi\hbar)^3}
K(\rr,\pp,\qq)  \delta n(\rr,\qq,\tau)
\label{eq:cin_lin}
\ee
the diagonal term comes from the fluctuation $ \delta n (\rr, \pp, \tau) $ in the right-hand side of equation (\ref{eq:Icoll}), 
and the non-local momentum term comes from fluctuations $ \delta n (\rr, \qq, \tau) $ and $ \delta n (\rr, \pp \pm \qq, \tau) $ whose contributions are collected by changing the variables $ \qq'= \pp \pm \qq $ in $ \int \dd^3q$. The expression of $ K (\rr, \pp, \qq) $ is not useful for the following, so let us only give the expression of the local damping rate of the Bogoliubov quasiparticles of momentum $ \pp $ at point $ \rr $: 
\bea
\Gamma(\rr,\pp) &=&\frac{4\pi \rho_0(\rr)g^2}{\hbar} \int \frac{\dd^3q}{(2\pi \hbar)^3} \; 
\left[ {\cal A}_{\qq , |\pp-\qq|}^{\pp}(\rr) \right]^2 \delta\left(\epsilon(\rr,\qq)+\epsilon(\rr,\pp-\qq)-\epsilon(\rr,\pp)\right)
 \left[ 1 + \bar{n}(\rr,\qq) + \bar{n}(\rr,\pp-\qq)  \right] \nonumber\\
&+& \frac{8\pi \rho_0(\rr)g^2}{\hbar} \int \frac{\dd^3q}{(2\pi \hbar)^3} \; 
\left[ {\cal A}_{\pp , \qq}^{|\pp+\qq|}(\rr) \right]^2 \delta\left(\epsilon(\rr,\pp)+\epsilon(\rr,\qq)-\epsilon(\rr,\pp+\qq)\right)
 \left[ \bar{n}(\rr,\qq) - \bar{n}(\rr,\pp+\qq) \right]
\label{eq:Gamrp}
\eea
This expression coincides with the damping rate of a momentum $ \pp $ mode in a spatially homogeneous condensed gas of density $ g \rho_0 (\rr) $ \cite{Giorgini}. Just like $ \delta n (\rr, \pp, \tau) $, 
$ \langle {\cal F} | \delta n (\rr, \pp, \tau) \delta n (\rr ' , \pp ', 0) | {\cal F} \rangle $ considered as a function of $ (\rr, \pp, \tau) $, obeys equation (\ref{eq:cin_lin}); the same is true for its average $ \langle \delta n (\rr, \pp, \tau) \delta n (\rr ', \pp', 0) \rangle $ over all initial Fock states $ | {\cal F} \rangle $, since the coefficients $ \Gamma $ and $ K $ do not depend on $ | {\cal F} \rangle $. Let's 
contract the latter by the quantity
\be
B(\rr',\pp')\equiv \frac{1}{\hbar}\overline{\frac{\dd\epsilon(\rr',\pp')}{\dd N}}
\label{eq:defB}
\ee
as in equation (\ref{eq:C_sc}) to form the auxiliary unknown 
\be
X(\rr,\pp,\tau)=\int \frac{\dd^3r'\dd^3p'}{(2\pi \hbar)^3} \; B(\rr',\pp') \; \langle  \delta n(\rr,\pp,\tau) \delta n(\rr',\pp',0) \rangle
\label{eq:inconnue_auxiliaire}
\ee
Then $ X (\rr, \pp, \tau) $ evolves according to the linear kinetic equations (\ref{eq:cin_lin}) with the initial condition 
\be
X(\rr,\pp,0)=\int \frac{\dd^3r'\dd^3p'}{(2\pi \hbar)^3} \; Q(\rr,\pp;\rr',\pp')\; B(\rr',\pp')
\ee
where the matrix of covariances at equal times of the number of quasiparticles has been introduced: 
\be
Q(\rr,\pp;\rr',\pp')= \langle  \delta n(\rr,\pp,0) \delta n(\rr',\pp',0) \rangle
\label{eq:defmatcov}
\ee
whose expression in the microcanonical ensemble will be connected to that in the canonical ensemble in due time, in sub-section \ref{subsec:solution}. The sought microcanonical correlation function of $ {\dd \hat {\theta} ^ {\rm sc}} / {\dd t} $ is then  
\be
C_{\rm mc}^{\rm sc}(\tau) = \int\frac{\dd^3r \dd^3p}{(2\pi \hbar)^3} \; B(\rr,\pp) X(\rr,\pp,\tau)
\label{eq:Cmcsc}
\ee

\subsection{Solution in the secular-ergodic approximation}
\label{subsec:solution}

Our study restricts to the collisionless regime $ \Gamma _ {\rm th} \ll \omega_ \alpha $ where $ \Gamma _ {\rm th} $ is 
the typical thermal value of the quasiparticles damping rate $ \Gamma (\rr, \pp) $ and $ \omega_ \alpha $ are the trapping angular frequencies. The quasiparticles then have time to perform a large number of Hamiltonian oscillations in the trap before undergoing a collision. We can therefore perform the secular approximation that consists in replacing the coefficients of the linearized kinetic equation (\ref{eq:cin_lin}) by their temporal average over a trajectory. Thus, 
\be
\Gamma(\rr,\pp) \stackrel{\mathrm{secular}}{\underset{\mbox{\scriptsize approx.}}{\rightarrow}} \overline{\Gamma(\rr,\pp)} = \lim_{t\to{\yvan +}\infty} \frac{1}{t} \int_0^t \dd\tau \, \Gamma(\rr(\tau),\pp(\tau))
\label{eq:approx_secu}
\ee 
the auxiliary unknown $ X (\rr, \pp, \tau) $ of equation (\ref{eq:inconnue_auxiliaire}), just like the fluctuations of the occupation numbers $ \delta n (\rr, \pp, t) $, depend only 
on the trajectory $ \tau \mapsto (\rr (\tau), \pp (\tau)) $ passing through $ (\rr, \pp) $ and on time. Still the problem remains formidable. 

Fortunately, as we have said, 
in a completely anisotropic trap, the Hamiltonian dynamics of quasiparticles should be highly 
chaotic, except within the limits of very low energy $ \epsilon \ll \mu _ {\rm TF} $ or very high energy $ \epsilon \gg \mu _ {\rm TF} $ 
\cite {Graham, GrahamD}. We thus use the ergodic hypothesis, by identifying the temporal average on an trajectory of energy $ \epsilon $ to the  \oge{uniform}\fge \, mean in the phase space on the energy shell $ \epsilon $: 
\be
\overline{\Gamma(\rr,\pp)} \stackrel{\mbox{\scriptsize ergodic}}{\underset{\mathrm{hypothesis}}{=}} 
\Gamma(\epsilon) = \langle \Gamma(\rr,\pp)\rangle_\epsilon \equiv\frac{1}{\rho(\epsilon)} \int \frac{\dd^3r\dd^3p}{(2\pi\hbar)^3} 
\Gamma(\rr,\pp) \delta(\epsilon-\epsilon(\rr,\pp))
\label{eq:defGamergo}
\ee
where the density of states $ \rho (\epsilon) $ is given by equation (\ref{eq:rhoeps}). 
We will come back to this hypothesis in section \ref{subsec:discuss_ergo}. In this case, the function $ X (\rr, \pp, \tau) $ 
depends only on the energy $ \epsilon = \epsilon (\rr, \pp) $ and time: 
\be
X(\rr,\pp,\tau) \stackrel{\mbox{\scriptsize ergodic}}{\underset{\mathrm{hypothesis}}{=}} X(\epsilon,\tau)
\ee
We obtain the evolution equation of $ X (\epsilon, \tau) $ by averaging that of $ X (\rr, \pp, \tau) $ on the energy shell $ \epsilon $: 
\footnote{The simplest is to average the complete kinetic equations (\ref{eq:eqcin}), then linearize the result around the stationary solution (\ref{eq:solstat}).} 
\begin{multline}
\label{eq:evolX}
\partial_\tau X(\epsilon,\tau)= -{\Gamma}(\epsilon) X(\epsilon,\tau)
-\frac{1}{2\rho(\epsilon)} \int_0^\epsilon \dd\epsilon' L(\epsilon-\epsilon',\epsilon')
\{X(\epsilon',\tau)[\bar{n}(\epsilon) -\bar{n}(\epsilon-\epsilon')]+X(\epsilon-\epsilon',\tau)[\bar{n}(\epsilon)-\bar{n}(\epsilon')]\} \\
-\frac{1}{\rho(\epsilon)} \int_0^{+\infty} \dd\epsilon'\, L(\epsilon,\epsilon') \{ X(\epsilon',\tau) [\bar{n}(\epsilon)-\bar{n}(\epsilon+\epsilon')]
-X(\epsilon+\epsilon',\tau)[1+\bar{n}(\epsilon)+\bar{n}(\epsilon')]
\}
\end{multline}
with 
\be
\label{eq:Gamma_eps}
{\Gamma}(\epsilon)= \frac{1}{2\rho(\epsilon)} \int_0^\epsilon \dd\epsilon'\, L(\epsilon-\epsilon',\epsilon') [1+\bar{n}(\epsilon')
+\bar{n}(\epsilon-\epsilon')]
+\frac{1}{\rho(\epsilon)} \int_0^{+\infty} \dd\epsilon'\, L(\epsilon,\epsilon') [\bar{n}(\epsilon')-\bar{n}(\epsilon+\epsilon')]
\ee
In these expressions, the first integral, limited to energies $ \epsilon '$ lower than the energy of the quasiparticle 
$ \epsilon $ considered, corresponds to Beliaev processes, and the second integral to Landau processes. The integral kernel 
\footnote{To get (\ref{eq:Lexpli}), we reduced equation (\ref{eq:defL}) to a single integral on the $ r $ modulus (after having 
formally reduced to the case of an isotropic trap as in note \ref{note:isotropisation}) 
by integrating in spherical coordinates 
on $p$, $q$ and $u$, the cosine of the angle between $\pp$ and $\qq$. In $\int_{-1}^{1} \dd u$, the argument 
of the third Dirac vanishes at a point $ u_0 $ and only one, given the inequalities 
$\epsilon_{|p-q|}^{\rm Bog} \leq \epsilon_p^{\rm Bog} + \epsilon_q^{\rm Bog} \leq \epsilon_{p+q}^{\rm Bog}$ 
satisfied by the Bogoliubov dispersion relation $\epsilon_p^{\rm Bog}=[\frac{p^2}{2m}(\frac{p^2}{2m}+2\mu_0)]^{1/2}$,
$\forall \mu_0\geq 0$.}
\bea
\label{eq:defL}
L(\epsilon,\epsilon') &=& \int\frac{\dd^3r\,\dd^3p\,\dd^3q}{(2\pi\hbar)^6} \frac{8\pi g^2\rho_0(\rr)}{\hbar} 
\left[A_{\epsilon,\epsilon'}^{\epsilon+\epsilon'}(\rr)\right]^2 \delta(\epsilon-\epsilon(\rr,\pp))
\delta(\epsilon'-\epsilon(\rr,\qq)) \delta(\epsilon+\epsilon'-\epsilon(\rr,\pp+\qq)) \\
&=& \frac{32\sqrt{2}}{\pi^{1/2}} \frac{[\rho_0(\mathbf{0})a^3]^{1/2}}{\hbar\mu_{\rm TF}} 
\left(\frac{\mu_{\rm TF}}{\hbar\bar{\omega}}\right)^3\int_0^{\mu_{\rm TF}}
\frac{\mu_0\dd\mu_0 (\mu_{\rm TF}-\mu_0)^{1/2}\epsilon\epsilon'(\epsilon+\epsilon')\left[A_{\epsilon,\epsilon'}^{\epsilon+\epsilon'}(\mu_0)\right]^2}
{\mu_{\rm TF}^{5/2}(\epsilon^2+\mu_0^2)^{1/2} (\epsilon'^2+\mu_0^2)^{1/2} [(\epsilon+\epsilon')^2+\mu_0^2]^{1/2}}
\label{eq:Lexpli}
\eea
uses the reduced coupling amplitude (\ref{eq:ampred}) at point $ \rr $, reparametrized in terms of energies 
$ \epsilon_i = \epsilon (\rr, \pp_i) $ ($ 1 \leq i \leq 3 $) or even in terms of the local Gross-Pitaevskii chemical potential 
$ \mu_0 = g \rho_0 (\rr) $. 
It has the symmetry property $ L (\epsilon, \epsilon ') = L (\epsilon', \epsilon) $. 

We write the result before giving some indications on its obtention (one will also consult reference \cite {PRAlong}). 
In the secularo-ergodic approximation, the microcanonical correlation function of $ {\dd \hat {\theta} ^ {\rm sc}} / {\dd t} $ is 
\be
C_{\rm mc}^{\rm ergo}(\tau) = \int_0^{+\infty} \dd\epsilon\, \rho(\epsilon) B(\epsilon) X(\epsilon,\tau)
\label{eq:Cmcergo}
\ee
Here $ B (\epsilon) $ is the ergodic average of the quantity $ B (\rr, \pp) $ introduced in equation (\ref{eq:defB}): 
\bea
B(\epsilon) &=& \frac{1}{\rho(\epsilon)} \int \frac{\dd^3r\, \dd^3p}{(2\pi\hbar)^3} 
\frac{\dd\epsilon(\rr,\pp)}{\hbar\dd N}\delta(\epsilon-\epsilon(\rr,\pp)) \\
\label{eq:Bergoexpli}
&=& \frac{\dd\mu_{\rm TF}/\dd N}{\hbar \pi f(\epsc)} \left[
2\epsc^{1/2}(\epsc+1)-\sqrt{2}(\epsc^2+1) \argsh\frac{(2\epsc)^{1/2}}{(1+\epsc^2)^{1/2}} 
-\epsc^{1/2}(\epsc-1)-(1+\epsc)^2\acos\frac{1}{(1+\epsc)^{1/2}}
\right] \\
B(\epsilon)& \underset{\epsilon\to 0}{=}& \frac{\dd\mu_{\rm TF}}{\hbar\dd N} \left[ -\frac{\epsc}{5}-\frac{3\pi}{40\sqrt{2}}\epsc^{3/2}+O(\epsc^2)\right],
\quad B(\epsilon) \underset{\epsilon\to+\infty}{=} \frac{\dd\mu_{\rm TF}}{\hbar\dd N} \left[-1 +\frac{32}{3\pi} \epsc^{-3/2} + O(\epsc^{-5/2})\right]
\label{eq:Bbashaut}
\eea
with $ \epsc = \epsilon / \mu _ {\rm TF} $ and $ f (\epsc) $ the reduced density of states (\ref{eq:deff}). 
The auxiliary unknown $ X (\epsilon, \tau) $ is a solution of the linear equation (\ref{eq:evolX}) with the initial condition 
\be
X(\epsilon,0) = \bar{n}(\epsilon) [1+\bar{n}(\epsilon)] [B(\epsilon)-\Lambda \epsilon]
\label{eq:condiX}
\ee
where $ \hbar \Lambda $ is the derivative of the microcanonical chemical potential with respect to to the total energy $ E $ of the gas 
\footnote {The deep reason for the appearance of this derivative is given in reference \cite {PRAlong}. It explains why 
the kinetic equations allow to find in the canonical ensemble the ballistic term $ At ^ 2 $ of equation (\ref{eq:vartot}) 
with the correct expression of the coefficient $ A = (\partial_E \mu_ { \rm mc} / \hbar) ^ 2 \mbox {Var} \, E $.}, 
as in equation (\ref{eq:acalc1}): 
\be
\label{eq:Lambda}
\Lambda = \frac{\int_0^{+\infty} \dd\epsilon\,\rho(\epsilon) \epsilon B(\epsilon) \bar{n}(\epsilon) [1+\bar{n}(\epsilon)]}
{\int_0^{+\infty} \dd\epsilon\, \rho(\epsilon)\epsilon^2 \bar{n}(\epsilon) [1+\bar{n}(\epsilon)]}
\ee
The equation (\ref{eq:Cmcergo}) is the ergodic rewriting of equation (\ref{eq:Cmcsc}). The initial condition 
(\ref{eq:condiX}) is the difference of two contributions: 
\begin{itemize} 
\item the first is the one that one would obtain in the canonical ensemble. The ergodic mean of the covariance matrix {\yvan (\ref{eq:defmatcov})}
would then be simply $ Q _ {\rm can} (\epsilon, \epsilon ') = \bar {n} (\epsilon) 
[1+ \bar {n} (\epsilon)] \delta (\epsilon -\epsilon') / \rho (\epsilon) $; 
\item the second comes from a projection of $ \delta n $ canonical fluctuations on the subspace of the $ \delta n $ fluctuations 
of zero energy, $ \int_0 ^ {+ \infty} \dd \epsilon \, \rho (\epsilon) \epsilon \delta n (\epsilon) = $ 0, the only one eligible in 
the microcanonical ensemble. Only subtle point, this projection must be carried out parallel to the stationary solution 
$ e_0 (\epsilon) = \epsilon \bar {n} (\epsilon) [1+ \bar {n} (\epsilon)] $ of the linearized kinetic equations (\ref{eq:evolX}). 
\footnote {For this projection to be compatible with linearized kinetic evolution, 
it is necessary that the projection direction and the hyperplane on which we project be invariant by 
temporal evolution, the second point being ensured by conservation of energy. The form of $ e_0 (\epsilon) $ 
derives from the fact that (\ref{eq:solstat}) remains a stationary solution for an infinitesimal variation of $ \beta $, 
$ \beta \to \beta + \delta \beta $, around its physical value.} We then check that, for the value of $ \Lambda $ given, $ X (\epsilon, 0) $ is 
in the subspace of zero energy fluctuations. 
\end{itemize} 

\subsection{{\yvan Results and discussion}}
\label{subsec:resetdiscus}

We present some results in graphic form, after a clever scaling making them independent of the
trapping {\yvan angular} frequencies (provided they are quite distinct two by two to allow the ergodic hypothesis) and of the strength of the 
interactions \footnote {In a first step, we show that the results 
can depend on the trapping frequencies $ \omega_ \alpha $ only through 
their geometric mean $ \bar {\omega} $. This is a direct consequence of the ergodic hypothesis and the fact 
that the observables involved here, including the Hamiltonian, depend only on the position $ \rr $ of the quasiparticles {\sl via} the 
trapping potential $ U (\rr) = \frac{1} {2} m \sum_ \alpha \omega_ \alpha ^ 2 r_ \alpha ^ 2 $. In the integral $ \int \dd ^ 3r $ participating in the 
ergodic mean, one can then perform the isotropising change of variables of note \ref{note:isotropisation}.}; 
it is enough to know the temperature in units of Thomas-Fermi's chemical potential $ \mu _ {\rm TF} $. 
These results illustrate the universality class of the completely anisotropic harmonic traps, different 
from that of the spatially homogeneous systems of reference \cite {PRAlong}.

\begin{figure} [t]
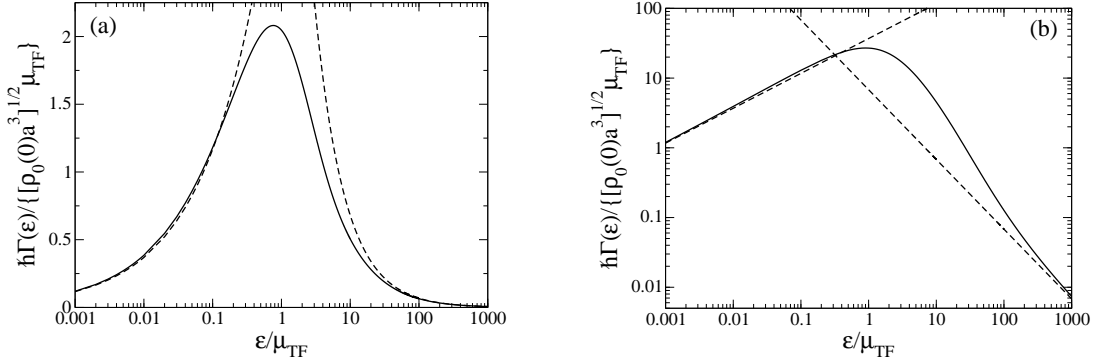
 
\centerline {\includegraphics [width = 0.4 \textwidth,clip=] {gammaeps_en.eps} \\\includegraphics [width = 0.4 \textwidth,clip=] {gammaeps10_en.eps}} 
\caption {Beliaev-Landau damping rate $ {\Gamma} (\epsilon) $ of Bogoliubov modes of a 
condensate in a completely anisotropic harmonic trap as a function of the mode energy $ \epsilon $, at the thermodynamic limit, in the 
secularo-ergodic approximation (\ref{eq:approx_secu}, \ref{eq:defGamergo}, {\yvan\ref{eq:Gamma_eps})}, at temperature (a) $ k_B T = \mu _ {\rm TF} $ and (b) $ k_B T = 10 \mu _ {\rm TF} $, 
where $ \mu _ {\rm TF} $ is the Thomas-Fermi's chemical potential of 
the condensate. Thanks to the chosen units, the curve is universal; in particular, it does not depend 
on the trapping angular frequencies $ \omega_ \alpha $. The Bogoliubov modes considered must be in the classical motion regime 
$ \epsilon \gg \hbar \omega_ \alpha $ 
and the system must be in the regime of an almost pure condensate, $ [\rho_0 (\mathbf {0}) a ^ 3] ^ {1/2} \ll 1 $ and 
$ T \ll T_c $, where $ \rho_0 (\mathbf {0}) = \mu _ {\rm TF} / g $ is the density of the condensate in the center of the trap and $ T_c $ the critical temperature. 
In dashed lines, the equivalents {\yvan (\ref{eq:gambassener})} and (\ref{eq:asymptGammaeps}) of $ {\Gamma} (\epsilon) $ at low and high energy. } 
\label {fig:Gamma} 
\end{figure}

An interesting by-product of our study is shown in figure \ref{fig:Gamma}: it is 
the damping rate $ {\Gamma} (\epsilon) $ in the secularo-ergodic approximation of the Bogoliubov modes of energy 
$ \epsilon $. In a cold atom experiment it is possible to excite such modes and to follow their decay 
in time. The rate we predict is then measurable and it can be compared to the experiments, at least in its validity regime
of classical motion $ \epsilon \gg \hbar \omega_ \alpha $ (deviations from the ergodic hypothesis are discussed in section 
\ref{subsec:discuss_ergo}). The limiting behaviors 
\bea
\label{eq:gambassener}
\hbar {\Gamma}(\epsilon) &\underset{\epsilon\to 0}{\sim}& 
\frac{3\mathcal{I}}{4} \left(\frac{\epsilon}{\mu_{\rm TF}}\right)^{1/2} k_B T [\rho_0(\mathbf{0})a^3]^{1/2}
\quad \mbox{with}\quad \mathcal{I}=4.921\, 208\, \ldots \\
\hbar\Gamma (\epsilon) &\underset{\epsilon\to +\infty}{\sim}& \frac{128\sqrt{2}}{15\sqrt{\pi}} \frac{\mu_{\rm TF}^2}{\epsilon} [\rho_0(\mathbf{0})a^3]^{1/2}
\label{eq:asymptGammaeps}
\eea
shown in dashed lines in figure \ref{fig:Gamma}, \footnote {For $ k_B T = \mu _ {\rm TF} $, $ \Gamma (\epsilon) / \epsilon ^ {1/2 } $ has 
a deceptive maximum in the neighborhood of $ \epsilon / \mu _ {\rm TF} = $ 0.02 of about $ 5 \% $ above its limit in $ \epsilon = $ 0.} 
are derived in Appendix A. 
They are very different from the spatially homogeneous case, 
where the damping rate vanishes linearly in $ \epsilon $ at low energy and diverges as $ \epsilon ^ {1/2} $ at high energy. 
In particular, the behavior (\ref{eq:gambassener}) in $ \epsilon ^ {1/2} $ results from the existence of the Thomas-Fermi edge of the condensate. 

\begin{figure} [t]
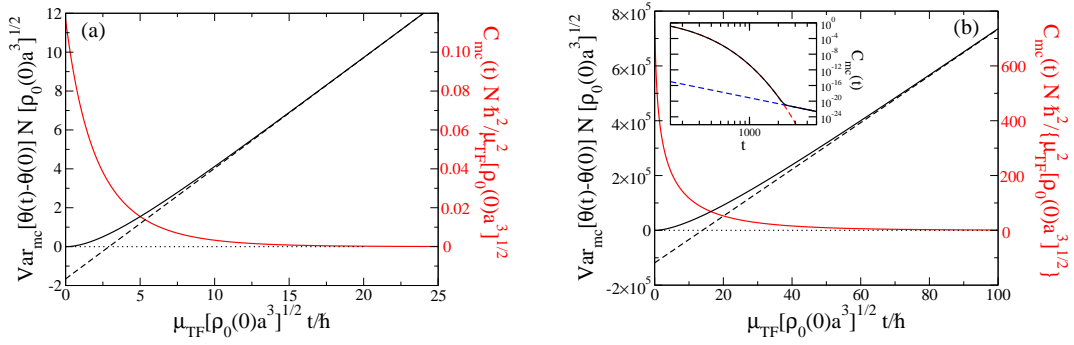
 
\centerline {\includegraphics [width = 0.4 \textwidth,clip=] {kergo4_1_en.eps} \quad \quad \includegraphics [width = 0.4 \textwidth,clip= ] {kergo4_10sf.eps}} 
\caption {In the conditions of figure \ref{fig:Gamma}, for a system prepared at $t=0$ in the microcanonical ensemble 
at temperature (a) $ k_B T = \mu _ {\rm TF} $ or (b) $ k_B T = 10 \mu _ {\rm TF} $, 
and isolated from its environment in its subsequent evolution, variance of the condensate phase shift $ \hat {\theta} (t) 
- \hat {\theta} (0) $ as a function of time $ t $ (solid line black) and its asymptotic diffusive behavior 
(\ref{eq:varasympt}) (dashed). The correlation function $ C _ {\rm mc} (t) $ of $ \dd \hat {\theta} / \dd t $ is shown on the same figure in 
the secular-ergodic approximation (\ref{eq:Cmcergo}) as a function of time (red solid line, ticks on the right) and, for (b), 
in an inset in log-log scale at long times (black solid line) to show that after a quasi-exponential decay in the square root of time 
(fit with $ t^6 \exp (-C \sqrt{t}) $ in red dashed line) it follows a power law $ \propto t ^ {- 5} $ (blue dashed). 
As in figure \ref{fig:Gamma}, the multiplication of the quantities on the axes by well-chosen factors makes these 
results universal.} 
\label{fig:VaretC} 
\end{figure} 

Let's go back to the condensate phase spreading  in the microcanonical ensemble. In figure \ref{fig:VaretC}, we represent in black solid line the variance of the phase shift $ \hat {\theta} (t) - \hat {\theta} (0) $ 
of the condensate as a function of time $ t $ in the ergodic approximation (\ref{eq:Cmcergo}) at the temperatures $ T = \mu _ {\rm TF} / k_B $ and $ T = 10 \mu _ {\rm TF} / k_B $. 
The variance has a parabolic departure in time, which corresponds to the precollisional regime $ t \ll t _ {\rm coll} $, where $ t _ {\rm coll} $ 
is the typical collision time between quasiparticles: we can then assume that $ C _ {\rm mc} (\tau) \simeq C _ {\rm mc} (0) $, 
so that the integral contribution to equation (\ref{eq:vartot}) is $ \simeq C _ {\rm mc} (0) t ^ 2 $. At long times, $ t \gg t _ {\rm coll} $, 
the correlation function of $ {\dd \hat {\theta}} / {\dd t} $ seems to quickly reach zero (red solid line); 
a more detailed numerical study (see the inset in figure \ref{fig:VaretC} b) reveals however the presence of a 
power law tail $ t ^ {- \alpha} $, 
\be
C_{\rm mc}(t) \underset{t\to +\infty}{\sim} \frac{\mathcal{C}}{t^5}
\label{eq:cmcloipuis}
\ee
the exponent $ \alpha = 5 $ is greater than the one $ \alpha_h = 3 $ of the decay law of $ C _ {\rm mc} (t) $ 
in the spatially homogeneous case \cite {PRAlong}. Its value can be found by a rough heuristic approximation, 
called {\sl rate approximation} or {\sl projected Gaussian} \cite {Genuine}, already used for $ \alpha_h $ with success 
in this same reference \cite {PRAlong}: 
we keep in the linearized kinetic equations (\ref{eq:evolX}) only the term of pure decay $ - \Gamma (\epsilon) X (\epsilon, \tau) $ 
in the right-hand side, 
which makes them immediately integrable and leads to the estimate \footnote{Care has been taken 
to account for the projection on the microcanonical subspace of 
zero energy fluctuations not only in the initial condition (\ref{eq:condiX}), but also in the contraction by $ B (\epsilon) $ in 
(\ref{eq:Cmcergo}), replacing $ B (\epsilon) $ with $ B (\epsilon) - \Lambda \epsilon $; this precaution, optional in the exact formulation, is 
necessary here since the rate approximation violates conservation of energy.} 
\be
C_{\rm mc}(t) \approx \int_0^{+\infty} \dd\epsilon\, \rho(\epsilon) [B(\epsilon)-\Lambda\epsilon]^2 \bar{n}(\epsilon) [1+\bar{n}(\epsilon)]
\eee^{-\Gamma(\epsilon)t}
\label{eq:approxtauxCmc}
\ee
The power law behavior of the density of states $ \rho (\epsilon)$ at low energy [see (\ref{eq:fbasse})], of the coefficients 
$ B (\epsilon) $ in $ \dd \hat {\theta} / \dd t $ [see (\ref{eq:Bbashaut})], of the occupation numbers $ n (\epsilon) \sim k_BT / \epsilon $ and of the damping rate 
$ \Gamma (\epsilon) $ [see (\ref{eq:gambassener})] then reproduce the exponent $ \alpha = $ 5 found numerically. \footnote {In contrast, 
the predicted value for the coefficient $ \mathcal {C} $ in 
(\ref{eq:cmcloipuis}) for $ k_BT = 10 \mu _ {\rm TF} $, that is $ \simeq 10 ^ {- 5} $, differs significantly from the 
numerical value $ \simeq 7 \times 10^{-5}$.} 
Since $ C _ {\rm mc} (t) $ tends to zero faster than $ 1 / t ^ {2+ \eta} $, for some $ \eta> 0 $, we obtain the following important result: 
the variance of the condensate phase shift $ \mbox{Var}_ {\rm mc} [\hat{\theta} (t) - \hat{\theta} (0)] $ 
exhibits at long times the typical affine growth of a diffusive regime with delay: 
\be
\mbox{Var}_{\rm mc}[\hat{\theta}(t)-\hat{\theta}(0)] \underset{t\gg t_{\rm coll}}{=}  2 D (t-t_0) + o(1)
\label{eq:varasympt}
\ee
represented by a dashed line on figure \ref{fig:VaretC}.
The delay $ t_0 $ is due to the non-zero width of the correlation function $ C _ {\rm mc} (\tau) $: 
\bea
\label{eq:Dint}
D &=& \int_0^{+\infty} \dd\tau\, C_{\rm mc}(\tau) \\
t_0 &=& \frac{\int_0^{+\infty} \dd\tau\, \tau C_{\rm mc}(\tau)}{\int_0^{+\infty}\dd\tau\, C_{\rm mc}(\tau)}
\label{eq:t0rapint}
\eea

\begin{figure} [t]
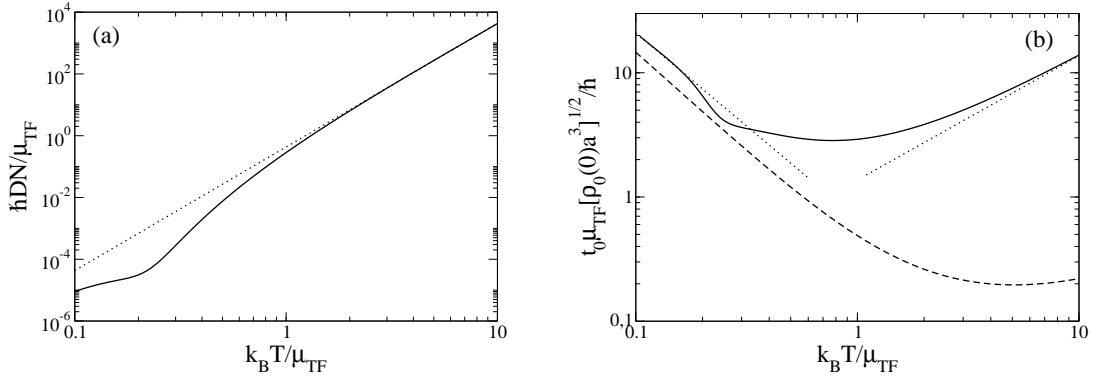
 
\centerline {\includegraphics [width = 0.4 \textwidth,clip=] {kergo3_en.eps} \quad \quad \includegraphics [width = 0.417 \textwidth,clip=] {kergo5_en.eps}} 
\caption {In the conditions of the previous figures \ref{fig:Gamma} and \ref{fig:VaretC}, (a) diffusion coefficient $ D $ of the phase 
of the harmonically trapped condensate and (b) delay time $ t_0 $ to the phase diffusion as a function of temperature (in black solid line) 
deduced from equations (\ref{eq:Dint}, \ref{eq:t0rapint}) and (\ref{eq:Cmcergo}). These quantities are independent of 
the statistical ensemble. With the chosen scalings, the curves are universal. 
Dotted lines: in (a) fit by a $ T ^ 4 $ law at high temperature 
and in (b) fit by a law linear in $ T $ at high temperature and by a $ T ^ {- 3/2} $ law at low temperature. 
The power laws represented at high temperature are only indicative because they certainly omit 
logarithmic factors in $ k_B T / \mu _ {\rm TF} $. 
In (b) the dashed line gives the estimate $ 1 / {\Gamma} (\epsilon = k_B T) $ of the typical collision time $ t _ {\rm coll} $ between quasiparticles. 
} 
\label {fig:Diff_et_t0} 
\end{figure} 

We represent the diffusion coefficient $ D $ of the condensate phase as a function of temperature in figure \ref{fig:Diff_et_t0} a. 
It exhibits a high temperature growth ($ k_B T> \mu _ {\rm TF} $) much faster than in the spatially homogeneous case: 
it was only linear (up to logarithmic factors), it seems to scale here as $ T ^ 4 $ (dotted in the figure). 
The 
diffusion delay time $ t_0 $ is plotted as a function of temperature in figure \ref{fig:Diff_et_t0} b. We compare it to the estimate 
$t_{\rm coll} \simeq 1 / {\Gamma} (\epsilon = k_B T) $ of the collision time between quasiparticles, in dashed line: this one gives a good account of the 
sudden rise of $t_0$ at low temperatures, but reproduces with much delay and greatly underestimating that 
at high temperatures. 
The rise of $t_0$ is well represented by a $T^{-3/2}$ low temperature law, and appears to be linear in $T$ 
at high temperature (see dashed line). 

Let us find by a simple reasoning the observed power laws. If a scaling law exists, it should survive the 
rate approximation on the linearized kinetic equations; we can therefore take the approximate expression (\ref{eq:approxtauxCmc}) of $ C _ {\rm mc} (t) $ as a starting point and put it 
in the expressions (\ref{eq:Dint}) and (\ref{eq:t0rapint}) of $ D $ and $ t_0 $. 

At high temperature, the integrals on $ \epsilon $ giving $ D $ and $ t_0 $ in the rate approximation are dominated by 
energies of order $ k_B T $; we set $ \epsilon = k_B T \bar {\epsilon} $ and send $ T $ to $ + \infty $ at fixed $ \bar {\epsilon} $ 
under the integral. The behaviors of $ \rho (\epsilon) $ and $ B (\epsilon) $ at high energy are known. 
Only that of $ \Gamma (k_B T \bar {\epsilon}) $ is missing; to get it, we notice on 
(\ref{eq:Lexpli}) that $ L (k_B T \bar {\epsilon}, k_B T \bar {\epsilon} ') $ tends to a constant when $ T \to + \infty $. The approximation 
$ L (\epsilon, \epsilon ') \simeq L (\epsilon- \epsilon', \epsilon ') \simeq \mbox {const} $, however, triggers a logarithmic infrared divergence 
in the integrals on $ \epsilon' $ in (\ref{eq:Gamma_eps}), which stops at $ \epsilon '\lesssim \mu _ {\rm TF} $, so that 
\footnote{A more accurate calculation leads to replace in (\ref{eq:equivgam}) the symbol $ \sim $ by $ = $ and the factor $ \ln \frac{k_B T} {\mu _ {\rm TF}} $ 
by $ [\ln \frac{k_B T} {\mu _ {\rm TF}} + \frac{\bar {\epsilon}} {4} + \ln (1- \eee ^ {- \bar {\epsilon}}) + \frac{31} {15} - 3 \ln 2+ O (\mu _ {\rm TF} / k_B T)] $.} 
\be
\frac{\hbar\Gamma(k_BT \bar{\epsilon})}{\mu_{\rm TF} [\rho_0(\mathbf{0})a^3]^{1/2}} \underset{k_BT/\mu_{\rm TF}\to +\infty}{\sim}
\frac{512\sqrt{2}}{15\pi^{1/2}} \frac{1}{\bar{\epsilon}^{2}} \frac{\mu_{\rm TF}}{k_B T} \ln \frac{k_B T}{\mu_{\rm TF}}
\label{eq:equivgam}
\ee
All this leads to the scaling laws $ D \approx T ^ 4 $ and $ t_0 \approx T $ at high temperature, up to logarithmic factors. 

At low temperatures, we proceed in the same way. The behavior of $ \Gamma (k_B T \bar {\epsilon}) $ is as  $ T ^ {3/2} $ when $ T \to 0 $ at fixed $ \bar {\epsilon} $, as one could expect it from the equivalent (\ref{eq:gambassener}) and as it is confirmed by
a calculation. 
The only trap to avoid is that $ B (k_B T \bar {\epsilon}) - \Lambda k_BT \bar {\epsilon} $ 
scales as $ T ^ {3/2} $ when $ T \to 0 $, not as $ T $ as one might think, because the dominant terms of $ B (k_B T \bar {\epsilon}) $ and 
$ \Lambda k_BT \bar {\epsilon} $, 
both linear in $ k_B T \bar {\epsilon} $, exactly compensate each other, see equation (\ref{eq:Lambda}). This leads 
to the exact power laws (without logarithmic corrections) $ D \propto T ^ 4 $ and $ t_0 \propto T ^ {- 3/2} $ at low temperature; 
only the second one is accessible 
on the temperature interval of figure \ref{fig:Diff_et_t0}, but we checked the first one numerically on a larger temperature range. 

{\yvan
To encourage an experimental study with cold atoms, let's finish with a small study of the fundamental limits to the observability of
phase diffusion of a trapped condensate. There are, of course, several practical difficulties to overcome, such as $ (i) $ significant reduction of the
fluctuations in the energy and number of particles in the gas to mitigate the ballistic blurring of the phase, which is a dangerous competitor of the diffusion,
$ (ii) $ the introduction of a sensitive and unbiased detection scheme for the condensate phase shift or coherence function $ g_1 (t) $,
of the Ramsey type as proposed in references \cite{chapitre,CohFer}, $ (iii) $ reduction of the technical noise of the experimental device,
$ (iv) $ trapping of the atoms in a cell with a sufficiently high vacuum to make cold atom losses by
collision with the residual hot gas negligible (one-body losses): lifetimes of the order of the hour are possible under cryogenic environment
\cite {Libbrecht, antiH}. These practical aspects vary according to the experimental groups and are beyond the scope of this article. In contrast, particle losses
due to three-body collisions, with the formation of a dimer and a fast atom, are intrinsic to alkaline atoms and
constitute a fundamental limit. Each atom loss changes, at a random time, the rate of variation of the phase
$ \frac{\dd} {\dd t} \hat {\theta} $, since this is a function of $ N $, which adds a stochastic component to its evolution \cite {PRAlong, Sinatra98}.
To calculate the variance of the phase shift of the condensate induced by the three-body losses, we place ourselves
at zeroth order in the uncondensed fraction, that is in the case of a pure condensate at zero temperature prepared at time $0$
with an initially well defined number of particles $N$, as in reference \cite {PRAlong} from which we can recycle
(adapting them to the trapped case and to three-body losses) expressions (G7) and (64):
\be
\mbox{Var}_{\rm losses} [\hat{\theta}(t)-\hat{\theta}(0)]= \left(\frac{\dd\mu_{\rm TF}}{\hbar\dd N}\right)^2 \int_0^t\dd\tau \int_0^t \dd\tau'
\left[\langle \hat{N}(\tau)\hat{N}(\tau')\rangle-\langle\hat{N}(\tau)\rangle \langle\hat{N}(\tau')\rangle\right]
\underset{\Gamma_3 t\to 0}{\sim} \left(\frac{\dd\mu_{\rm TF}}{\hbar\dd N}\right)^2 N \Gamma_3 t^3
\label{eq:varpertes1}
\ee
We introduced the $\Gamma_3$ particle number decay rate, related as follows to the $K_3$ rate constant of the three-body losses and to
the Thomas-Fermi $\rho_0(\rr)$ density profile of the condensate:
\be
\frac{\dd}{\dd t}N \equiv -\Gamma_3 N = -K_3 \int\dd^3r \,[\rho_0(\rr)]^3
\ee
We obtain a more meaningful writing, directly comparable to our results without losses, by writing (\ref{eq:varpertes1}) in adimensioned form:
\be
\overline{\mbox{Var}}_{\rm losses} [\hat{\theta}(t)-\hat{\theta}(0)] \underset{\Gamma_3 t\to 0}{\sim} \frac{8}{525\pi} \bar{K}_3 {\bar{t}}\,^3
\ee
where $\overline {\mbox{Var}}$ and $\bar{t}$ are the variance of the phase shift and the elapsed time in the units of figure \ref{fig:VaretC}, 
and $\bar{K}_3 = m K_3/(\hbar a^4)$. The reduced constant $\bar{K}_3$ is an intrinsic property of the atomic species used in the experiment 
(even if it can be varied using a magnetic Feshbach resonance \cite{Lev}). To estimate the order of magnitude of $\bar{K}_3$ in a gas of cold atoms, 
consider the example of rubidium 87 in the ground hyperfine sublevel $|F=1, m_F = -1 \rangle$ at vanishing magnetic field:
measurements give $ K_3=6 \times 10^{- 42} $ m $ {} ^ 6 $/s and $a= 5.31 $nm \cite {Egorov} so $\bar{K}_3\simeq 10$. In figure \ref{fig:VaretC}a 
($k_B T=\mu_{\rm TF} $), at the reduced entrance time $\bar{t}=5$ in the asymptotic regime of phase diffusion we see that the loss-induced parasitic 
variance for this value of $\bar{K}_3 $ is about three times the useful variance;
their very different time dependencies should, however, make it possible to separate them. 
The situation is much more favorable at higher temperature, $ k_B T \gg \mu_{\rm TF} $, the effect of the losses on the phase shift variance 
being for example still negligible at the reduced time $\bar{t}=$ 100 in figure \ref{fig:VaretC}b ($ k_B T = 10 \mu_{\rm TF} $).
}

\subsection{Discussion of ergodic hypothesis}
\label{subsec:discuss_ergo}

\begin{figure}[t]
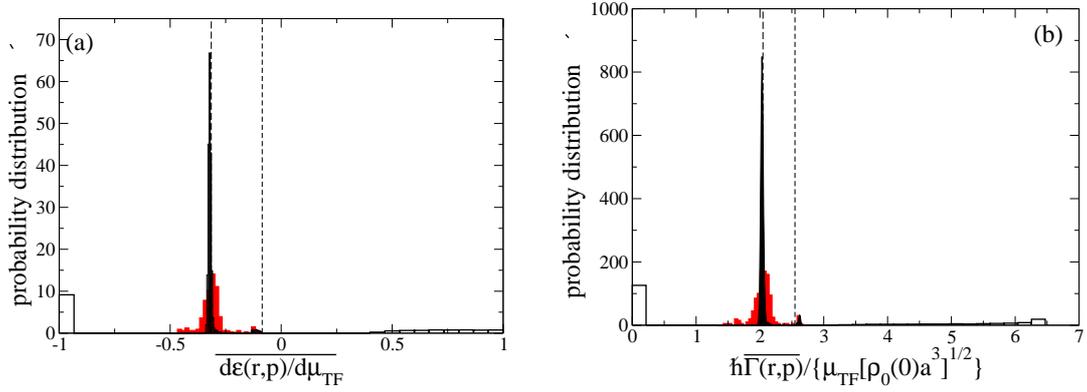

\centerline{\includegraphics[width=0.4\textwidth,clip=]{histoergo2_dedmu_en.eps}\quad \quad \includegraphics[width=0.418\textwidth,clip=]{histoergo2_taux_en.eps}}
\caption{Histogram of time averages of the physical quantities (a) $ {\dd \epsilon (\rr, \pp)} / {\dd \mu _ {\rm TF}} $ and (b) 
$ \Gamma (\rr, \pp) $ for $ 10^4 $ independent initial values $ (\rr (0), \pp (0)) $ drawn uniformly on the energy shell 
$ \epsilon = \mu _ {\rm TF} $ and a Hamiltonian evolution (\ref{eq:drdt}, \ref{eq:dpdt}) 
of Bogoliubov quasiparticles for a variable duration $ t $: $ t = 0 $ (hollow bars, black line), 
$ t = 5 \times 10^3 / \bar {\omega}$ (solid red bars), $ t = 2.5 \times 10^5 / \bar {\omega} $ (black solid bars). 
The harmonic potential is completely anisotropic, 
with incommensurate trapping frequencies (ratios $ \omega_x: \omega_y: \omega_z = 1: \sqrt {3}: \sqrt {5} -1 $). 
 The temperature, which enters in the 
$ \Gamma (\rr, \pp) $ damping rate, is $ T = \mu _ {\rm TF} / k_B $. Black vertical dashed lines: on the left, the ergodic value (average of the physical quantity 
on the energy shell $ \epsilon $); on the right, the temporal average of the quantity over a period of the linear trajectory of energy  
$ \epsilon $ 
along the direction $ Oy $ (most confining axis of the trap), obtained analytically for (a) (see note \ref{note:trajlin}) and 
numerically for (b). }
\label {fig:histoergo} 
\end{figure} 

\begin{figure}[ht] 
\begin{center} 
\includegraphics[width=0.4\textwidth,clip=]{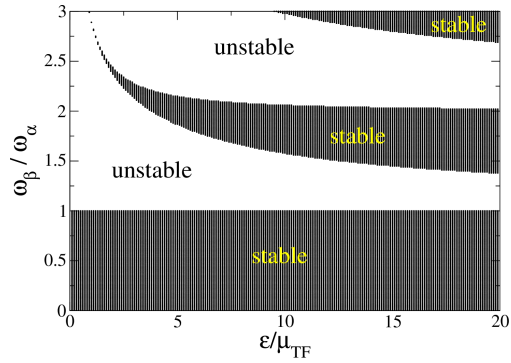} 
\end{center} 
\caption {For the classical Hamiltonian dynamics of Bogoliubov quasiparticles in a harmonic potential, 
stability of the linear motion along an  eigenaxis $ O \alpha $ of the trap for an infinitesimal initial perturbation (a 
displacement) along another eigenaxis $ O \beta $, as a function of the energy $ \epsilon $ of the trajectory and the ratio $ \omega_ \beta / \omega_ \alpha $ 
of the trapping frequencies (the shaded areas are stable).} 
\label {fig:stabil} 
\end{figure} 

\begin{figure} [t] 
\begin{center} 
\includegraphics [width = \textwidth,clip=] {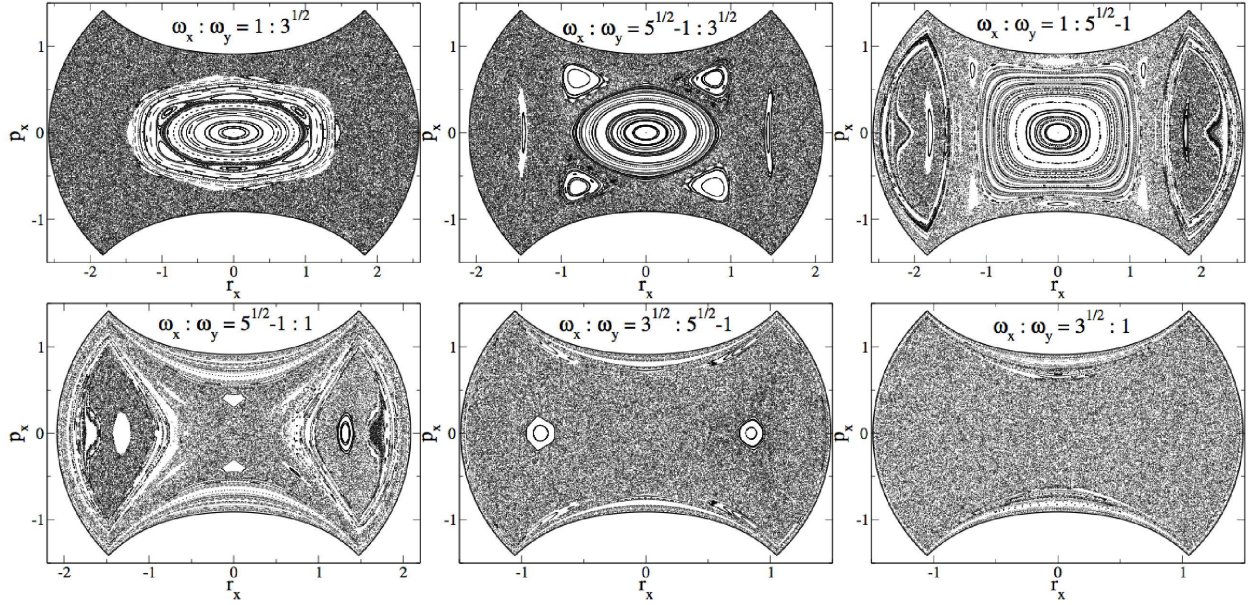} 
\end{center} 
\caption {For the classical Hamiltonian dynamics of Bogoliubov quasiparticles in a harmonic potential, 
Poincaré sections in the plane $ (r_y = 0, p_y (\epsilon)> 0) $ of planar trajectories in $ xOy $ 
(200 independent trajectories, evolution time $ 5000 / \bar { \omega} $), 
with a ratio $ \omega_x: \omega_y $ taking all 
possible values in the trap of figure \ref{fig:histoergo}: $ 1: \sqrt {3} $, $ \sqrt {3 }: 1 $, $ 1: \sqrt {5} -1 $, $ \sqrt {5} -1: 1 $, 
$ \sqrt {3}: \sqrt {5} -1 $ and $ \sqrt {5} -1: \sqrt {3} $. 
{\yvan The sections are ordered by increasing ratio $\omega_x/\omega_y$ from left to right and from top to bottom
(given that $1/\sqrt{3} < (\sqrt{5}-1)/\sqrt{3} < 1/(\sqrt{5}-1) < 1$). This shows that}
the Poincaré section is more chaotic as the ratio $ \omega_x / \omega_y $ is larger. 
$r_x$ is in units of $(\mu_{\rm TF} / m \bar {\omega} ^ 2) ^ {1/2} $ and $ p_x $ in units of $(m\mu_{\rm TF})^{1/2}$.} 
\label{fig:poinc} 
\end{figure} 

\begin{figure}[t]
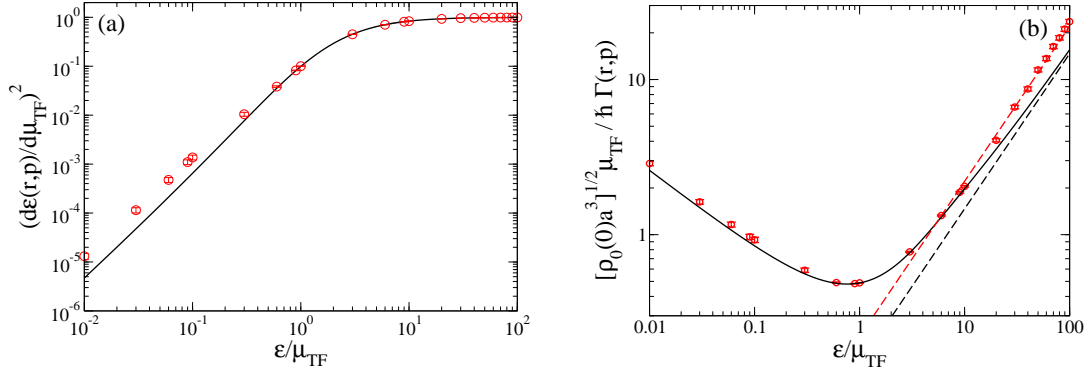
 
\centerline{\includegraphics[width=0.418\textwidth,clip=]{dedm2_ener_log.eps}\quad \quad \includegraphics[width=0.4\textwidth,clip=]{invtaux_ener_log_en.eps}}
\caption{Visualization of the error introduced by the ergodic hypothesis $ \bar {O} = \langle O \rangle_ \epsilon $ 
on two physical quantities $ O (\rr, \pp) $ involved in the diffusion of the condensate phase, as a function of the energy $ \epsilon $: 
(a) for $ O (\rr, \pp) = {\dd \epsilon (\rr, \pp)} / {\dd \mu _ {\rm TF}} $, 
we compare as in equation (\ref{eq:entre1}) $ \langle \bar {O} ^ 2 \rangle_\epsilon $ (red circles) 
to its ergodic approximation $ \langle O \rangle^ 2_\epsilon $ from (\ref{eq:Bergoexpli}) (solid line black); 
(b) for $ O (\rr, \pp) = \Gamma (\rr, \pp) $, we compare as in equation (\ref{eq:entre2}) $ \langle 1 / \bar {O } \rangle_ \epsilon $ (red circles) 
to its ergodic approximation $ 1 / \langle O \rangle_\epsilon $ (solid line black). 
The time average $ \bar {O} $ is computed over an evolution time $ t = 5 \times 10^4 / \bar{\omega} $ 
{\yvan of the quasiparticles in the trap of figure~\ref{fig:histoergo}}; 
the average on the energy shell $ \epsilon $ is taken on 200 independent trajectories, 
with initial conditions $ (\rr, \pp) $ drawn according to the uniform law $\delta(\epsilon-\epsilon(\rr,\pp))/\rho(\epsilon)$, 
which leads to a statistical uncertainty represented by the error bars in the figure. 
Dashed lines in (b): asymptotic equivalents (\ref{eq:invGamasympt}) of $\langle 1/\bar{\Gamma}\rangle_\epsilon$ (in red)  
and (\ref{eq:asymptGammaeps}) of $1/\langle \Gamma\rangle_\epsilon$ (in black).} 
\label{fig:compergo} 
\end{figure} 

As shown by references \cite{Graham, GrahamD} in the case of a 
harmonic trap with cylindrical symmetry, the classical motion of Bogoliubov quasiparticles is highly chaotic at energies $ \epsilon \simeq \mu _ {\rm TF} $ but even at this 
energy, the Poincaré sections reveal patches of stability in the phase space, which are not crossed 
by the trajectories of the chaotic sea: there is no ergodicity in the strict sense. 

What about the case of a completely anisotropic trap ~? We want to test the ergodic hypothesis for two physical quantities. 
The first one appears in our linearized kinetic equations, it is the $ \Gamma (\rr, \pp) $ damping rate. 
The second appears in the initial conditions of the $ C _ {\rm mc} (\tau) $ correlation function of $ {\dd \hat {\theta}} / {\dd t} $, 
it is $ {\dd \epsilon (\rr, \pp)} / {\dd \mu _ {\rm TF}} $. For a uniform sampling of the energy surface $ \epsilon $, 
that is with the probability distribution $ \delta (\epsilon- \epsilon (\rr, \pp)) / \rho (\epsilon) $ in the 
phase space, we show in figure \ref{fig:histoergo} the histograms of these quantities after time averaging 
on each trajectory {\yvan over} times $ t = 0 $, $ t = 5000 / \bar {\omega} $ and $ t = 250 \, 000 / \bar {\omega} $, 
at {\yvan the energy} $ \epsilon = k_B T = \mu _ {\rm TF} $ for incommensurable trapping frequencies. 
\footnote{Equations of motion (\ref{eq:drdt}, \ref{eq:dpdt}), 
put together as $ \frac{\dd} {\dd t} \mathbf {X} = \mathbf {f} (\mathbf {X}) $, are 
numerically integrated with a semi-implicit scheme of the second order, $ \mathbf {X} (t + \dd t) = \mathbf {X} (t) + \dd t [1- \frac {\dd t} {2} M] ^ {- 1} \mathbf {f} (\mathbf {X} (t)) $ 
where $ M $ is the first differential of $ \mathbf {f} (\mathbf {X}) $ in $ \mathbf {X} (t) $ \cite {Recipes}. If the trajectory crosses the surface 
of the condensate between $ t $ and $ t + \dd t $, we must determine the crossing time $ t_s $ with an error $ O (\dd t) ^ 3 $, then 
apply the diagram semi-implicit successively on $ [t, t_s] $ and $ [t_s, t + \dd t] $, to overcome the discontinuity of $ \mathbf {f} (\mathbf {X}) $ 
and its derivatives.} The temporal averaging 
leads to a spectacular narrowing of the probability distribution, which peaks around the 
ergodic mean (dashed line on the left), which goes in the direction of the ergodic hypothesis. This narrowing dynamics continues 
over very long times, but never eliminates a small lateral peak far from the ergodic average. 

An inspection of the trajectories contributing to the lateral peak shows that they are small perturbations of the stable linear trajectories along the most confining
trap axis. The temporal average value of the two quantities considered on these linear trajectories 
is represented by the right dashed vertical lines in figure \ref{fig:histoergo}, it is actually close to the peak in question. 
The stability diagram of a linear trajectory along an eigenaxis $ \alpha $ of the trap, with respect 
to a perturbation along another eigenaxis $ \beta $ is shown in figure \ref{fig:stabil}, in the plane (energy, ratio 
$ \omega_ \beta / \omega_ \alpha $). It shows that the linear trajectory along the most confining axis is stable at all energies. 
\footnote {The linear trajectory of a quasiparticle of energy $ \epsilon $ along the proper axis $ O \alpha $ of the trap is written 
$ m ^ {1/2} \omega_ \alpha r_ \alpha (t) = | \mu _ {\rm TF} + \ii \epsilon | \sin (\sqrt {2} \omega_ \alpha t) / \sqrt {G (t)} $ and 
$ p_ \alpha (t) / (2m) ^ {1/2} = \epsilon / \sqrt {G (t)} $ with $ G (t) = \mu _ {\rm TF} + | \mu _ {\rm TF} + \ii \epsilon | \cos (\sqrt {2} \omega_ \alpha t) $. This 
corresponds to the choice $ r_ \alpha (0) = 0, p_ \alpha (0) \geq 0 $ and is for $ -t_s \leq t \leq t_s $, where the time to reach the surface of the condensate 
is given by $ \sqrt {2} \omega_ \alpha t_s = \acos \frac{\epsilon- \mu _ {\rm TF}} {| \mu_ {\rm TF} + \ii \epsilon |} $. Outside the condensate, 
the quasiparticle oscillates harmonically like a free particle for a time $ 2 \omega_ \alpha ^ {- 1} \atan [(\epsilon / \mu _ {\rm TF}) ^ {1 / 2}] $ 
before regaining the condensate, to cross it in a time $ 2 t_s $, and so on. The knowledge of the trajectory 
makes immediate the linear analysis of numerical stability. It also allows to calculate analytically the time average 
of $ \dd \epsilon (\rr, \pp) / \dd \mu _ {\rm TF} $ on the linear trajectory; if we put $ \epsc = \epsilon / \mu_ {\rm TF} $, the result is written 
\[
\overline{\frac{\dd\epsilon(\rr,\pp)}{\dd\mu_{\rm TF}}} = \frac{\ln\frac{1+\epsc+\sqrt{2\epsc}}{(1+\epsc^2)^{1/2}}-\sqrt{2}\atan\sqrt{\epsc}}
{\acos\frac{\epsc-1}{(1+\epsc^2)^{1/2}} +\sqrt{2}\atan\sqrt{\epsc}}
\]
\label{note:trajlin} 
} 
The Poincaré sections of the planar trajectories in the $ \alpha O \beta$ planes in figure \ref{fig:poinc} 
specify the width of the stability island and reveal 
the existence of secondary islands, etc. There is therefore no full ergodicity of our classical dynamics, 
even at the energies $ \epsilon \approx \mu _ {\rm TF} $, even in the completely anisotropic case. 

To quantitatively measure the error made by the ergodic hypothesis in the computation of $ C _ {\rm mc} (0) $ and $ C _ {\rm mc} (\tau> 0) $, we 
consider the differences between 
\bea
\label{eq:entre1}
\left\langle\overline{\frac{\dd\epsilon(\rr,\pp)}{\dd\mu_{\rm TF}}}^2\right\rangle_\epsilon &\mbox{and}& 
\left\langle\frac{\dd\epsilon(\rr,\pp)}{\dd\mu_{\rm TF}}\right\rangle_\epsilon^2 \\
\left\langle \frac{1}{\overline{\Gamma(\rr,\pp)}} \right\rangle_\epsilon &\mbox{and}& \frac{1}{\left\langle{\Gamma(\rr,\pp)}\right\rangle}_\epsilon
= \frac{1}{\Gamma(\epsilon)}
\label{eq:entre2}
\eea
where we recall that the horizontal bar $ \overline {O (\rr, \pp)} $ above a physical quantity represents the time average on the 
trajectory going through $ (\rr, \pp) $ in phase space, {\yvan as in equation (\ref{eq:approx_secu})},
and the brackets $ \langle O (\rr, \pp) \rangle_ \epsilon $ represent 
the uniform mean over the energy shell $ \epsilon $ as in equation (\ref{eq:defGamergo}). In equations (\ref{eq:entre1}, \ref{eq:entre2}), the left column contains the quantities appearing in $ C _ {\rm mc} (0) $ 
or in the secular kinetic equations before the ergodic approximation, and the right column what they become 
after ergodic approximation. Importantly, we consider in equation (\ref{eq:entre2}) 
$ 1 / \bar {\Gamma} $ rather than $ \bar {\Gamma} $ because it is the inverse $ M ^ {- 1} $ and $ M ^ {- 2} $ that appear in expressions
(\ref{eq:Dint}, \ref{eq:t0rapint}) 
of the diffusion coefficient $ D $ and the delay time $ t_0 $, $ M $ being the operator representing the right-hand side of 
linearized kinetic equations (\ref{eq:evolX}). \footnote{Should it be recalled, $ \langle \overline {\Gamma (\rr, \pp)} \rangle_ \epsilon = \Gamma (\epsilon) $, the 
uniform mean being invariant by time evolution. So, the inequality between arithmetic mean and harmonic mean 
imposes $ \langle 1 / \overline {\Gamma (\rr, \pp)} \rangle_ \epsilon \ge 1 / \Gamma (\epsilon) $.} 
The quantities to be compared (\ref{eq:entre1}, \ref{eq:entre2}) are represented as functions of the energy $ \epsilon $ in figure 
\ref{fig:compergo} at temperature $ T = \mu_ { \rm TF} / k_B $. There is a remarkable agreement on a wide range of energies 
around $ \epsilon = \mu _ {\rm TF} $. 
Deviations from the ergodic {\yvan approximation} at very low energy and very high energy were expected: within these limits, 
the classical dynamics becomes integrable \cite{Graham}. At high energy, we obtain for the quantity $ \Gamma (\rr, \pp) $ the 
following analytic prediction: \footnote{ 
At the dominant order in $ \epsilon $, we get $ \overline { \Gamma (\rr, \pp)} $ using the equivalent (\ref{eq:Gamhomasympt}) 
(in which $ \mu_0 = g \rho_0 (\rr) $) on a harmonic trajectory undisturbed by the condensate, $ r_ \alpha (t) = A_ \alpha \cos (\omega_ \alpha t + \phi_ \alpha) $, 
$ \forall \alpha \in \{x, y, z \} $. Let us consider cleverly the quantity $ g \rho_0 (\rr) $ to be averaged as a function $ f (\boldsymbol {\theta}) $ 
of the angles $ \theta_ \alpha = \omega_ \alpha t + \phi_ \alpha $. It is a periodic function of period $ 2 \pi $ in each direction, expandable 
in Fourier series, $ f (\boldsymbol {\theta}) = \sum_ {\mathbf {n} \in \mathbb {Z} ^ 3 } c _ {\mathbf {n}} \eee ^ {\ii \mathbf {n} \cdot \boldsymbol {\theta}} $. 
In the incommensurable case, $ \mathbf {n} \cdot \boldsymbol {\omega} \neq 0 $ and the time average of $ \eee ^ {\ii \mathbf {n} \cdot \boldsymbol {\theta}} $ is 
zero $ \forall \mathbf {n} \in \mathbb {Z} ^ {3 *} $, so that $ \overline {f (\boldsymbol {\theta})} = c _ {\mathbf {0}} $. 
In the usual integral expression of $ c _ {\mathbf {0}} $, we make the change of variable $ x_ \alpha = X_ \alpha \cos \theta_ \alpha $, 
where $ X_ \alpha = (\epsilon_ \alpha / \mu _ {\rm TF}) ^ {1/2} $ and $ \epsilon_ \alpha $ is the energy 
of the motion along $ O \alpha $. It remains to take the limit of all $ X_ \alpha $ tending to $ + \infty $ under the integral sign 
to get 
\[
\frac{\hbar\overline{\Gamma(\rr,\pp)}}{\mu_{\rm TF}[\rho_0(\mathbf{0})a^3]^{1/2}} \underset{\epsilon\to +\infty}{\sim} 
\frac{32\sqrt{2}}{15\pi^{3/2}} \left(\frac{\epsilon}{\mu_{\rm TF}}\right)^{1/2} 
\prod_\alpha \left(\frac{\mu_{\rm TF}}{\epsilon_\alpha}\right)^{1/2}
\]
By averaging {\yvan the inverse of} this equivalent 
over the probability distribution $ 2 \epsilon ^ {- 2} \delta (\epsilon- \sum_ \alpha \epsilon_ \alpha) $ of the energies 
per direction for a harmonic oscillator of total energy $ \epsilon $, we find (\ref{eq:invGamasympt}).} 
\be
\left\langle \frac{1}{\hbar\overline{\Gamma(\rr,\pp)}}\right\rangle_\epsilon 
\underset{\epsilon\to+\infty}{\sim}\frac{\pi^{5/2}}{56\sqrt{2}}\frac{\epsilon}{\mu_{\rm TF}^2} \frac{1}{[\rho_0(\mathbf{0})a^3]^{1/2}}
\label{eq:invGamasympt}
\ee
It differs from the ergodic prediction (\ref{eq:asymptGammaeps}) by a numerical coefficient, and reproduces well the results of the 
numerical simulations (see the red dashed line in figure \ref{fig:compergo} b). This prevents us from calculating the diffusion coefficient 
$ D $ and the diffusion delay $ t_0 $ in 
the secularo-ergodic approximation at a too high temperature. 
Concerning $ {\dd \epsilon (\rr, \pp)} / {\dd \mu _ {\rm TF}} $, which tends to $ -1 $ at high energy, 
the deviation can only be significant at low energy; in fact it does so only at very low energy, and it would be a 
problem for our ergodic calculation of $ D $ and $ t_0 $ only at temperatures $ k_B T \ll \mu _ {\rm TF} $ rarely reached 
in cold atom experiments.

\section{Conclusion}
\label{sec:conclusion}

Motivated by recent experimental advances in the manipulation of trapped cold atom gases 
\cite {Treutlein, Oberthaler, Schmiedmayer}, 
we theoretically studied the coherence time and the phase dynamics of a Bose-Einstein condensate in an isolated and harmonically
trapped boson gas, a fundamental problem important for interferometric applications.
The variance of the phase shift experienced by the condensate after a time $ t $ increases indefinitely with $ t $, which limits the intrinsic coherence time of the gas. For $ t \gg t _ {\rm coll} $, where $ t _ {\rm coll} $ is the typical collision time between Bogoliubov quasiparticles, it becomes a quadratic function of time, 
\begin{equation} 
\mbox {Var} [\hat {\theta} (t) - \hat {\theta} (0)] = At^ 2 + 2D (t-t_0) + o (1) 
\end{equation}
where $ \hat {\theta} $ is the phase operator of the condensate. This asymptotic law has the same form as in the spatially homogeneous case previously studied \cite {PRAlong}, which was not guaranteed, but the coefficients of course differ. To calculate them, we consider the thermodynamic limit in the trap, in which the number of particles tends to infinity, 
$ N \to + \infty $, at fixed temperature $ T $ and fixed Gross-Pitaevskii chemical potential $ \mu _ {\rm GP} $. This requires that the reduced trapping frequencies tend to zero, 
$ \hbar \omega_ \alpha / \mu_ {\rm GP} \to 0 $, which we reinterpret as a classical limit $ \hbar \to0 $. 

The dominant term $ At ^ 2 $ is due to the fluctuations in the initial state of the quantities conserved by temporal evolution, $ N $ and $ E $, where $ E $ is the total energy of the gas. We give an explicit expression (\ref{eq:grandA}) - (\ref{eq:acalc1}) - (\ref{eq:acalc2}) 
of the coefficient $ A $ in a generalized ensemble, any statistical mixture of microcanonical ensembles with at most normal fluctuations of $ N $ and $ E $. 
In this case, $ A = O (1 / N) $. We obtain a simpler form (\ref{eq:Apart}) in the case of a statistical mixture of canonical ensembles of the same temperature but of variable number of particles. At usual temperatures, larger than $ \mu _ {\rm GP} / k_B $, and for 
Poissonian particle number fluctuations, the contribution to $A$ of thermal fluctuations of $ E $ is rendered negligible by a factor of order the non-condensed fraction $ \propto (T / T_c) ^ {3} $. The variance of $ N $ must be reduced to see the effect of thermal fluctuations on the ballistic spread of the 
condensate phase. 

The subdominant term $ 2D (t-t_0) $ does not depend on the ensemble in which the system is prepared, at least to the first non-zero order $ 1 / N $ at the thermodynamic limit, and it is the only one that remains in the microcanonical ensemble. 
The calculation of its two ingredients, the diffusion coefficient $ D $ of the phase and the diffusion delay $ t_0 $, 
requires the knowledge at all times of the correlation function of $ \dd {\hat {\theta}} / \dd t $ in the microcanonical ensemble, 
and thus the resolution of linearized kinetic equations on the Bogoliubov quasiparticle occupation numbers. 
It is indeed the temporal fluctuations of these occupation numbers for a given realization of the system which stochastise the evolution of the phase of the condensate. 
To this end, we adopt a semiclassical description, in which the motion of quasiparticles in the trapped gas is 
treated classically in the phase space $ (\rr, \pp) $, but the quasiparticle bosonic field is still quantum, 
through the occupation number operators $ \hat {n} (\rr, \pp) $. In quantum observables of the form 
$ \hat {A} = \sum_k a_k \hat {n} _k $, such as $ \dd {\hat {\theta}} / \dd t $, the average $ a_k $ and the sum on the Bogoliubov quantum modes $k$
are then replaced, according to a correspondence principle, by a temporal mean and an integral on the classical trajectories (see equations (\ref{eq:A})-(\ref{eq:Asc})). The linearized kinetic equations on the fluctuations 
$ \delta \hat {n} (\rr, \pp) $ include a transport part, according to the classical Hamiltonian motion of quasiparticles, 
and a collision integral, local in position, which describes the Beliaev-Landau interaction processes among three quasiparticles. 
{\yvan They take the same form as the linearized quantum Boltzmann equations on the semiclassical distribution function $n(\rr, \pp, t)$ of quasiparticles in phase space.}
We simplify them in the secular limit $ \omega_ \alpha t _ {\rm coll} \gg1 $ and under the assumption of a classical ergodic motion of quasiparticles. This hypothesis, according to which the fluctuations $ \delta \hat {n} (\rr, \pp) $ averaged on a trajectory depend only on the energy of the trajectory, only holds if the trap is completely anisotropic; in this case we give a careful numerical justification.

The desired quantities $ D $ and $ t_0 $, correctly adimensioned, are universal functions 
of $ k_BT / \mu _ {\rm TF} $ where $ \mu _ {\rm TF} $ is the Thomas-Fermi limit of $ \mu_ {\rm GP} $, 
and are in particular independent of the ratios $ \omega_ \alpha / \omega _ {\beta} $ of the trapping frequencies. 
They are represented on figure \ref{fig:Diff_et_t0}. 
An interesting and more directly measurable by-product of our study are the $ \Gamma (\epsilon)$ damping rates of the 
Bogoliubov modes of energy $ \epsilon $ in the trap. Once the adimensioned temperature $ k_BT / \mu _ {\rm TF} $ is fixed, the rate is also described by a universal function of $ \epsilon / \mu _ {\rm TF} $ independent of the trapping frequencies,  
see figure \ref{fig:Gamma}. 
These results are part of a new class of universality, that of the completely anisotropic harmonic traps, 
very different from that, theoretically better explored, of spatially homogeneous systems, and will hopefully receive an experimental confirmation soon.

\section*{Acknowledgments} 
We thank the members of the cold fermions and the atom chips teams of LKB, especially Christophe Salomon, 
for useful discussions.

\section*{Appendix A. Behavior of $ \Gamma (\epsilon) $ at low and high energy}

To get the limiting behaviors (\ref{eq:gambassener}) and (\ref{eq:asymptGammaeps}) of the $ \Gamma (\epsilon)$ damping rates 
of the Bogoliubov modes in a trap in the secularo-ergodic approximation, 
we rewrite the integral in phase space (\ref{eq:defGamergo}) as an average on the local Gross-Pitaevskii 
chemical potential $ \mu_0 = g \rho_0 (\rr) $ of the damping rate $ \Gamma_h (\epsilon, \mu_0, k_B T) $ of a mode of
energy $ \epsilon $ in a homogeneous system of density $ \mu_0 / g $ and temperature $ T $: 
\be
{\Gamma}(\epsilon) = \int_0^{\mu_{\rm TF}} \dd\mu_0 P_\epsilon(\mu_0) \Gamma_h(\epsilon,\mu_0,k_B T)
\label{eq:GamepsP0}
\ee
with
\be
P_\epsilon(\mu_0) \equiv \frac{1}{\rho(\epsilon)} \int \frac{\dd^3r\,\dd^3p}{(2\pi\hbar)^3} \delta(\epsilon-\epsilon(\rr,\pp))
\delta(\mu_0-g\rho_0(\rr)) = \frac{4}{\pi\rho(\epsilon)} \frac{1}{(\hbar\bar{\omega})^3} 
\frac{\epsilon^2(\mu_{\rm TF}-\mu_0)^{1/2}}
{(\mu_0^2+\epsilon^2)^{1/2} [(\mu_0^2+\epsilon^2)^{1/2}+\mu_0]^{1/2}}
\ee

In the $ \epsilon \to 0 $ limit, we first heuristically replace the integrand in equation (\ref{eq:GamepsP0}) with a 
low energy equivalent, using: 
\bea
\label{eq:equivPeps}
P_\epsilon(\mu_0) &\underset{\epsilon\to 0}{\sim}& \frac{3}{8\sqrt{2}} \frac{\epsilon^{1/2}(\mu_{\rm TF}-\mu_0)^{1/2}} {\mu_{\rm TF}^{1/2}\mu_0^{3/2}} \\
\frac{\hbar\Gamma_h(\epsilon,\mu_0,k_B T)}{2} &\underset{\epsilon\to 0}{\sim}& \epsilon \left(\frac{\mu_0 a^3}{g}\right)^{1/2} F(k_B T/\mu_0)
\label{eq:equivGammaeps}
\eea
In equation (\ref{eq:equivPeps}), we used equation (\ref{eq:fbasse}); the result (\ref{eq:equivGammaeps}) is 
in reference \cite {Giorgini}, where the function $ F $ is computed and studied. As $ F (\theta) \underset {\theta \to + \infty} {\sim} 
\frac{3 \pi ^ {3/2}} {4} \theta $, this causes in equation (\ref{eq:GamepsP0}) 
the divergent integral $ {\epsilon ^ {3/2}} \int_0 ^ {\mu_ {\rm TF}} \dd \mu_0 / {\mu_0 ^ 2} $ to appear. It is clear, however, that
one should cut this integral to $ \mu_0> \epsilon $ so that the equivalent 
(\ref{eq:equivGammaeps}) remains usable, hence the scaling law $ {\Gamma} (\epsilon) \approx \epsilon ^ {1/2} $, 
dominated by the edge of the trapped condensate and very different from the linear law of the homogeneous case. 
To find the prefactor in the law, we simply make the change of scale $ \mu_0 = \epsilon \nu_0 $ in the integral and use 
the \oge high temperature \fge \, approximation of reference \cite{PRAGora} on $ \Gamma_h $, 
uniformly valid near the edge of the trapped condensate, 
\be
\frac{\hbar \Gamma_h(\epsilon,\mu_0,k_B T)}{2} \underset{k_B T\gg \epsilon,\mu_0}{\sim} k_B T \left(\frac{\mu_0 a^3}{g}\right)^{1/2} \phi(\epsilon/\mu_0)
\ee
before going to the $ \epsilon \to 0 $ limit under the integral sign, 
which leads to the sought equation (\ref{eq:gambassener}) 
with \footnote {In practice, the function $ \phi $ is deduced from equation (\ref{eq:Gamrp}) by doing the 
classical field approximation $ 1 + \bar {n} (\rr , \qq) \simeq \bar {n} (\rr, \qq) \simeq k_B T / \epsilon (\rr, \qq) $. In the numerical calculation of $ \mathcal {I} $, done by taking $ \check {\epsilon} 
= 1 / \nu_0 $ as the integration variable, we reduce the effect of digital truncation with the help of the development asymptotic 
$ \phi (\epsc) \underset {\epsc \to + \infty} {=} 4 \left (\frac{2 \pi} {\epsc} \right) ^ {1/2} \left [2 \ln \frac{\epsc} {2} + \frac{1- \ln (\epsc / 2)} {\epsc} 
+ \frac{23 + 6 \ln (\epsc / 2)} {8 \epsc ^ 2} + O (\frac{\ln \epsc} {\epsc ^ 3}) \right] $, which corrects and improves equation (35) 
of reference \cite {PRAGora}.} 
\be
\mathcal{I}=\int_0^{+\infty} \dd\nu_0 \frac{\nu_0^{1/2} \phi(1/\nu_0)}{(1+\nu_0^2)^{1/2} [\nu_0+(1+\nu_0^2)^{1/2}]^{1/2}}
=4.921\, 208\, \ldots
\ee

In the limit $ \epsilon \to + \infty $, we use the fact that, in the homogeneous case, the damping rate of quasiparticles 
is reduced to the collision rate $ \rho_0 \sigma v $ of a particle of velocity $ v = (2 \epsilon / m) ^ {1/2} $ with condensate particles, 
with zero velocity and density $ \rho_0 $, with the cross section $ \sigma = 8 \pi a ^ 2 $ for indistinguishable bosons (this is a 
Beliaev process): 
\be
\frac{\hbar \Gamma_h(\epsilon,\mu_0, k_B T)}{2} \underset{\epsilon\to +\infty}{\sim} \mu_0 a \frac{(2m\epsilon)^{1/2}}{\hbar}
\label{eq:Gamhomasympt}
\ee
Using the same high energy expansion (\ref{eq:fhaute}) for $ \rho (\epsilon) $, we find that 
$ P_ \epsilon (\mu_0) \sim (8 / \pi) (\mu _ {\rm TF} - \mu_0) ^ {1/2} \epsilon ^ {- 3/2} $. The insertion of these equivalents in equation 
(\ref{eq:GamepsP0}) gives (\ref{eq:asymptGammaeps}).


\end{document}


\begin{frontmatter}


\selectlanguage{francais}
\title{Temps de cohérence d'un condensat de Bose-Einstein dans un gaz isolé harmoniquement piégé \\
{\it 
Coherence time of a Bose-Einstein condensate in an isolated harmonically trapped gas
}}


\author{Yvan Castin}
\author{Alice Sinatra}

\address{Laboratoire Kastler Brossel, ENS-PSL, CNRS, Sorbonne Universit\'e et Coll\`ege de France, Paris, France}



\selectlanguage{english}
\begin{abstract}
We study the condensate phase dynamics in a low-temperature equilibrium gas of weakly interacting bosons, harmonically trapped and isolated from the environment. 
We find that at long times, much longer than the collision time between Bogoliubov quasi-particles, 
the variance of the phase accumulated by the condensate grows with a ballistic term quadratic in time and a diffusive term affine in time.
We give the corresponding analytical expressions in the limit of a large system, in the collisionless regime and 
in the ergodic approximation for the quasi-particle motion. When properly rescaled, they are described
by universal functions of the temperature divided by the Thomas-Fermi chemical potential. The same conclusion holds for the mode damping rates. 
Such universality class differs from the previously studied one of the homogeneous gas.  
\\
\noindent{\small {\it Keywords:} Bose gases; Bose-Einstein condensate; temporal coherence; trapped gases; ultracold atoms}
\\
\selectlanguage{francais}
\vskip 0.25\baselineskip
\noindent{\bf R\'esum\'e}
\vskip 0.5\baselineskip
\noindent
Nous \'etudions la dynamique de phase \`a l'\'equilibre d'un condensat dans un gaz de bosons en interaction faible harmoniquement
pi\'eg\'e et isol\'e de l'environnement. Nous trouvons qu'au bout d'un temps long devant le temps de collision typique 
entre les quasi-particules de Bogolioubov,
la variance du d\'ephasage du condensat comporte en g\'en\'eral un terme balistique quadratique en temps et un terme diffusif
affine en temps. Nous donnons des expressions analytiques des coefficients correspondants, \`a la limite d'un grand syst\`eme,
dans le r\'egime faiblement collisionnel et dans l'approximation ergodique pour le mouvement des quasi-particules. Correctement adimensionn\'es,
ils sont d\'ecrits, tout comme les taux d'amortissement des modes, par des fonctions universelles de la temp\'erature ramen\'ee au potentiel
chimique de Thomas-Fermi du condensat. Cette classe d'universalit\'e diff\`ere de celle pr\'ec\'edemment \'etudi\'ee du gaz spatialement
homog\`ene.
\\
\noindent{\small{\it Mots-cl\'es~:} gaz de bosons~; condensat de Bose-Einstein~; coh\'erence temporelle~; gaz pi\'eg\'es~; atomes froids}

\noindent 
\vskip 0.5\baselineskip

\selectlanguage{francais}
\end{abstract} 
\end{frontmatter}

\selectlanguage{francais}


\section{Introduction et vue d'ensemble}

Nous consid\'erons ici un probl\`eme encore non r\'esolu de la th\'eorie des gaz quantiques, celui du temps de coh\'erence d'un gaz de bosons 
sans spin en interaction faible de port\'ee n\'egligeable,
pr\'epar\'e dans un pi\`ege harmonique \`a l'\'equilibre thermique \`a une temp\'erature $T$ tr\`es inf\'erieure \`a la temp\'erature critique $T_c$,
c'est-\`a-dire dans un r\'egime fortement condens\'e de Bose, et parfaitement isol\'e dans son \'evolution ult\'erieure. 
Le temps de coh\'erence du champ bosonique est alors intrins\`eque et domin\'e par celui du condensat. Au vu de progr\`es techniques r\'ecents 
\cite{Treutlein,Oberthaler,Schmiedmayer}, 
cette question pourrait recevoir bient\^ot une r\'eponse exp\'erimentale dans les gaz d'atomes froids confin\'es dans des potentiels 
magn\'etiques non dissipatifs \cite{Cornell,Ketterle,Stringari_revue} et, contrairement \`a d'autres
syst\`emes de gaz de bosons en physique du solide \cite{JBloch,LeSiDang,Bramati,Dubin}, 
bien d\'ecoupl\'es de leur environnement et ne pr\'esentant que de faibles pertes de particules. Aussi notre \'etude
th\'eorique est-elle importante pour les applications \`a venir en optique atomique et interf\'erom\'etrie \`a ondes de mati\`ere.

Faisant suite aux travaux pionniers des r\'ef\'erences \cite{Zoller,Grahamdiff,Kuklov},
nos \'etudes th\'eoriques \cite{Superdiff,Genuine,PRAlong,chapitre}, 
effectu\'ees dans un gaz de bosons spatialement homog\`ene, s'appuyent sur la m\'ethode de Bogolioubov,
qui r\'eduit le syst\`eme \`a un gaz de quasi-particules en interaction faible. Elles ont identifi\'e deux m\'ecanismes limitant le temps
de coh\'erence et faisant intervenir tous deux la dynamique de l'op\'erateur phase $\hat{\theta}(t)$ du condensat~:
\begin{itemize}
\item {\sl le brouillage de phase}~: lorsque les quantit\'es conserv\'ees (l'\'energie $E$ du gaz et son nombre de particules $N$) fluctuent
d'une r\'ealisation exp\'erimentale \`a l'autre, la vitesse moyenne d'\'evolution de la phase $[\hat{\theta}(t)-\hat{\theta}(0)]/t$ sur une r\'ealisation,
fonction de ces quantit\'es conserv\'ees, fluctue elle aussi. Apr\`es moyenne sur les r\'ealisations, ceci induit un \'etalement balistique du d\'ephasage $\hat{\theta}(t)-\hat{\theta}(0)$, c'est-\`a-dire une divergence quadratique de sa variance, avec un coefficient balistique $A$ \cite{Superdiff}~:
\be
\mbox{Var}[\hat{\theta}(t)-\hat{\theta}(0)] \sim A t^2
\label{eq:bali}
\ee
ceci aux temps longs devant $\gamma_{\rm coll}^{-1}$, o\`u $\gamma_{\rm coll}$ est le taux de collision typique entre les quasi-particules de Bogolioubov
thermiques~;
\item {\sl la diffusion de phase}~: m\^eme si le syst\`eme est pr\'epar\'e dans l'ensemble microcanonique, o\`u $E$ et $N$ sont fix\'es,
les interactions entre quasi-particules font fluctuer leurs nombres d'occupation, et donc la vitesse instantan\'ee $\frac{\dd}{\dd t} \hat{\theta}$
de la phase, qui en d\'epend. Ceci induit un \'etalement diffusif de $\hat{\theta}(t)-\hat{\theta}(0)$ aux temps
$t\gg \gamma_{\rm coll}^{-1}$, avec un coefficient de diffusion $D$ \cite{Genuine,PRAlong}~:
\be
\mbox{Var}_{\rm mc} [\hat{\theta}(t)-\hat{\theta}(0)] \sim 2 D t
\label{eq:diff} 
\ee
\end{itemize}
Dans le cas g\'en\'eral, les deux m\'ecanismes sont pr\'esents et la variance du d\'ephasage admet (\ref{eq:bali}) comme terme dominant,
(\ref{eq:diff}) comme terme sous-dominant. L'\'etalement de la phase du condensat renseigne directement sur sa fonction de coh\'erence temporelle
du premier ordre,
\be
g_1(t) = \langle \hat{a}_0^\dagger(t) \hat{a}_0(0)\rangle
\ee
o\`u $\hat{a}_0$ est l'op\'erateur d'annihilation d'un boson dans le mode du condensat, en vertu de la relation approch\'ee
\be
g_1(t) \simeq \eee^{-\ii \langle\hat{\theta}(t)-\hat{\theta}(0)\rangle} \eee^{-\mathrm{Var}[\hat{\theta}(t)-\hat{\theta}(0)]/2}
\label{eq:g1}
\ee
admise dans la r\'ef\'erence {\yvan \cite{chapitre} sous l'hypoth\`ese d'une distribution gaussienne de $\hat{\theta}(t)-\hat{\theta}(0)$},
puis justifi\'ee dans la r\'ef\'erence \cite{CohFer} \`a suffisamment basse temp\'erature {\yvan sous des conditions assez g\'en\'erales.
\footnote{{\yvan Rappelons les hypoth\`eses utilis\'ees dans la r\'ef\'erence \cite{CohFer} pour \'etablir l'\'equation (\ref{eq:g1}). $(i)$ Les fluctuations
relatives du module de $\hat{a}_0$ sont faibles, le syst\`eme \'etant fortement condens\'e de Bose. $(ii)$ Le syst\`eme est suffisamment proche de
la limite thermodynamique, avec des fluctuations normales et des lois asymptotiquement gaussiennes pour l'\'energie et le nombre de particules. 
Ceci sert en 
particulier \`a mettre la contribution balistique du d\'ephasage \`a $g_1(t)$ sous la forme (\ref{eq:g1}). $(iii)$ Le coefficient de diffusion
de la phase (d'ordre $1/N$) doit \^etre beaucoup plus faible que le taux de collision typique $\gamma_{\rm coll}$ entre quasi-particules
de Bogolioubov  (d'ordre $N^0$) mais beaucoup plus grand que l'espacement des niveaux d'\'energie (d'ordre $N^{-2}$) des paires de quasi-particules
cr\'e\'ees ou annihil\'ees lors des processus de collision Beliaev-Landau. Ceci sert, pour un syst\`eme pr\'epar\'e dans l'ensemble
microcanonique, \`a montrer que $g_1(t)$ est de la forme (\ref{eq:g1}) sur les intervalles de temps $t=O(N^0)$ et $t=O(N^1)$, avec le m\^eme
coefficient de diffusion. $(iv)$ La fonction de corr\'elation de $\frac{\dd}{\dd t}\hat{\theta}$ est r\'eelle, comme le pr\'edisent les 
\'equations cin\'etiques. $(v)$ On n\'eglige le commutateur de $\hat{\theta}(t)$ avec $\hat{\theta}(0)$, ce qui introduit une erreur de phase
$O(t/N)$ dans le facteur $\exp[-\ii\langle\hat{\theta}(t)-\hat{\theta}(0)\rangle]$. C'est une erreur d'ordre unit\'e aux temps $t\approx N$ mais $g_1(t)$ a alors
commenc\'e \`a d\'ecro\^\i tre sous l'effet de la diffusion de phase dans l'ensemble microcanique (et s'est sinon d\'ej\`a tr\`es fortement amortie
sous l'effet du brouillage de phase balistique au bout d'un temps $t\approx N^{1/2}$).}}}

Nous nous proposons ici de g\'en\'eraliser ces premi\`eres \'etudes au cas exp\'erimentalement plus habituel d'un syst\`eme harmoniquement
pi\'eg\'e (voir cependant la r\'ef\'erence \cite{ZoranBoite}). 
Les lois de d\'ependance des taux d'amortissement des modes de Bogolioubov en l'\'energie propre des modes ou en la temp\'erature
sont d\'ej\`a tr\`es diff\'erentes de celles du cas homog\`ene, comme l'a montr\'e la r\'ef\'erence \cite{PRLGora}. Il en ira certainement
de m\^eme pour l'\'etalement de la phase du condensat. Le cas pi\'eg\'e est non trivial, puisque les modes de Bogolioubov
ne sont pas connus analytiquement, et qu'il n'y a pas d'approximation d'homog\'en\'eit\'e locale applicable {\yvan \`a l'\'evolution de la phase}
(comme l'a v\'erifi\'e la r\'ef\'erence \cite{EPLspin}).  Nous disposons heureusement d'\'echappatoires~:
\begin{itemize}
\item la limite classique pour le mouvement des quasi-particules de Bogolioubov dans le gaz pi\'eg\'e. En effet, \`a la limite thermodynamique 
($N\to +\infty$ \`a potentiel chimique de Gross-Pitaevskii $\mu_{\rm GP}$ et temp\'erature fix\'es), les pulsations de pi\'egeage
$\omega_\alpha$, $\alpha\in \{x,y,z\}$, tendent vers z\'ero {\yvan comme $1/N^{1/3}$} si bien que
\be
\hbar\omega_\alpha \ll \mu_{\rm GP}, k_B T
\label{eq:condval1}
\ee
ce que l'on peut r\'einterpr\'eter astucieusement comme une limite classique $\hbar\to 0$~;
\item la limite d'interactions tr\`es faibles entre les quasi-particules de Bogolioubov~:
\be
\gamma_{\rm coll} \ll \omega_\alpha
\label{eq:condval2}
\ee
Ceci implique que tous les modes du condensat, m\^eme ceux de plus basse pulsation $\approx \omega_\alpha$, sont dans le r\'egime
faiblement collisionnel (par opposition \`a hydrodynamique), et permet d'effectuer une approximation s\'eculaire sur les \'equations cin\'etiques
d\'ecrivant les collisions entre les quasi-particules~;
\item l'ergodicit\'e dans un pi\`ege compl\`etement anisotrope~: comme l'ont montr\'e les r\'ef\'erences \cite{Graham,GrahamD}, le mouvement
classique des quasi-particules dans un pi\`ege harmonique non isotrope \`a sym\'etrie de r\'evolution
est fortement chaotique aux \'energies $\epsilon\approx \mu_{\rm GP}$ mais
quasi int\'egrable lorsque $\epsilon\to 0$ ou $\epsilon\to +\infty$.  Dans un pi\`ege compl\`etement anisotrope,
aux temp\'eratures ni trop petites ni trop grandes devant $\mu_{\rm GP}/k_B$,
on peut esp\'erer compl\'eter l'approximation s\'eculaire par l'hypoth\`ese d'ergodicit\'e, ce que nous nous attacherons \`a montrer.
\end{itemize}

Notre article est articul\'e comme suit. Dans la section \ref{sec:formalisme_resultats}, apr\`es quelques rappels minimaux sur la th\'eorie
de Bogolioubov dans un pi\`ege, nous sp\'ecifions l'\'etat du syst\`eme et introduisons les quantit\'es permettant de d\'ecrire
formellement l'\'etalement de la phase, \`a savoir la d\'eriv\'ee de l'op\'erateur phase du condensat et sa fonction de corr\'elation temporelle. 
Dans la section \ref{sec:balistique}, nous donnons une expression du coefficient balistique $A$ \`a la limite thermodynamique dans un pi\`ege harmonique
quelconque (y compris isotrope), d'abord dans l'\'etat du syst\`eme le plus g\'en\'eral consid\'er\'e ici puis dans le cas plus simple d'un m\'elange 
statistique d'ensembles canoniques de m\^eme temp\'erature $T$. Dans la longue section \ref{sec:micro}, nous nous attaquons au c\oe ur du probl\`eme,
le calcul de la fonction de corr\'elation $C_{\rm mc}(\tau)$ de $\dd\hat{\theta}/\dd t$ dans l'ensemble microcanonique, 
qui donne acc\`es en toute g\'en\'eralit\'e
aux termes d'\'etalement sous-balistiques de la phase, puisqu'ils sont ind\'ependants de l'\'etat du syst\`eme \`a la limite thermodynamique 
\`a \'energie moyenne et nombre moyen de particules fix\'es.
Nous passons d'abord \`a la limite semi-classique dans la sous-section \ref{subsec:formesemiclassique}, 
le mouvement des quasi-particules de Bogolioubov \'etant trait\'e
classiquement mais le champ des quasi-particules restant quantique bosonique~; la forme semi-classique de $\dd\hat{\theta}/\dd t$ est d\'eduite d'un 
principe de correspondance. Nous \'ecrivons ensuite, dans la sous-section \ref{subsec:equacin}, 
des \'equations cin\'etiques sur les nombres d'occupation des quasi-particules dans l'espace des phases classique $(\rr,\pp)$
et nous montrons comment,
une fois lin\'earis\'ees, elles conduisent formellement \`a $C_{\rm mc}(\tau)$. Le probl\`eme reste formidable, puisque les nombres d'occupation
d\'ependent des six variables ($\rr,\pp)$ et du temps. \`A la limite s\'eculaire $\gamma_{\rm coll}\ll\omega_\alpha$ et dans l'approximation ergodique
sur le mouvement des quasi-particules (ce qui exclut le pi\`ege isotrope ou \`a sym\'etrie de r\'evolution), nous nous ramenons dans la sous-section
\ref{subsec:solution} \`a des nombres d'occupation fonctions de la seule \'energie $\epsilon$ du mouvement classique et du temps, ce qui conduit \`a
des r\'esultats explicites sur $C_{\rm mc}(\tau)$, sur la diffusion de phase et, sous-produit int\'eressant, sur le taux d'amortissement
des modes de Bogolioubov dans le pi\`ege, {\yvan dans la sous-section \ref{subsec:resetdiscus} o\`u nous \'evaluons \'egalement le d\'ephasage
du condensat d\^u aux pertes de particules}. Enfin, nous proc\'edons \`a une discussion critique de l'approximation ergodique dans la sous-section
\ref{subsec:discuss_ergo}, en estimant en particulier l'erreur qu'elle introduit sur les quantit\'es pilotant la diffusion de phase du condensat. 
Nous concluons dans la section \ref{sec:conclusion}.

\section{Rappels sur le formalisme et les résultats}
\label{sec:formalisme_resultats}

{\sl La d\'eriv\'ee de la phase} --
Comme nous l'avons rappel\'e dans l'introduction, le temps de coh\'erence d'un condensat est contr\^ol\'e par la dynamique de son
op\'erateur phase $\hat{\theta}(t)$ aux temps longs devant le temps de collision $\gamma_{\rm coll}^{-1}$ typique des quasi-particules.
Le point de d\'epart de notre \'etude est donc l'expression de la d\'eriv\'ee temporelle de $\hat{\theta}(t)$,  liss\'ee temporellement
(c'est-\`a-dire moyenn\'ee sur un temps court devant $\gamma_{\rm coll}^{-1}$ mais long devant l'inverse de la pulsation
typique $\epsilon_k^{\rm th}/\hbar$ des quasi-particules thermiques), telle que l'a \'etablie en toute g\'en\'eralit\'e
la r\'ef\'erence \cite{CohFer} \`a l'ordre un en la fraction non condens\'ee~:
\be
-\hbar \frac{\dd\hat{\theta}}{\dd t} = \mu_0(\hat{N}) + \sum_{k\in {\cal F}_+} \frac{\dd\epsilon_k}{\dd N} \:  \hat{n}_k \equiv \hat{\mu}
\label{eq:thetadot}
\ee
Ici $\mu_0(N)$ est le potentiel chimique du gaz dans l'\'etat fondamental et $\hat{N}$ est l'op\'erateur nombre total de particules. 
La somme sur le nombre quantique g\'en\'erique $k$ (ce n'est pas un nombre d'onde)
porte sur les modes de Bogolioubov d'\'energie propre $\epsilon_k$, et $\hat{n}_k$ est l'op\'erateur nombre de quasi-particules
dans le mode $k$. L'expression (\ref{eq:thetadot}) est une version quantique de la seconde relation de Josephson~:
son second membre est un op\'erateur potentiel chimique $\hat{\mu}$ du gaz, 
puisque c'est la d\'eriv\'ee adiabatique (aux nombres d'occupation $\hat{n}_k$ fix\'es) par rapport \`a $N$
du hamiltonien de Bogolioubov
\be
\hat{H}_{\rm Bog} = E_0(\hat{N}) + \sum_{k\in {\cal F}_+} \epsilon_k \hat{n}_k
\label{eq:H_Bog}
\ee
Les modes de Bogolioubov sont de la famille $\mathcal{F}_+$, suivant la terminologie de la r\'ef\'erence \cite{CastinDum}, au sens
o\`u leurs fonctions modales $(u_k(\rr),v_k(\rr))$ sont solutions de l'\'equation aux valeurs propres
\be
\epsilon_k \binom{|u_k\rangle}{|v_k\rangle} =  \left(\begin{array}{cc}
 H_{\rm GP}+Q g\rho_0(\hat{\rr}) Q &  Q g\rho_0(\hat{\rr}) Q \\
 -Q g\rho_0(\hat{\rr}) Q & -[ H_{\rm GP}+Q g\rho_0(\hat{\rr}) Q ]
 \end{array} \right)  \binom{|u_k\rangle}{|v_k\rangle} \equiv {\cal L}(\hat{\rr},\hat{\pp}) \binom{|u_k\rangle}{|v_k\rangle}
\label{eq:calL}
\ee
avec la condition de normalisation $\int \dd^3r\, (|u_k(\rr)|^2-|v_k(\rr)|^2) =1 >0$. On a pris la fonction d'onde $\phi_0(\rr)$
du condensat r\'eelle, normalis\'ee \`a l'unit\'e ($\int \dd^3r\,\phi_0^2(\rr)=1$),
et \'ecrite \`a l'ordre z\'ero en la fraction non condens\'ee, c'est-\`a-dire \`a l'approximation
de Gross-Pitaevskii~:
\be
H_{\rm GP} |\phi_0\rangle = 0 \ \ \mbox{avec} \ \ H_{\rm GP}=\frac{\hat{\pp}^2}{2m}+U(\hat{\rr})+g\rho_0(\hat{\rr})-\mu_{\rm GP}
\ee
si bien qu'\`a cet ordre, la densit\'e condens\'ee vaut $\rho_0(\rr)=N \phi^2_0(\rr)$. Ici, $g=4\pi\hbar^2 a/m$ est la constante de couplage,
proportionnelle \`a la longueur de diffusion $a$ dans l'onde $s$ entre les bosons de masse $m$, et $U(\rr)=\sum_\alpha  m \omega_\alpha^2 r_\alpha^2/2$
est leur potentiel de pi\'egeage. Le projecteur $Q$ projette orthogonalement \`a $|\phi_0\rangle$ 
et assure que $|\phi_0\rangle\bot |u_k\rangle$ et $|\phi_0\rangle\bot |v_k\rangle$ comme il
se doit \cite{CastinDum}. Comme le condensat est dans son mode fondamental ($\phi_0$ minimise la fonctionnelle \'energie de Gross-Pitaevskii),
les $\epsilon_k$ sont positifs.

{\sl L'\'etat du syst\`eme} -- Le refroidissement des gaz d'atomes froids par \'evaporation ne conduit {\sl a priori} \`a aucun des ensembles 
habituels de la physique statistique. Pour couvrir tous les cas raisonnables, nous supposons donc que le gaz est pr\'epar\'e \`a
l'instant $0$ dans un ensemble g\'en\'eralis\'e, m\'elange statistique d'\'etats propres $|\psi_\lambda\rangle$
du hamiltonien complet $\hat{H}$ \`a $N_\lambda$ corps et d'\'energie $E_\lambda$, donc d'op\'erateur densit\'e
\be
\hat{\sigma} = \sum_\lambda \Pi_\lambda |\psi_\lambda\rangle \langle \psi_\lambda|
\label{eq:OpDens}
\ee
avec comme seule restriction l'existence de lois \'etroites sur $E_\lambda$ et $N_\lambda$, de variances et covariance ne croissant pas plus vite
que les moyennes $\bar{E}$ et $\bar{N}$ \`a la limite thermodynamique.

{\sl D\'ephasage moyen} -- Moyennons l'expression (\ref{eq:thetadot}) dans l'\'etat stationnaire $|\psi_\lambda\rangle$. Au second membre appara\^\i t
l'esp\'erance de l'op\'erateur potentiel chimique dans $|\psi_\lambda\rangle$. 
\`A cause des interactions entre les quasi-particules de Bogolioubov, on s'attend
\`a ce que le syst\`eme \`a $N$ corps soit ergodique au sens quantique du terme, c'est-\`a-dire qu'il souscrive au principe de microcanonicit\'e
des \'etats propres ({\sl Eigenstate Thermalisation Hypothesis} dans la litt\'erature anglo-am\'ericaine, 
voir les r\'ef\'erences \cite{Deutsch1991,Srednicki1994,Olshanii2008}), soit
\be
\langle\psi_\lambda| \hat{\mu} | \psi_\lambda\rangle = \mu_{\rm mc} (E_\lambda,N_\lambda) 
\ee
o\`u $\mu_{\rm mc}(E,N)$ est le potentiel chimique dans l'ensemble microcanonique d'\'energie $E$ \`a $N$ particules. 
Pour un grand syst\`eme, il suffit de d\'evelopper au premier ordre en les fluctuations, \'etant donn\'e la faiblesse de leurs valeurs relatives~:
\be 
\mu_{\rm mc}(E_\lambda,N_\lambda) = \mu_{\rm mc}(\bar{E},\bar{N}) + (E_\lambda-\bar{E}) \partial_E \mu_{\rm mc}(\bar{E},\bar{N})
+(N_\lambda-\bar{N}) \partial_N \mu_{\rm mc}(\bar{E},\bar{N}) + O(1/\bar{N})
\ee
Il reste \`a moyenner sur les \'etats $|\psi_\lambda\rangle$ avec les poids $\Pi_\lambda$ comme dans l'\'equation (\ref{eq:OpDens}) pour
obtenir la premi\`ere brique \`a la fonction de coh\'erence temporelle (\ref{eq:g1}), le d\'ephasage moyen~:
\be
\langle\hat{\theta}(t)-\hat{\theta}(0)\rangle = -\mu_{\rm mc}(\bar{E},\bar{N}) t/\hbar
\label{eq:dephmoy}
\ee
avec une erreur $O(1/\bar{N})$ sur le coefficient de $t$.

{\sl D\'ephasage quadratique moyen} -- En proc\'edant de la m\^eme mani\`ere pour le second moment du d\'ephasage du condensat, nous
trouvons comme il est \'ecrit un peu implicitement dans \cite{PRAlong,CohFer} que
\be
\mbox{Var}\,[\hat{\theta}(t)-\hat{\theta}(0)] = A t^2 + 2\int_0^t \dd\tau\, (t-\tau)\, {\yvan \re}\, C_{\rm mc}(\tau)
\label{eq:vartot}
\ee
avec le coefficient balistique
\be
A = \mbox{Var}[(N_\lambda-\bar{N})\partial_N\mu_{\rm mc}(\bar{E},\bar{N}) + (E_\lambda-\bar{E}) \partial_E \mu_{\rm mc}(\bar{E},\bar{N})]
/\hbar^2
\label{eq:grandA}
\ee
et la fonction de corr\'elation de la d\'eriv\'ee de la phase dans l'ensemble
microcanonique d'\'energie $\bar{E}$ \`a $\bar{N}$ particules~:
\be
C_{\rm mc}(\tau) = \left\langle \frac{\dd\hat{\theta}}{\dd t}(\tau) \frac{\dd\hat{\theta}}{\dd t}(0)\right\rangle_{\rm mc} - 
\left\langle \frac{\dd\hat{\theta}}{\dd t}\right\rangle_{\rm mc}^2
\label{eq:defCmc}
\ee
Voici qui compl\`ete notre connaissance formelle de $g_1(t)$.

En vue des observations exp\'erimentales \`a venir, il reste cependant \`a calculer {\sl explicitement} $A$ et $C_{\rm mc}(\tau)$
pour un syst\`eme harmoniquement pi\'eg\'e. Il faudra en particulier v\'erifier que $C_{\rm mc}(\tau)$ dans le cas pi\'eg\'e
d\'ecro\^\i t assez vite pour que l'on retrouve une loi diffusive (\ref{eq:diff}) comme dans le cas spatialement homog\`ene.

\section{Calcul du coefficient balistique dans la variance du déphasage}
\label{sec:balistique}

{\sl Dans l'ensemble statistique g\'en\'eralis\'e} -- 
Pour calculer le d\'ephasage moyen (\ref{eq:dephmoy}) et le coefficient balistique (\ref{eq:grandA}) dans le cas g\'en\'eral,
nous devons conna\^\i tre le potentiel chimique
microcanonique $\mu_{\rm mc}(\bar{E},\bar{N})$ et ses d\'eriv\'ees dans le pi\`ege harmonique. 
\`A la limite thermodynamique, $\mu_{\rm mc}$ co\"\i ncide avec le potentiel chimique
$\mu_{\rm can}$ dans l'ensemble canonique de temp\'erature $T$ et nombre de particules $\bar{N}$, plus commode \`a calculer,
si la temp\'erature $T$ est ajust\'ee pour qu'il y ait \'egalit\'e des \'energies moyennes $E_{\rm can}(T,\bar{N})$ et $\bar{E}$. 
En d'autres termes,
\be
\mu_{\rm mc}(E_{\rm can}(T,\bar{N}),\bar{N}) \sim \mu_{\rm can}(T,\bar{N})
\ee
Il suffit de d\'eriver cette relation par rapport \`a $T$ ou $\bar{N}$ pour obtenir les d\'eriv\'ees utiles de $\mu_{\rm mc}$,
puis de remplacer $E_{\rm can}$ par $\bar{E}$, ce qui donne~:
\bea
\label{eq:acalc1}
\partial_E \mu_{\rm mc}(\bar{E},\bar{N}) &\sim &  \frac{\partial_T \mu_{\rm can}(T,\bar{N})}{\partial_T E_{\rm can}(T,\bar{N})} \\
\partial_N \mu_{\rm mc}(\bar{E},\bar{N}) &\sim & \partial_N \mu_{\rm can}(T,\bar{N}) 
-\frac{\partial_N E_{\rm can}(T,\bar{N})}{\partial_T E_{\rm can}(T,\bar{N})} \partial_T \mu_{\rm can}(T,\bar{N})
\label{eq:acalc2}
\eea

Au premier ordre en la fraction non condens\'ee, le potentiel chimique canonique se d\'eduit de l'\'energie libre $F$ du gaz parfait
de quasi-particules de Bogolioubov de hamiltonien (\ref{eq:H_Bog}) par la relation thermodynamique habituelle $\mu_{\rm can}=\partial_N
F$. L'\'energie libre est une fonctionnelle simple de la densit\'e d'\'etats $\rho(\epsilon)$ des quasi-particules,
\be
F(T,\bar{N})=E_0(\bar{N}) + k_B T \int_0^{+\infty} \dd\epsilon\,\rho(\epsilon) \ln\left(1-e^{-\beta\epsilon}\right)
\ee
avec $\beta=1/k_B T$.  
\`A la limite thermodynamique, l'\'energie {\yvan $E_0$} de l'\'etat fondamental {\yvan du gaz} dans le pi\`ege harmonique se d\'eduit de celle du syst\`eme homog\`ene
\cite{LY} par une approximation d'homog\'en\'eit\'e locale, et la densit\'e d'\'etats $\rho(\epsilon)$ s'obtient par prise
de la limite classique $\hbar\to 0$, en vertu de {\yvan l'in\'egalit\'e} (\ref{eq:condval1}) \cite{Stringari_revue}~:
\be
\rho(\epsilon)= \int \frac{\dd^3r\dd^3p}{(2\pi\hbar)^3} \delta(\epsilon-\epsilon(\rr,\pp))
\label{eq:rhoeps}
\ee
Le hamiltonien classique $\epsilon(\rr,\pp)$ est la valeur propre positive de la matrice $2\times 2$ de Bogolioubov de l'\'equation
(\ref{eq:calL}) avec la position $\rr$ et l'impulsion $\pp$ trait\'ees classiquement \footnote{Le projecteur $Q$, 
projetant sur un espace de codimension un, peut \^etre omis \`a la limite thermodynamique.} et la densit\'e condens\'ee $\rho_0(\rr)$ \'ecrite 
\`a la limite classique c'est-\`a-dire dans l'approximation de Thomas-Fermi~:
\be
g \rho_0^{\rm TF}(\rr) = \left\{ 
\begin{array}{lll} 
\mu_{\rm TF}-U(\rr) \equiv \mu_{\rm loc}(\rr) \quad \quad \quad \quad & \mbox{si} & \quad U(\rr) <  \mu_{\rm TF} \\
0 \quad \quad & \mbox{sinon} &
\end{array}  \right.
\ee
Ici, le potentiel chimique de Thomas-Fermi, limite classique de celui $\mu_{\rm GP}$ de Gross-Pitaevskii, vaut
\be
\mu_{\rm TF}=\frac{1}{2}\hbar\bar{\omega}[15\bar{N}a(m\bar{\omega}/\hbar)^{1/2}]^{2/5}
\label{eq:mutf}
\ee
et $\bar{\omega}=(\omega_x \omega_y \omega_z)^{1/3}$ est la moyenne g\'eom\'etrique des pulsations de pi\'egeage.
On en d\'eduit que
\be
\epsilon(\rr,\pp) =\left\{ 
\begin{array}{lll} 
\left\{ {\displaystyle\frac{p^2}{2m}}\left[{\displaystyle\frac{p^2}{2m}}+2 \mu_{\rm loc}(\rr)\right] \right\}^{1/2} \quad \quad & \mbox{si} & \quad U(\rr) <  \mu_{\rm TF} \\
{\displaystyle\frac{p^2}{2m}}+ U(\rr)-\mu_{\rm TF} \quad \quad & \mbox{sinon} &
\end{array} \right.
\label{eq:eps_rp}
\ee
L'int\'egrale sextuple (\ref{eq:rhoeps}) a \'et\'e calcul\'ee dans la r\'ef\'erence \cite{PRLLincoln}.\footnote{Le cas d'un pi\`ege harmonique anisotrope se ram\`ene au cas
isotrope trait\'e dans \cite{PRLLincoln} par le changement de variable de jacobien unit\'e $r_\alpha = \lambda_\alpha r'_\alpha$, avec
$\omega_\alpha \lambda_\alpha=\bar{\omega}$, tel que $U(\rr)=\frac{1}{2} m\bar{\omega}^2 r'^{2}$. \label{note:isotropisation}} 
Nous donnons ici le r\'esultat sous une forme un peu plus compacte~:
\bea
\rho(\epsilon) &=& \frac{\mu_{\rm TF}^2}{(\hbar\bar{\omega})^3} f(\check{\epsilon}\equiv \epsilon/\mu_{\rm TF}) \\
f(\check{\epsilon}) &=& \frac{1}{\pi}\left[-2\sqrt{2}\epsc^2 \acos\frac{\epsc-1}{(1+\epsc^2)^{1/2}} 
              +2\sqrt{2}\epsc\ln\frac{1+\sqrt{2\epsc}+\epsc}{(1+\epsc^2)^{1/2}}
              +\sqrt{\epsc}(5\epsc-1)+(1+\epsc)^2\acos\frac{1}{(1+\epsc)^{1/2}}\right]
\label{eq:deff}
\eea
Nous obtenons finalement le potentiel chimique canonique
\be
\mu_{\rm can}(T,\bar{N})=\mu_0(\bar{N})+\frac{6k_B T}{5\bar{N}} \left(\frac{\mu_{\rm TF}}{\hbar\bar{\omega}}\right)^3
\int_0^{+\infty} \dd\epsc f(\epsc) \ln\left(1-\eee^{-\check{\beta}\epsc}\right)
+\frac{2\mu_{\rm TF}}{5\bar{N}} \left(\frac{\mu_{\rm TF}}{\hbar\bar{\omega}}\right)^3 
\int_0^{+\infty} \dd\epsc \frac{f(\epsc)\epsc}{\eee^{\check{\beta}\epsc}-1} 
\label{eq:mucanfin}
\ee
avec la contribution de l'\'etat fondamental \cite{Stringari_revue}
\be
\mu_0(\bar{N})= \mu_{\rm TF} \left[1+\pi^{1/2} \left({\mu_{\rm TF}a^3}/{g}\right)^{1/2}\right]
\ee
Lorsqu'on d\'erivera (\ref{eq:mucanfin}) par rapport \`a $T$ et $\bar{N}$ pour \'evaluer les expressions (\ref{eq:acalc1}) et (\ref{eq:acalc2}), 
on se souviendra que $\check{\beta}=\mu_{\rm TF}/k_B T$ d\'epend de $\bar{N}$ au travers de $\mu_{\rm TF}$. Pour abr\'eger, nous ne donnons
pas le r\'esultat ici.

\begin{figure}[t]
\centerline{\includegraphics[width=0.4\textwidth,clip=]{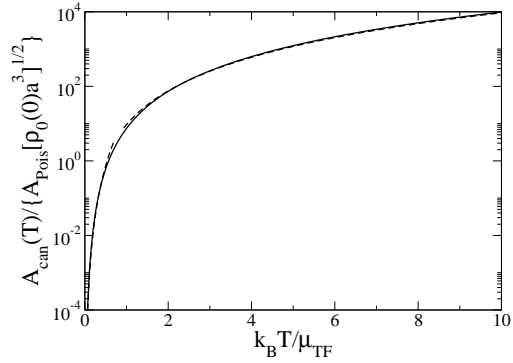}}
\caption{Coefficient de l'\'etalement balistique (\ref{eq:bali}) de la phase du condensat aux temps longs devant le temps de collision
$\gamma_{\rm coll}^{-1}$ des quasi-particules, pour un gaz de $\bar{N}$ bosons pr\'epar\'e dans l'ensemble canonique dans un pi\`ege harmonique
isotrope ou pas, en fonction de la temp\'erature. Le r\'esultat vaut \`a la limite thermodynamique o\`u 
les pulsations de pi\'egeage $\omega_\alpha$ sont n\'egligeables par rapport au potentiel chimique $\mu_{\rm TF}$ de Thomas-Fermi (\ref{eq:mutf}). 
Trait plein~: second terme de l'\'equation 
(\ref{eq:Apart}), d\'eduit du potentiel chimique canonique (\ref{eq:mucanfin}) 
\`a l'approximation de Bogolioubov (interactions faibles, $T\ll T_c$). Tiret\'es~:
\'equivalents \`a basse et \`a haute temp\'erature (termes dominants des \'equations (\ref{eq:equivbT},\ref{eq:equivhT})). 
La division de $A_{\rm can}(T)$
par le petit param\`etre (\ref{eq:rhoa3}) de la th\'eorie de Bogolioubov et 
par la valeur (\ref{eq:APois}) du coefficient balistique pour des fluctuations de $N$ poissoniennes
conduit \`a une fonction universelle de $k_B T/\mu_{\rm TF}$. }
\label{fig:Acan}
\end{figure}

{\sl Dans un ensemble un peu moins g\'en\'eral} -- Une expression plus simple
\footnote{L'expression g\'en\'erale (\ref{eq:grandA}) de $A$ est un peu d\'elicate \`a appr\'ehender. En effet,
comme l'\'energie du fondamental d\'epend de $N$, des fluctuations de $N$ entra\^\i nent m\'ecaniquement des fluctuations
d'\'energie. Par exemple, si $N$ fluctue \`a $T=0$ (dans chaque sous-espace \`a $N$ fix\'e, le syst\`eme est dans l'\'etat
fondamental), on peut, pour retrouver $A(T=0)$ de l'\'equation (\ref{eq:Apart}) \`a partir de l'\'equation (\ref{eq:grandA}), 
utiliser le fait que $E_\lambda-\bar{E}=(N_\lambda-\bar{N})\mu_0(\bar{N})+O(\bar{N}^0)$ 
et que $\partial_E\mu_{\rm mc}(\bar{E},\bar{N})\underset{T\to 0}{\sim} -2/(25 \bar{N})$, dont le report dans (\ref{eq:acalc2})
donne $\partial_N\mu_{\rm mc}(\bar{E},\bar{N})\underset{T\to 0}{\sim} \partial_N \mu_0(\bar{N})+2\mu_0(\bar{N})/(25 \bar{N})$.
}
du coefficient balistique $A$ peut \^etre obtenue lorsque 
l'\'etat du syst\`eme est un m\'elange statistique d'ensembles canoniques de m\^eme temp\'erature $T$ mais de nombre de particules
variable. En exprimant les diff\'erents coefficients dans (\ref{eq:grandA},\ref{eq:acalc1},\ref{eq:acalc2}) 
comme des d\'eriv\'ees de l'\'energie libre $F(T,\bar{N})$ par rapport \`a $\bar{N}$ et $T$, et en nous souvenant de l'expression
$\mbox{Var}_{\rm can} E = k_B T^2 \partial_T E_{\rm can}$ de la variance de l'\'energie dans l'ensemble canonique, nous trouvons 
\`a l'ordre dominant $1/\bar{N}$ que 
\be
A(T) = (\mbox{Var}\,N) \left(\frac{\partial_N\mu_{\rm can}(T,\bar{N})}{\hbar}\right)^2
+\frac{k_B T^2 \left[\partial_T\mu_{\rm can}(T,\bar{N})\right]^2}{\hbar^2 \partial_T E_{\rm can}(T,\bar{N})}
\label{eq:Apart}
\ee
\`A temp\'erature nulle, seul le premier terme contribue, et l'on retrouve la pr\'ediction des r\'ef\'erences \cite{Walls,DalibardCastin} pouss\'ee
\`a l'ordre un en la fraction non condens\'ee $f_{\rm nc}$.
\`A $T\neq 0$ mais en l'absence de fluctuations de $N$, seul le second terme contribue~; il n'est autre que le coefficient balistique
$A_{\rm can}(T)$ dans l'ensemble canonique.
Dans le r\'egime de validit\'e de l'approximation de Bogolioubov, $f_{\rm nc}\ll 1$, le potentiel
chimique $\mu_{\rm can}(T,\bar{N})$ du gaz reste proche de celui de Thomas-Fermi du condensat pur, si bien que
\be
\partial_N \mu_{\rm can}(T,\bar{N}) = \partial_N \mu_{\rm TF} + O\left(\frac{f_{\rm nc}}{\bar{N}}\right)
\ee
En revanche, $\partial_T \mu_{\rm can}(T,\bar{N})$ est imm\'ediatement du premier ordre en $f_{\rm nc}$, et il en va de m\^eme pour 
le second terme dans l'\'equation (\ref{eq:Apart}). C'est donc seulement pour des fluctuations de $N$ fortement subpoissoniennes
($\mbox{Var}\, N\ll \mbox{Var}_{\rm Pois} N \equiv \bar{N}$) que le second terme de (\ref{eq:Apart}), c'est-\`a-dire l'effet
des fluctuations thermiques, n'est pas domin\'e par le premier.  En supposant cette condition satisfaite dans l'exp\'erience,
nous repr\'esentons sur la figure \ref{fig:Acan} le coefficient canonique $A_{\rm can}(T)$ adimensionn\'e par la valeur $A_{\rm Pois}$
de $A$ dans un condensat pur avec des fluctuations de $N$ poissoniennes,
\be
A_{\rm Pois} = \bar{N} \left(\frac{\partial_N \mu_{\rm TF}}{\hbar}\right)^2
\label{eq:APois}
\ee
le tout divis\'e par le petit param\`etre de la th\'eorie de Bogolioubov \`a temp\'erature nulle,\footnote{On pr\'ef\`ere parfois prendre
comme petit param\`etre $1/[\rho_0(\mathbf{0})\xi^3]$, o\`u la longueur de relaxation $\xi$ du condensat au centre du pi\`ege
est telle que $\hbar^2/(m \xi^2)=\mu_{\rm TF}$. On peut passer ais\'ement d'un petit param\`etre \`a l'autre \`a l'aide de la
relation $[\rho_0(\mathbf{0})a^3]^{1/2} \rho_0(\mathbf{0})\xi^3=1/(8\pi^{3/2})$.} 
proportionnel \`a $f_{\rm nc}(T=0)$~:
\be
[\rho_0(\mathbf{0})a^3]^{1/2} = \frac{2\sqrt{2}}{15\pi^{1/2}\bar{N}} \left(\frac{\mu_{\rm TF}}{\hbar\bar{\omega}}\right)^3
\label{eq:rhoa3}
\ee
Le rapport ainsi form\'e est une fonction universelle de $k_B T/\mu_{\rm TF}$.
\`A partir des d\'eveloppements \`a basse et \`a haute \'energie de la densit\'e d'\'etats des quasi-particules,
\bea
\label{eq:fbasse}
f(\epsc) &\underset{\epsc\to 0}{=}& \frac{32}{3\pi} \epsc^{3/2} -2\sqrt{2}\,\epsc^2 + O(\epsc^{5/2}) \\
f(\epsc) &\underset{\epsc\to+\infty}{=}& \frac{1}{2} \epsc^2 + \epsc + \frac{1}{2} +O(\epsc^{-1/2})
\label{eq:fhaute}
\eea
nous obtenons les d\'eveloppements \`a basse et \`a haute temp\'erature ($\check{T}=k_B T/\mu_{\rm TF}=1/\check{\beta}$)
\bea
\label{eq:equivbT}
\frac{A_{\rm can}(T)}{A_{\rm Pois}[\rho_0(\mathbf{0})a^3]^{1/2}} &\underset{\check{T}\to 0}{=}& \frac{21\zeta(7/2)}{\sqrt{2}}
\check{T}^{9/2} \left[1+\frac{4\sqrt{2}\pi^{9/2}}{525\zeta(7/2)}\check{T}^{1/2}+O(\check{T})\right] \\
&\underset{\check{T}\to+\infty}{=}&\frac{15\pi^{1/2}}{2\sqrt{2}}\frac{3\zeta(3)^2}{4\zeta(4)}\check{T}^3 
\left[1+\check{\beta}\left(\frac{4\zeta(2)}{3\zeta(3)}-\frac{\zeta(3)}{2\zeta(4)}\right)+O(\check{\beta}^{3/2})\right]
\label{eq:equivhT}
\eea
dont les termes dominants \footnote{Dans la fen\^etre de valeurs de la figure \ref{fig:Acan}, en pratique $1/10\leq \check{T}\leq 10$,
l'inclusion des termes sous-dominants ne rapproche pas utilement du r\'esultat exact.} 
sont repr\'esent\'es en tiret\'e sur la figure \ref{fig:Acan}. Signalons une r\'ecriture particuli\`erement simple et belle de l'\'equivalent
\`a haute temp\'erature, accidentellement op\'erationnel d\'ej\`a \`a $k_B T/\mu_{\rm TF}\geq 2$~:
\be
\frac{A_{\rm can}(T)}{A_{\rm Pois}}\underset{k_BT\gg\mu_{\rm TF}}{\sim}\frac{3\zeta(3)}{4\zeta(4)}\left(\frac{T}{T_c^{(0)}}\right)^3
\ee
o\`u $T_c^{(0)}$ est la temp\'erature critique du gaz parfait de bosons dans un pi\`ege harmonique \`a la limite thermodynamique,
$k_B T_c^{(0)}=\hbar\bar{\omega}[\bar{N}/\zeta(3)]^{1/3}$. Dans cette limite, $A_{\rm can}(T)$ est donc plus faible que $A_{\rm Pois}$ par un facteur
proportionnel \`a la fraction non condens\'ee $(T/T_c^{(0)})^3\ll 1$.

\section{Variance du déphasage du condensat dans l'ensemble microcanonique}
\label{sec:micro}
Nous calculons ici la fonction de corrélation de $\dd\hat{\theta}/\dd t$, à savoir $C_{\rm mc}(\tau)$, pour un système préparé dans l'ensemble microcanonique, en utilisant à la limite thermodynamique $\frac{\hbar \omega_\alpha}{\mu_{\rm TF}}\to0$ une description semi-classique des quasi-particules et en prenant en compte l'effet de leur interaction par des équations cinétiques de 
type Boltzmann quantique sur leur distribution dans l'espace des phases classique $(\rr,\pp)$.

\subsection{Forme semi-classique du Hamiltonien de Bogolioubov et de ${\dd\hat{\theta}}/{\dd t}$}
\label{subsec:formesemiclassique}
Dans la description semi-classique, le mouvement des quasi-particules de Bogolioubov est traité {\it classiquement}, c'est-à-dire qu'elles ont à chaque instant une position $\rr$ et une impulsion $\pp$ bien définies \cite{Stringari_revue}, dont l'évolution dans l'espace des phases dérive du hamiltonien $\epsilon(\rr,\pp)$ donné dans l'équation (\ref{eq:eps_rp}) \cite{Graham}~:
\bea
\label{eq:drdt}
\frac{\dd\rr}{\dd t} &=& \partial_{\pp} \, \epsilon(\rr,\pp) \\
\frac{\dd\pp}{\dd t} &=& -\partial_{\rr} \, \epsilon(\rr,\pp) 
\label{eq:dpdt}
\eea
mais l'on traite {\it quantiquement} le champ bosonique des quasi-particules en introduisant leurs opérateurs nombres d'occupation
$\hat{n}(\rr,\pp)$ dans l'espace des phases, {\yvan ce qui permet de prendre en compte la nature discr\`ete des nombres de quasi-particules
et les effets de statistique quantique (loi de Bose plut\^ot que loi d'\'equipartition du champ classique \`a l'\'equilibre).}

Dans cette limite semi-classique, le hamiltonien de Bogolioubov (\ref{eq:H_Bog}) (sans interaction entre les quasi-particules) s'écrit immédiatement 
\be
H_{\rm Bog}^{\rm sc} = E_0(\hat{N}) + \int \frac{\dd^3r\, \dd^3p}{(2\pi \hbar)^3} \: \epsilon(\rr,\pp) \, \hat{n}(\rr,\pp)
\label{eq:H_Bog_sc}
\ee
On pourrait croire, au vu de la formule (\ref{eq:thetadot}), que ${\dd\hat{\theta}}/{\dd t}$ admet une écriture similaire, avec 
$\epsilon(\rr,\pp)$ remplacé par $\frac{\dd}{\dd N}\epsilon(\rr,\pp)$. Il n'en est rien, la raison étant que {\yvan la d\'eriv\'ee}
$\frac{\dd}{\dd N}\epsilon(\rr,\pp)$ n'est pas constante sur la trajectoire classique. L'op\'erateur ${\dd\hat{\theta}}/{\dd t}$ fait partie d'une classe générale d'observables quantiques dites de Fock (diagonales dans la base de Fock des quasi-particules donc fonctionnelles - ici lin\'eaires - des nombres d'occupation des modes de
Bogolioubov)~:
\be
\hat{A}=\sum_{k\in {\cal F}_+} a_k \hat{n}_k \quad \quad \mbox{avec} \quad 
a_k = (\langle u_k|, \langle v_k| ) \, {\cal A}(\hat{\rr},\hat{\pp}) \, \binom{|u_k\rangle}{|v_k\rangle}
\label{eq:A}
\ee 
où ${\cal A}(\hat{\rr},\hat{\pp}) $ est un opérateur matriciel $2\times2$ hermitien et $a_k$ sa moyenne dans le mode de Bogolioubov d'énergie propre $\epsilon_k$. L'observable ${\dd\hat{\theta}}/{\dd t}$ correspond au choix ${\cal A}_{\dot{\theta}}=\sigma_z \frac{\dd}{\dd N}{\cal L}$
où $\sigma_z$ est la troisi\`eme matrice de Pauli et ${\cal L}(\hat{\rr},\hat{\pp})$ est l'opérateur apparaissant dans l'équation (\ref{eq:calL})~: en vertu du théorème de Hellmann-Feynman\footnote{Le th\'eor\`eme est ici généralisé au cas d'un opérateur ${\cal L}$ non hermitien, $(\langle u_k|, -\langle v_k| )$ \'etant le vecteur dual du vecteur propre $(|u_k \rangle, |v_k \rangle )$ de ${\cal L}$.}, on a en effet
\be
(\langle u_k|, -\langle v_k| ) \left( \frac{\dd}{\dd N} \, {\cal L}\right)  \binom{|u_k\rangle}{|v_k\rangle} = \frac{\dd\epsilon_k}{\dd N}
\ee
Pour ces opérateurs de Fock nous utilisons le principe de correspondance semi-classique 
\be
\hat{A}^{\rm sc} = \int \frac{\dd^3r\, \dd^3p}{(2\pi \hbar)^3} \: \overline{a(\rr,\pp)} \, \hat{n}(\rr,\pp)
\label{eq:Asc}
\ee
où $a(\rr,\pp)=(U(\rr,\pp),V(\rr,\pp)) \, {\cal A}(\rr,\pp) \, \binom{U(\rr,\pp)}{V(\rr,\pp)}$, ${\cal A}(\rr,\pp)$ étant l'équivalent classique 
 de ${\cal A}(\hat{\rr},\hat{\pp})$, et $\overline{a(\rr,\pp)} $ représente la moyenne temporelle de $a(\rr,\pp)$ sur l'unique trajectoire classique passant par $(\rr,\pp)$ à l'instant $t=0$~:
\be
\overline{a(\rr,\pp)} \equiv \lim_{t \to {\yvan +}\infty} \frac{1}{t} \int_0^t \, \dd\tau \,  a(\rr(\tau),\pp(\tau))
\ee
Le vecteur $(U(\rr,\pp),V(\rr,\pp))$, normalisé selon la condition $U^2-V^2=1$, est vecteur propre de l'équivalent classique 
${\cal L}(\rr,\pp)$ de ${\cal L}(\hat{\rr},\hat{\pp})$ avec la valeur propre $\epsilon(\rr,\pp)$~;
d'où
\be
\binom{U(\rr,\pp)}{V(\rr,\pp)} = \left\{ 
\begin{array}{lll} 
\left(
\begin{array}{l}
\frac{1}{2} \left[ \left( {\displaystyle\frac{p^2/2m}{\epsilon(\rr,\pp)}} \right)^{1/2} + 
\left( {\displaystyle\frac{p^2/2m}{\epsilon(\rr,\pp)}} \right)^{-1/2}\right] \\ \\
\frac{1}{2} \left[ \left( {\displaystyle\frac{p^2/2m}{\epsilon(\rr,\pp)}} \right)^{1/2} -  \left( {\displaystyle\frac{p^2/2m}{\epsilon(\rr,\pp)}} \right)^{-1/2}\right]
\end{array} 
\right)
\quad \quad \quad \quad & \mbox{si} & \quad U(\rr) <  \mu_{\rm TF} \\ \\
{\displaystyle\binom{1}{0}} & \mbox{sinon} &
\end{array}
\right.
\ee
\`A la base de ce principe de correspondance réside l'idée que l'équivalent d'un mode quantique stationnaire 
$(|u_k\rangle,|v_k\rangle )$ dans le monde classique est une trajectoire classique de même énergie, elle aussi stationnaire dans son ensemble par évolution temporelle. \`A l'espérance quantique $a_k$ de l'observable ${\cal A}(\hat{\rr},\hat{\pp})$
dans le mode $(|u_k\rangle,|v_k\rangle )$ il faut donc associer {\it une moyenne} sur une trajectoire de l'espérance
$a(\rr,\pp)$ de l'équivalent classique ${\cal A}(\rr,\pp)$ dans le mode local $(U(\rr,\pp),V(\rr,\pp))$. Nous retenons donc pour la version semi-classique de la dérivée de l'opérateur phase du condensat~:
\be
-\hbar \frac{\dd\hat{\theta}^{\rm sc}}{\dd t} = \mu_0(\hat{N}) + \int \frac{\dd^3r\, \dd^3p}{(2\pi \hbar)^3} \: 
\overline{ \frac{\dd\epsilon(\rr,\pp)}{\dd N} }\:  \hat{n}(\rr,\pp)
\label{eq:thetadot_sc}
\ee
Ici, r\'ep\'etons-le, l'espérance $a(\rr,\pp)= \frac{\dd\epsilon(\rr,\pp)}{\dd N}$ n'est pas une constante du mouvement, au contraire de 
$\epsilon(\rr,\pp)$, donc on ne peut pas faire comme dans (\ref{eq:H_Bog_sc}) l'économie de la moyenne temporelle.

\subsection{De l'utilité des équations cinétiques dans le calcul de la fonction de corrélation de ${\dd\hat{\theta}}/{\dd t}$}
\label{subsec:equacin}

\begin{figure}[t]
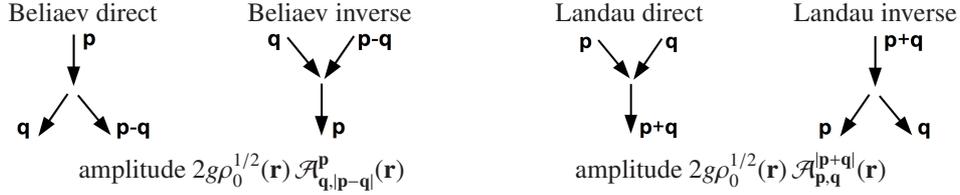

\begin{center}
\begin{tabular}{ccccccc}
Beliaev direct & \quad \quad & Beliaev inverse &  \quad \quad  \quad \quad & Landau direct &  \quad \quad & Landau inverse \\
\includegraphics[width=0.11\textwidth,clip=]{BelD.eps} & &\includegraphics[width=0.1\textwidth,clip=]{BelI.eps} & &
\includegraphics[width=0.08\textwidth,clip=]{LanD.eps}& & \includegraphics[width=0.09\textwidth,clip=]{LanI.eps} 
\end{tabular}
\begin{tabular}{ccc}
amplitude $2g\rho_0^{1/2}(\rr) \, {\cal A}_{\qq , |\pp-\qq|}^{\pp}(\rr)$ &\quad \quad \quad \quad \quad &amplitude $2g\rho_0^{1/2}(\rr)\, {\cal A}_{\pp , \qq}^{|\pp+\qq|}(\rr)$
\end{tabular}
\end{center}
\caption{Processus de Beliaev et de Landau à trois quasi-particules et amplitudes de couplage correspondantes.}
\label{fig:LK}
\end{figure}

Nous devons déterminer, dans la limite semi-classique, la fonction de corrélation de ${\dd\hat{\theta}}/{\dd t}$ pour un système préparé dans l'ensemble microcanonique. Compte tenu {\yvan des équations (\ref{eq:defCmc}) et (\ref{eq:thetadot_sc})} il faut calculer 
\be
C_{\rm mc}^{\rm sc}(\tau)=\int \frac{\dd^3r\, \dd^3p}{(2\pi \hbar)^3} \int  \frac{\dd^3r'\, \dd^3p'}{(2\pi \hbar)^3} \: 
\overline{ \frac{\dd\epsilon(\rr,\pp)}{\hbar \dd N} }\:  \overline{ \frac{\dd\epsilon(\rr',\pp')}{\hbar \dd N} }\:  
\langle  \delta \hat{n}(\rr,\pp,\tau) \,  \delta \hat{n}(\rr',\pp',0)\rangle
\label{eq:C_sc}
\ee 
où $\langle \ldots \rangle$ repr\'esente la moyenne dans l'état du système et où l'on a introduit les fluctuations des opérateurs nombres d'occupation dans l'espace des phases à l'instant $\tau$,
\be
 \delta \hat{n}(\rr,\pp,\tau) =  \hat{n}(\rr,\pp,\tau) -  \bar{n}(\rr,\pp)
\ee
L'ensemble microcanonique peut être vu semi-classiquement comme un mélange statistique à énergie constante d'états de Fock 
$|{\cal F}\rangle=|n(\rr'',\pp'')_{(\rr'',\pp'')\in \mathbb{R}^6}\rangle$ dans l'espace des phases, états propres de $H_{\rm Bog}^{\rm sc}$,
où tous les $n(\rr'',\pp'')$ sont entiers. On suppose dans un premier temps que le système est préparé dans un tel état de Fock 
$|{\cal F}\rangle$ à l'instant initial $t=0$. C'est un état propre de $\delta \hat{n}(\rr',\pp',0)$ avec la valeur propre 
$n(\rr',\pp')-\bar{n}(\rr',\pp')$~; il reste donc à calculer dans l'équation (\ref{eq:C_sc}) la quantit\'e
\be
\langle {\cal F} | \delta \hat{n}(\rr,\pp,\tau) |{\cal F}\rangle=n(\rr,\pp,\tau)-\bar{n}(\rr,\pp)\equiv  \delta n(\rr,\pp,\tau)
\ee
à $\tau>0$, c'est-à-dire l'évolution des nombres moyens d'occupation {\yvan $n(\rr,\pp,\tau)$} dans l'espace des phases, leurs valeurs initiales étant connues, 
en tenant compte $(i)$ du transport hamiltonien 
des quasi-particules et $(ii)$ de l'effet des collisions entre quasi-particules par les processus à trois quasi-particules de Beliaev ou de Landau 
\footnote{Les processus \`a quatre quasi-particules, d'ordre sup\'erieur en la fraction non condens\'ee, sont suppos\'es ici n\'egligeables.}
repr\'esent\'es sur la figure~\ref{fig:LK}.
C'est exactement ce que les équations cinétiques {\yvan habituelles de type Boltzmann quantique} savent faire, {\yvan \`a la diff\'erence que
la fonction de distribution semi-classique $n(\rr,\pp,\tau)$ ne correspond pas ici \`a un \'etat d'\'equilibre thermique local du syst\`eme, mais au nombre
moyen d'occupation \`a l'instant $\tau$ conditionn\'e au fait que l'\'etat initial du syst\`eme est un \'etat de Fock de quasi-particules. }
L'équation d'évolution des nombres d'occupation moyens $n(\rr,\pp,\tau)$ est de la forme
\be
\frac{\rm D}{{\rm D}\tau} n(\rr,\pp,\tau) + I_{\rm coll}(\rr,\pp,\tau) =0
\label{eq:eqcin}
\ee
Le premier terme est la dérivée convective résultant des équations de Hamilton classiques~:
\be
\frac{\rm D}{{\rm D}\tau} = \partial_\tau + \partial_\pp \epsilon(\rr,\pp) \cdot \partial_\rr - \partial_\rr \epsilon(\rr,\pp) \cdot \partial_\pp
\ee
Il conserve la densité dans l'espace des phases le long d'une trajectoire classique (théorème de Liouville). Le second terme décrit l'effet des collisions entre les quasi-particules, locales dans l'espace des positions, et qui ne peuvent se produire, \`a l'ordre de Beliaev-Landau,
qu'aux points où la densité de Thomas-Fermi du condensat $\rho_0(\rr)$ est non nulle (voir les diagrammes sur la figure \ref{fig:LK})~:
\footnote{Ces diagrammes font intervenir de mani\`ere
cach\'ee des processus d'absorption ou d'\'emission stimul\'ee dans le mode du condensat.}
\bea
I_{\rm coll}(\rr,\pp,\tau) &=& \frac{1}{2} \int \frac{\dd^3q}{(2\pi \hbar)^3} \; \frac{2\pi}{\hbar} \left[ 2g\rho_0^{1/2}(\rr) \, {\cal A}_{\qq , |\pp-\qq|}^{\pp}(\rr) \right]^2 \delta\left( \epsilon(\rr,\qq)+\epsilon(\rr,\pp-\qq)-\epsilon(\rr,\pp)\right) \nonumber\\
&&\times \left\{ -n(\rr,\pp,\tau)[1+n(\rr,\qq,\tau)][1+n(\rr,\pp-\qq,\tau)]+ n(\rr,\qq,\tau) n(\rr,\pp-\qq,\tau)[1+n(\rr,\pp,\tau)]\right\} \nonumber\\
&+& \phantom{\frac{1}{2} } \int \frac{\dd^3q}{(2\pi \hbar)^3} \; \frac{2\pi}{\hbar} \left[ 2g\rho_0^{1/2}(\rr) \, {\cal A}_{\pp ,\qq}^{|\pp+\qq|}(\rr) \right]^2 \delta\left( \epsilon(\rr,\pp)+\epsilon(\rr,\qq)-\epsilon(\rr,\pp+\qq)\right) \nonumber\\
&&\times \left\{ -n(\rr,\pp,\tau)n(\rr,\qq,\tau)[1+n(\rr,\pp+\qq,\tau)]+ n(\rr,\pp+\qq,\tau)[1+n(\rr,\pp,\tau)][1+n(\rr,\qq,\tau)]\right\} 
\label{eq:Icoll}
\eea
Dans ce processus sont mis en jeu, au point $\rr$, une quasi-particule d'impulsion $\pp$ (dont il faut déterminer l'évolution du nombre moyen $n(\rr,\pp\,\tau)$), une deuxième quasi-particule sortante ou entrante d'impulsion $\qq$ sur laquelle il faut intégrer, et une 
troisième quasi-particule dont l'impulsion est fixée par la conservation de la quantité de mouvement. Dans l'équation (\ref{eq:Icoll})
la première intégrale prend en compte les processus de Beliaev~; elle est affecté d'un facteur $1/2$ pour éviter le double comptage des états finals ou initiaux à deux quasi-particules $(\qq,\pp-\qq)$ et $(\pp-\qq,\qq)$~; la deuxième intégrale prend en compte les processus de Landau. On notera dans les deux cas : $(i)$ le facteur $\frac{2\pi}{\hbar}$, provenant de la règle d'or de Fermi,
$(ii)$ la prise en compte des processus directs avec un signe $-$ (ils dépeuplent le mode $\pp$ au point $\rr$) et des processus 
inverses avec un signe $+$, avec les facteurs d'amplification bosonique $1+n$, $(iii)$ la présence d'un Dirac de conservation de l'énergie au point $\rr$. Les amplitudes de couplage réduites à trois quasi-particules sont données au point $\rr$ par \cite{Superdiff,Giorgini}
\be
{\cal A}_{\pp_2,\pp_3}^{\pp_1}(\rr)=\frac{s^2(\rr,\pp_2)+s^2(\rr,\pp_3)-s^2(\rr,\pp_1)}{4s(\rr,\pp_1)s(\rr,\pp_2)s(\rr,\pp_3)}
+\frac{3}{4}s(\rr,\pp_1)s(\rr,\pp_2)s(\rr,\pp_3)
\label{eq:ampred}
\ee
avec $s(\rr,\pp)=U(\rr,\pp)+V(\rr,\pp)$. Les équations cinétiques admettent bien comme solution stationnaire les nombres d'occupation moyens de l'équilibre thermique \footnote{\`A strictement parler, cette solution stationnaire correspond aux nombres d'occupation moyens dans l'ensemble canonique, plutôt que dans l'ensemble microcanonique. La différence, calculable comme dans l'appendice C de la référence \cite{PRAlong}, mais hors de portée de nos équations cinétiques, tend vers zéro à la limite thermodynamique et est négligeable ici. On notera aussi que la non-conservation du nombre total de quasi-particules par les processus Beliaev-Landau impose à la loi de Bose $\bar{n}$ d'avoir une fugacité {\yvan unit\'e}.} 
\be
\bar{n}(\rr,\pp) = \frac{1}{\eee^{\beta \epsilon(\rr,\pp)}-1}
\label{eq:solstat}
\ee
La propriété bien connue de la loi de Bose $1+\bar{n}=\eee^{\beta \epsilon} \bar{n}$ permet de le vérifier aisément~: jointe à la conservation de l'énergie, elle conduit à la compensation parfaite en tout point des processus directs et inverses, c'est-à-dire à l'annulation des quantités entre accolades dans l'équation (\ref{eq:Icoll}), suivant le principe de micror\'eversibilité~; on a également
$\frac{\rm D}{{\rm D}\tau} \bar{n}=0$ puisque $\bar{n}(\rr,\pp) $ est une fonction de $\epsilon(\rr,\pp)$, quantité conservée par le transport hamiltonien. 

Comme notre système fluctue faiblement autour de l'équilibre, nous linéarisons les équations cinétiques autour de $n=\bar{n}$
comme dans la r\'ef\'erence \cite{PRAlong} pour obtenir
\be
\frac{\rm D}{{\rm D}\tau} \delta n(\rr,\pp,\tau) =-\Gamma(\rr,\pp,\tau) \delta n(\rr,\pp,\tau) + \int \frac{\dd^3q}{(2\pi\hbar)^3}
K(\rr,\pp,\qq)  \delta n(\rr,\qq,\tau)
\label{eq:cin_lin}
\ee
Le terme diagonal provient de la fluctuation $\delta n(\rr,\pp,\tau)$ au second membre de l'équation (\ref{eq:Icoll}), 
et le terme non local en impulsion provient des fluctuations $\delta n(\rr,\qq,\tau)$ et $\delta n(\rr,\pp\pm\qq,\tau)$ dont on regroupe les contributions grâce aux changements de variables $\qq'=\pp\pm\qq$ dans $\int\dd^3q$. L'expression de $K(\rr,\pp,\qq)$ n'est pas utile pour la suite, donnons donc seulement celle du taux d'amortissement local des quasi-particules de Bogolioubov d'impulsion $\pp$ au point $\rr$~:
\bea
\Gamma(\rr,\pp) &=&\frac{4\pi \rho_0(\rr)g^2}{\hbar} \int \frac{\dd^3q}{(2\pi \hbar)^3} \; 
\left[ {\cal A}_{\qq , |\pp-\qq|}^{\pp}(\rr) \right]^2 \delta\left( \epsilon(\rr,\qq)+\epsilon(\rr,\pp-\qq)-\epsilon(\rr,\pp)\right)
 \left[ 1 + \bar{n}(\rr,\qq) + \bar{n}(\rr,\pp-\qq)  \right] \nonumber\\
&+& \frac{8\pi \rho_0(\rr)g^2}{\hbar} \int \frac{\dd^3q}{(2\pi \hbar)^3} \; 
\left[ {\cal A}_{\pp , \qq}^{|\pp+\qq|}(\rr) \right]^2 \delta\left( \epsilon(\rr,\pp)+\epsilon(\rr,\qq)-\epsilon(\rr,\pp+\qq)\right)
 \left[ \bar{n}(\rr,\qq) - \bar{n}(\rr,\pp+\qq) \right]
\label{eq:Gamrp}
\eea
Cette expression co\"incide avec le taux d'amortissement d'un mode d'impulsion $\pp$ dans un gaz spatialement homogène de densité condensée $g\rho_0(\rr)$ \cite{Giorgini}. Tout comme $\delta n(\rr,\pp,\tau)$, 
$\langle {\cal F}| \delta n(\rr,\pp,\tau) \delta n(\rr',\pp',0) | {\cal F}\rangle$ considéré comme une fonction de $(\rr,\pp,\tau)$, obéit à l'équation (\ref{eq:cin_lin})~; il en va de même pour sa moyenne $\langle  \delta n(\rr,\pp,\tau) \delta n(\rr',\pp',0) \rangle$ sur tous les états de Fock initiaux $| {\cal F}\rangle$, puisque les coefficients $\Gamma$ et $K$ ne dépendent pas de $| {\cal F}\rangle$.
Contractons cette dernière par 
\be
B(\rr',\pp')\equiv \frac{1}{\hbar}\overline{\frac{\dd\epsilon(\rr',\pp')}{\dd N}}
\label{eq:defB}
\ee
comme dans l'équation (\ref{eq:C_sc}) pour former l'inconnue auxiliaire 
\be
X(\rr,\pp,\tau)=\int \frac{\dd^3r'\dd^3p'}{(2\pi \hbar)^3} \; B(\rr',\pp') \; \langle  \delta n(\rr,\pp,\tau) \delta n(\rr',\pp',0) \rangle
\label{eq:inconnue_auxiliaire}
\ee
Alors $X(\rr,\pp,\tau)$ évolue selon les équations cinétiques linéarisées (\ref{eq:cin_lin}) avec la condition initiale
\be
X(\rr,\pp,0)=\int \frac{\dd^3r'\dd^3p'}{(2\pi \hbar)^3} \; Q(\rr,\pp;\rr',\pp')\; B(\rr',\pp')
\ee
où l'on a introduit la matrice des covariances des nombres de quasi-particules aux temps égaux~:
\be
Q(\rr,\pp;\rr',\pp')= \langle  \delta n(\rr,\pp,0) \delta n(\rr',\pp',0) \rangle
\label{eq:defmatcov}
\ee
dont l'expression dans l'ensemble microcanonique sera reliée à celle dans l'ensemble canonique en temps utile, dans la sous-section \ref{subsec:solution}. La fonction de corrélation microcanonique de ${\dd\hat{\theta}^{\rm sc}}/{\dd t}$ cherchée vaut alors
\be
C_{\rm mc}^{\rm sc}(\tau) = \int\frac{\dd^3r \dd^3p}{(2\pi \hbar)^3} \; B(\rr,\pp) X(\rr,\pp,\tau)
\label{eq:Cmcsc}
\ee

\subsection{Solution dans l'approximation sécularo-ergodique}
\label{subsec:solution}

Notre étude se place dans le régime faiblement collisionnel $\Gamma_{\rm th} \ll \omega_\alpha$ où $\Gamma_{\rm th}$ est 
la valeur thermique typique du taux d'amortissement $\Gamma(\rr,\pp)$ des quasi-particules et $\omega_\alpha$ sont les pulsations de piégeage. Les quasi-particules ont alors le temps d'effectuer un grand nombre d'oscillations hamiltoniennes dans le piège avant de subir une collision. Nous pouvons donc effectuer l'approximation séculaire consistant à remplacer les coefficients de l'équation cinétique linéarisée (\ref{eq:cin_lin}) par leur moyenne temporelle sur une trajectoire. Ainsi
\be
\Gamma(\rr,\pp) \stackrel{\mathrm{approx.}}{\underset{\mbox{\scriptsize s\'eculaire}}{\rightarrow}} \overline{\Gamma(\rr,\pp)} = \lim_{t\to{\yvan +}\infty} \frac{1}{t} \int_0^t \dd\tau \, \Gamma(\rr(\tau),\pp(\tau))
\label{eq:approx_secu}
\ee 
et l'inconnue auxiliaire $X(\rr,\pp,\tau)$ de l'\'equation (\ref{eq:inconnue_auxiliaire}), tout comme les fluctuations des nombres d'occupation $\delta n(\rr,\pp,t)$, d\'ependent seulement
de la trajectoire $\tau\mapsto (\rr(\tau),\pp(\tau))$ passant par $(\rr,\pp)$ et du temps. Le probl\`eme reste formidable.

Heureusement, comme nous l'avons dit, 
dans un pi\`ege compl\`etement anisotrope, la dynamique hamiltonienne des quasi-particules devrait \^etre fortement 
chaotique, sauf dans les limites de tr\`es basse \'energie $\epsilon\ll\mu_{\rm TF}$ ou de tr\`es haute \'energie $\epsilon\gg\mu_{\rm TF}$
\cite{Graham,GrahamD}. Nous effectuons donc l'hypoth\`ese ergodique, en identifiant la moyenne temporelle sur une trajectoire d'\'energie $\epsilon$ 
\`a la moyenne \og{uniforme}\fg\, dans l'espace des phases sur la couche d'\'energie $\epsilon$~:
\be
\overline{\Gamma(\rr,\pp)} \stackrel{\mbox{\scriptsize hypoth\`ese}}{\underset{\mathrm{ergodique}}{=}} 
\Gamma(\epsilon) = \langle \Gamma(\rr,\pp)\rangle_\epsilon \equiv\frac{1}{\rho(\epsilon)} \int \frac{\dd^3r\dd^3p}{(2\pi\hbar)^3} 
\Gamma(\rr,\pp) \delta(\epsilon-\epsilon(\rr,\pp))
\label{eq:defGamergo}
\ee
o\`u la densit\'e d'\'etats $\rho(\epsilon)$ est donn\'ee par l'\'equation (\ref{eq:rhoeps}).
Nous reviendrons sur cette hypoth\`ese dans la section \ref{subsec:discuss_ergo}. Dans ce cas, la fonction $X(\rr,\pp,\tau)$
d\'epend seulement de l'\'energie $\epsilon=\epsilon(\rr,\pp)$ et du temps~:
\be
X(\rr,\pp,\tau) \stackrel{\mbox{\scriptsize hypoth\`ese}}{\underset{\mathrm{ergodique}}{=}} X(\epsilon,\tau)
\ee
Nous obtenons l'\'equation d'\'evolution de $X(\epsilon,\tau)$ en moyennant celle de $X(\rr,\pp,\tau)$ sur la couche d'\'energie $\epsilon$~:
\footnote{Le plus simple est de moyenner les \'equations cin\'etiques compl\`etes (\ref{eq:eqcin}), puis de lin\'eariser 
le r\'esultat autour de la solution stationnaire (\ref{eq:solstat}).}
\begin{multline}
\label{eq:evolX}
\partial_\tau X(\epsilon,\tau)= -{\Gamma}(\epsilon) X(\epsilon,\tau)
-\frac{1}{2\rho(\epsilon)} \int_0^\epsilon \dd\epsilon' L(\epsilon-\epsilon',\epsilon')
\{X(\epsilon',\tau)[\bar{n}(\epsilon) -\bar{n}(\epsilon-\epsilon')]+X(\epsilon-\epsilon',\tau)[\bar{n}(\epsilon)-\bar{n}(\epsilon')]\} \\
-\frac{1}{\rho(\epsilon)} \int_0^{+\infty} \dd\epsilon'\, L(\epsilon,\epsilon') \{ X(\epsilon',\tau) [\bar{n}(\epsilon)-\bar{n}(\epsilon+\epsilon')]
-X(\epsilon+\epsilon',\tau)[1+\bar{n}(\epsilon)+\bar{n}(\epsilon')]
\}
\end{multline}
avec
\be
\label{eq:Gamma_eps}
{\Gamma}(\epsilon)= \frac{1}{2\rho(\epsilon)} \int_0^\epsilon \dd\epsilon'\, L(\epsilon-\epsilon',\epsilon') [1+\bar{n}(\epsilon')
+\bar{n}(\epsilon-\epsilon')]
+\frac{1}{\rho(\epsilon)} \int_0^{+\infty} \dd\epsilon'\, L(\epsilon,\epsilon') [\bar{n}(\epsilon')-\bar{n}(\epsilon+\epsilon')]
\ee
Dans ces expressions, la premi\`ere int\'egrale, limit\'ee \`a des \'energies $\epsilon'$ inf\'erieures \`a l'\'energie de la quasi-particule
$\epsilon$ consid\'er\'ee, correspond aux processus de Beliaev, et la deuxi\`eme int\'egrale aux processus de Landau. Le noyau int\'egral
\footnote{Pour obtenir (\ref{eq:Lexpli}), on a r\'eduit l'\'equation (\ref{eq:defL}) \`a une int\'egrale simple sur le module $r$ (apr\`es s'\^etre ramen\'e
formellement au cas d'un pi\`ege isotrope comme dans la note \ref{note:isotropisation})
en int\'egrant en coordonn\'ees
sph\'eriques sur $p$, $q$ et sur $u$, le cosinus de l'angle entre $\pp$ et $\qq$. Dans $\int_{-1}^{1} \dd u$, l'argument
du troisi\`eme Dirac s'annule en un point $u_0$ et un seul, compte tenu des in\'egalit\'es 
$\epsilon_{|p-q|}^{\rm Bog} \leq \epsilon_p^{\rm Bog} + \epsilon_q^{\rm Bog} \leq \epsilon_{p+q}^{\rm Bog}$ 
satisfaites par la relation de dispersion de Bogolioubov $\epsilon_p^{\rm Bog}=[\frac{p^2}{2m}(\frac{p^2}{2m}+2\mu_0)]^{1/2}$,
$\forall \mu_0\geq 0$.}
\bea
\label{eq:defL}
L(\epsilon,\epsilon') &=& \int\frac{\dd^3r\,\dd^3p\,\dd^3q}{(2\pi\hbar)^6} \frac{8\pi g^2\rho_0(\rr)}{\hbar} 
\left[A_{\epsilon,\epsilon'}^{\epsilon+\epsilon'}(\rr)\right]^2 \delta(\epsilon-\epsilon(\rr,\pp))
\delta(\epsilon'-\epsilon(\rr,\qq)) \delta(\epsilon+\epsilon'-\epsilon(\rr,\pp+\qq)) \\
&=& \frac{32\sqrt{2}}{\pi^{1/2}} \frac{[\rho_0(\mathbf{0})a^3]^{1/2}}{\hbar\mu_{\rm TF}} 
\left(\frac{\mu_{\rm TF}}{\hbar\bar{\omega}}\right)^3\int_0^{\mu_{\rm TF}}
\frac{\mu_0\dd\mu_0 (\mu_{\rm TF}-\mu_0)^{1/2}\epsilon\epsilon'(\epsilon+\epsilon')\left[A_{\epsilon,\epsilon'}^{\epsilon+\epsilon'}(\mu_0)\right]^2}
{\mu_{\rm TF}^{5/2}(\epsilon^2+\mu_0^2)^{1/2} (\epsilon'^2+\mu_0^2)^{1/2} [(\epsilon+\epsilon')^2+\mu_0^2]^{1/2}}
\label{eq:Lexpli}
\eea
fait intervenir l'amplitude de couplage r\'eduite (\ref{eq:ampred}) au point $\rr$, reparam\'etr\'ee en termes des \'energies 
$\epsilon_i=\epsilon(\rr,\pp_i)$ ($1\leq i\leq 3$) ou m\^eme du potentiel chimique de Gross-Pitaevskii local $\mu_0=g\rho_0(\rr)$.
Il jouit de la propri\'et\'e de sym\'etrie $L(\epsilon,\epsilon')=L(\epsilon',\epsilon)$.

\'Ecrivons d'abord le r\'esultat avant de donner quelques indications sur son obtention (on consultera aussi la r\'ef\'erence \cite{PRAlong}).
Dans l'approximation s\'ecularo-ergodique, la fonction de corr\'elation microcanonique de ${\dd\hat{\theta}^{\rm sc}}/{\dd t}$ vaut
\be
C_{\rm mc}^{\rm ergo}(\tau) = \int_0^{+\infty} \dd\epsilon\, \rho(\epsilon) B(\epsilon) X(\epsilon,\tau)
\label{eq:Cmcergo}
\ee
Ici $B(\epsilon)$ est la moyenne ergodique de la quantit\'e $B(\rr,\pp)$ introduite dans l'\'equation (\ref{eq:defB})~:
\bea
B(\epsilon) &=& \frac{1}{\rho(\epsilon)} \int \frac{\dd^3r\, \dd^3p}{(2\pi\hbar)^3} 
\frac{\dd\epsilon(\rr,\pp)}{\hbar\dd N}\delta(\epsilon-\epsilon(\rr,\pp)) \\
\label{eq:Bergoexpli}
&=& \frac{\dd\mu_{\rm TF}/\dd N}{\hbar \pi f(\epsc)} \left[
2\epsc^{1/2}(\epsc+1)-\sqrt{2}(\epsc^2+1) \argsh\frac{(2\epsc)^{1/2}}{(1+\epsc^2)^{1/2}} 
-\epsc^{1/2}(\epsc-1)-(1+\epsc)^2\acos\frac{1}{(1+\epsc)^{1/2}}
\right] \\
B(\epsilon)& \underset{\epsilon\to 0}{=}& \frac{\dd\mu_{\rm TF}}{\hbar\dd N} \left[ -\frac{\epsc}{5}-\frac{3\pi}{40\sqrt{2}}\epsc^{3/2}+O(\epsc^2)\right],
\quad B(\epsilon) \underset{\epsilon\to+\infty}{=} \frac{\dd\mu_{\rm TF}}{\hbar\dd N} \left[-1 +\frac{32}{3\pi} \epsc^{-3/2} + O(\epsc^{-5/2})\right]
\label{eq:Bbashaut}
\eea
avec $\epsc=\epsilon/\mu_{\rm TF}$ et $f(\epsc)$ la densit\'e d'\'etats r\'eduite (\ref{eq:deff}).
L'inconnue auxiliaire $X(\epsilon,\tau)$ est solution de l'\'equation lin\'eaire (\ref{eq:evolX})  avec la condition initiale
\be
X(\epsilon,0) = \bar{n}(\epsilon) [1+\bar{n}(\epsilon)] [B(\epsilon)-\Lambda \epsilon]
\label{eq:condiX}
\ee
o\`u $\hbar\Lambda$ est la d\'eriv\'ee du potentiel chimique microcanonique par rapport \`a l'\'energie totale $E$ du gaz
\footnote{La raison profonde de l'apparition de cette d\'eriv\'ee est donn\'ee dans la r\'ef\'erence \cite{PRAlong}. Elle explique pourquoi
les \'equations cin\'etiques permettent de retrouver dans l'ensemble canonique le terme balistique $At^2$ de l'\'equation (\ref{eq:vartot})
avec la bonne expression du coefficient $A=(\partial_E\mu_{\rm mc}/\hbar)^2\mbox{Var}\,E$.},
comme dans l'\'equation (\ref{eq:acalc1})~:
\be
\label{eq:Lambda}
\Lambda = \frac{\int_0^{+\infty} \dd\epsilon\,\rho(\epsilon) \epsilon B(\epsilon) \bar{n}(\epsilon) [1+\bar{n}(\epsilon)]}
{\int_0^{+\infty} \dd\epsilon\, \rho(\epsilon)\epsilon^2 \bar{n}(\epsilon) [1+\bar{n}(\epsilon)]}
\ee
L'\'equation (\ref{eq:Cmcergo}) est la r\'ecriture ergodique de l'\'equation (\ref{eq:Cmcsc}). La condition initiale 
(\ref{eq:condiX}) est la diff\'erence de deux contributions~:
\begin{itemize}
\item la premi\`ere est celle qu'on obtiendrait dans l'ensemble canonique. La moyenne ergodique de la matrice des covariances 
{\yvan (\ref{eq:defmatcov})} serait en effet simplement $Q_{\rm can}(\epsilon,\epsilon')=\bar{n}(\epsilon)
[1+\bar{n}(\epsilon)]\delta(\epsilon-\epsilon')/\rho(\epsilon)$~;
\item la seconde provient d'une projection des fluctuations $\delta n$ canoniques sur le sous-espace des fluctuations $\delta n$
d'\'energie nulle, $\int_0^{+\infty} \dd\epsilon\, \rho(\epsilon) \epsilon \delta n(\epsilon)=0$, seules admissibles dans
l'ensemble microcanonique. Seul point subtil, cette projection doit \^etre effectu\'ee parall\`element \`a la solution  stationnaire
$e_0(\epsilon)=\epsilon \bar{n}(\epsilon)[1+\bar{n}(\epsilon)]$ des \'equations cin\'etiques lin\'earis\'ees (\ref{eq:evolX}).
\footnote{Pour que cette projection soit compatible avec l'\'evolution cin\'etique lin\'earis\'ee, 
il faut en effet que la direction de projection ainsi que l'hyperplan sur lequel on projette soient invariants par \'evolution
temporelle, le second point \'etant assur\'e par la conservation de l'\'energie. La forme de $e_0(\epsilon)$
d\'ecoule du fait que (\ref{eq:solstat}) reste une solution stationnaire pour une variation infinit\'esimale de $\beta$,
$\beta\to \beta+\delta\beta$, autour de sa valeur physique.} On v\'erifie alors que, pour la valeur de $\Lambda$ donn\'ee, $X(\epsilon,0)$ est bien
dans le sous-espace des fluctuations d'\'energie nulle.
\end{itemize}

\subsection{{\yvan R\'esultats et discussion}}
\label{subsec:resetdiscus}

Nous pr\'esentons quelques r\'esultats sous forme graphique, apr\`es un adimensionnement astucieux les rendant ind\'ependants des {\yvan pulsations} de 
pi\'egeage (pourvu qu'elles soient assez distinctes deux \`a deux pour autoriser l'hypoth\`ese ergodique) et de la force des 
interactions\footnote{Dans une premi\`ere \'etape, on montre que les r\'esultats 
ne peuvent d\'ependre des pulsations de pi\'egeage $\omega_\alpha$ que par l'interm\'ediaire
de leur moyenne g\'eom\'etrique $\bar{\omega}$. Ceci est une cons\'equence assez directe de l'hypoth\`ese ergodique et du fait
que les observables mises en jeu ici, dont le hamiltonien, d\'ependent seulement de la position $\rr$ des quasi-particules {\sl via} le potentiel
de pi\'egeage $U(\rr)=\frac{1}{2} m \sum_\alpha \omega_\alpha^2 r_\alpha^2$. Dans l'int\'egrale $\int\dd^3r$ participant \`a la
moyenne ergodique, on peut alors effectuer le changement de variables isotropisant de la note \ref{note:isotropisation}.}~; 
il suffit de conna\^\i tre la temp\'erature en unit\'es du potentiel chimique de Thomas-Fermi $\mu_{\rm TF}$. 
Ces r\'esultats t\'emoignent donc de la classe d'universalit\'e des pi\`eges harmoniques compl\`etement anisotropes, diff\'erente
de celle des syst\`emes spatialement homog\`enes de la r\'ef\'erence \cite{PRAlong}.

\begin{figure}[t]
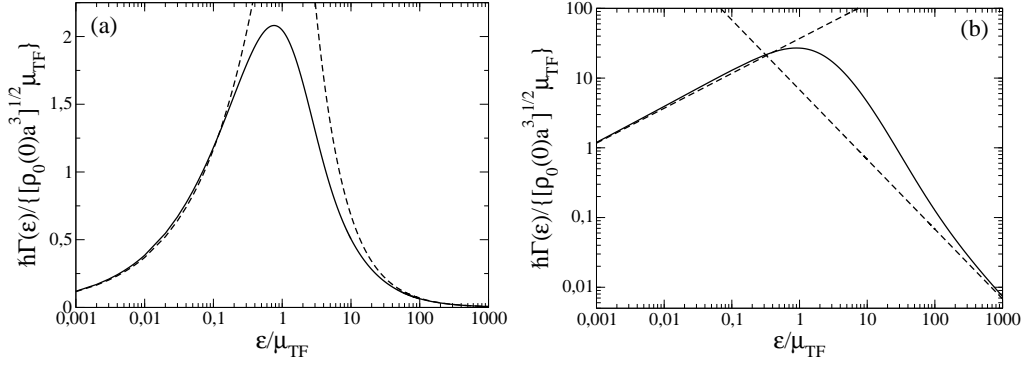

\centerline{\includegraphics[width=0.4\textwidth,clip=]{gammaeps.eps}\ \ \includegraphics[width=0.4\textwidth,clip=]{gammaeps10.eps}}
\caption{\`A la limite thermodynamique, taux d'amortissement Beliaev-Landau ${\Gamma}(\epsilon)$ des modes de Bogolioubov d'un
condensat dans un pi\`ege harmonique compl\`etement anisotrope en fonction de leur \'energie $\epsilon$, dans l'approximation
s\'ecularo-ergodique (\ref{eq:approx_secu},\ref{eq:defGamergo},{\yvan\ref{eq:Gamma_eps})}, \`a la temp\'erature (a) $k_B T=\mu_{\rm TF}$ et (b) $k_B T=10\mu_{\rm TF}$,
o\`u $\mu_{\rm TF}$ est le potentiel chimique de Thomas-Fermi
du condensat. Gr\^ace \`a l'adimensionnement choisi, la courbe est universelle~; en particulier, elle ne d\'epend pas des pulsations
de pi\'egeage $\omega_\alpha$. Les modes de Bogolioubov consid\'er\'es doivent \^etre dans le r\'egime de mouvement classique 
$\epsilon\gg\hbar\omega_\alpha$
et le syst\`eme doit \^etre dans le r\'egime d'un condensat presque pur, $[\rho_0(\mathbf{0}) a^3]^{1/2}\ll 1$ et 
$ T\ll T_c$, o\`u $\rho_0(\mathbf{0})=\mu_{\rm TF}/g$ est la densit\'e du condensat au centre du pi\`ege et $T_c$ la temp\'erature critique.
En tiret\'e, les \'equivalents {\yvan (\ref{eq:gambassener})} et (\ref{eq:asymptGammaeps}) de ${\Gamma}(\epsilon)$ \`a basse et \`a haute \'energie. }
\label{fig:Gamma}
\end{figure}

Un sous-produit int\'eressant de notre \'etude est pr\'esent\'e sur la figure \ref{fig:Gamma}~: il
s'agit du taux d'amortissement ${\Gamma}(\epsilon)$ \`a l'approximation s\'ecularo-ergodique des modes de Bogolioubov d'\'energie
$\epsilon$. Comme on sait dans une exp\'erience d'atomes froids exciter de tels modes et suivre leur d\'ecroissance
en temps, ce taux est mesurable et notre pr\'ediction peut \^etre compar\'ee aux exp\'eriences, au moins dans son r\'egime de validit\'e,
en particulier de mouvement classique $\epsilon\gg\hbar\omega_\alpha$ (les \'ecarts \`a l'hypoth\`ese ergodique sont discut\'es dans la section
\ref{subsec:discuss_ergo}).  Les comportements aux limites
\bea
\label{eq:gambassener}
\hbar {\Gamma}(\epsilon) &\underset{\epsilon\to 0}{\sim}& 
\frac{3\mathcal{I}}{4} \left(\frac{\epsilon}{\mu_{\rm TF}}\right)^{1/2} k_B T [\rho_0(\mathbf{0})a^3]^{1/2}
\ \ \ \ \mbox{avec}\ \ \ \ \mathcal{I}=4,921\, 208\, \ldots \\
\hbar\Gamma (\epsilon) &\underset{\epsilon\to +\infty}{\sim}& \frac{128\sqrt{2}}{15\sqrt{\pi}} \frac{\mu_{\rm TF}^2}{\epsilon} [\rho_0(\mathbf{0})a^3]^{1/2}
\label{eq:asymptGammaeps}
\eea
repr\'esent\'es en tiret\'e sur la figure \ref{fig:Gamma},\footnote{Pour $k_B T=\mu_{\rm TF}$, $\Gamma(\epsilon)/\epsilon^{1/2}$ pr\'esente
un maximum trompeur au voisinage de $\epsilon/\mu_{\rm TF}=0,02$ d'environ $5\%$ sup\'erieur \`a sa limite en $\epsilon=0$.}
sont \'etablis dans l'Annexe A.
Ils sont fort diff\'erents du cas spatialement homog\`ene,
o\`u le taux d'amortissement s'annule lin\'eairement en $\epsilon$ \`a basse \'energie et diverge comme $\epsilon^{1/2}$ \`a haute \'energie.
En particulier, le comportement (\ref{eq:gambassener}) en $\epsilon^{1/2}$ r\'esulte de l'existence du bord Thomas-Fermi du condensat.

\begin{figure}[t]
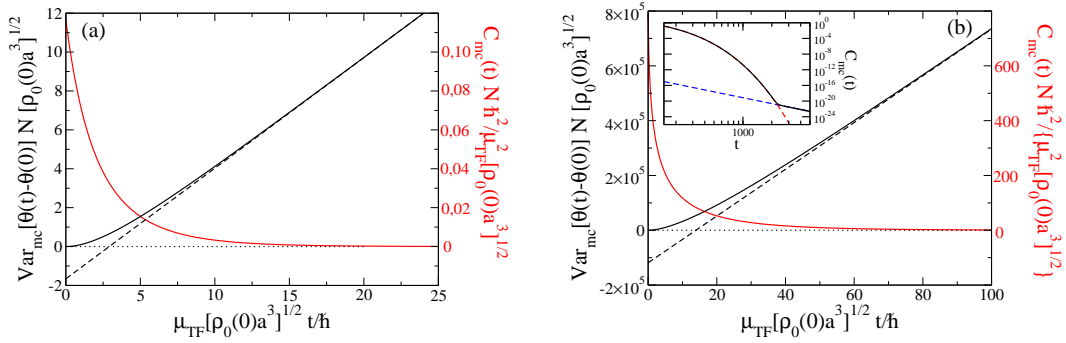

\centerline{\includegraphics[width=0.4\textwidth,clip=]{kergo4_1.eps}\quad\quad\includegraphics[width=0.4\textwidth,clip=]{kergo4_10sf.eps}}
\caption{Dans les conditions de la figure \ref{fig:Gamma}, pour un syst\`eme pr\'epar\'e \`a l'instant $0$ dans l'ensemble microcanonique
\`a la temp\'erature (a) $k_B T=\mu_{\rm TF}$ ou (b) $k_B T=10\mu_{\rm TF}$,
et isol\'e de son environnement dans son \'evolution ult\'erieure, variance du d\'ephasage $\hat{\theta}(t)
-\hat{\theta}(0)$ du condensat en fonction du temps $t$ (trait plein noir) et son comportement diffusif asymptotique
(\ref{eq:varasympt}) (tiret\'e). On a repr\'esent\'e sur la m\^eme figure la fonction de corr\'elation $C_{\rm mc}(t)$ de $\dd\hat{\theta}/\dd t$ dans 
l'approximation s\'ecularo-ergodique (\ref{eq:Cmcergo}) en fonction du temps (trait plein rouge, graduations \`a droite) et, pour (b),
dans une vignette en \'echelle log-log aux temps longs (trait plein noir) pour montrer qu'\`a une loi de d\'ecroissance quasi-exponentielle en la racine du temps
(ajustement en $t^6\exp(-C\sqrt{t})$ en tiret\'e rouge) succ\`ede une loi de puissance $\propto t^{-5}$ (tiret\'e bleu).
Comme dans la figure \ref{fig:Gamma}, la multiplication des quantit\'es sur les axes par des facteurs bien choisis rend ces 
r\'esultats universels.}
\label{fig:VaretC}
\end{figure}

Revenons \`a l'\'etalement de phase du condensat dans  l'ensemble
microcanonique. Sur la figure \ref{fig:VaretC}, nous repr\'esentons en trait plein noir la variance du d\'ephasage $\hat{\theta}(t)-\hat{\theta}(0)$
du condensat en fonction du temps $t$ dans l'approximation ergodique (\ref{eq:Cmcergo}) aux temp\'eratures $T=\mu_{\rm TF}/k_B$ et $T=10\mu_{\rm TF}/k_B$. 
La variance a un d\'epart parabolique en temps, qui correspond au r\'egime pr\'ecollisionnel $t\ll t_{\rm coll}$, o\`u $t_{\rm coll}$
est le temps de collision typique entre les quasi-particules~: on peut alors supposer que $C_{\rm mc}(\tau)\simeq C_{\rm mc}(0)$,
si bien que la contribution int\'egrale \`a l'\'equation (\ref{eq:vartot}) est $\simeq C_{\rm mc}(0) t^2$. Aux temps longs, $t\gg t_{\rm coll}$,
la fonction de corr\'elation de ${\dd\hat{\theta}}/{\dd t}$ semble tendre rapidement vers z\'ero (trait  plein rouge)~; 
une \'etude num\'erique plus pouss\'ee (voir la vignette incluse dans la figure \ref{fig:VaretC}b) r\'ev\`ele cependant la pr\'esence d'une queue
en loi de puissance $t^{-\alpha}$, 
\be
C_{\rm mc}(t) \underset{t\to +\infty}{\sim} \frac{\mathcal{C}}{t^5}
\label{eq:cmcloipuis}
\ee
L'exposant $\alpha=5$ est sup\'erieur \`a celui, $\alpha_h=3$, de la loi de d\'ecroissance de $C_{\rm mc}(t)$ 
dans le cas spatialement homog\`ene \cite{PRAlong}. Sa valeur peut \^etre retrouv\'ee par l'approximation heuristique grossi\`ere,
dite {\sl de taux} ou {\sl gaussienne projet\'ee} \cite{Genuine}, d\'ej\`a utilis\'ee pour $\alpha_h$ avec succ\`es 
dans cette m\^eme r\'ef\'erence \cite{PRAlong}~:
on ne garde dans les \'equations cin\'etiques lin\'earis\'ees (\ref{eq:evolX}) que le terme de d\'ecroissance pure $-\Gamma(\epsilon) X(\epsilon,\tau)$
au second membre,
ce qui rend leur int\'egration imm\'ediate et conduit \`a l'estimation\footnote{On a pris soin 
de tenir compte de la projection sur le sous-espace microcanonique des fluctuations 
d'\'energie nulle non seulement dans la condition initiale (\ref{eq:condiX}), mais aussi dans la contraction par $B(\epsilon)$ dans 
(\ref{eq:Cmcergo}), en y rempla\c{c}ant $B(\epsilon)$ par $B(\epsilon)-\Lambda\epsilon$~; cette pr\'ecaution, optionnelle dans la formulation exacte,
ne l'est plus ici puisque l'approximation viole la conservation de l'\'energie.}
\be
C_{\rm mc}(t) \approx \int_0^{+\infty} \dd\epsilon\, \rho(\epsilon) [B(\epsilon)-\Lambda\epsilon]^2 \bar{n}(\epsilon) [1+\bar{n}(\epsilon)]
\eee^{-\Gamma(\epsilon)t}
\label{eq:approxtauxCmc}
\ee
Les comportements en loi de puissance \`a basse \'energie de la densit\'e d'\'etats $\rho(\epsilon)$ [voir (\ref{eq:fbasse})], des coefficients 
$B(\epsilon)$ dans $\dd\hat{\theta}/\dd t$ [voir (\ref{eq:Bbashaut})], des nombres d'occupation $n(\epsilon)\sim k_BT/\epsilon$ et du taux d'amortissement
$\Gamma(\epsilon)$ [voir (\ref{eq:gambassener})] reproduisent alors l'exposant $\alpha=5$ constat\'e num\'eriquement.\footnote{En revanche,
la valeur pr\'edite pour le coefficient $\mathcal{C}$ dans 
(\ref{eq:cmcloipuis}) pour $k_BT=10\mu_{\rm TF}$, soit $\simeq 10^{-5}$, diff\`ere significativement de la valeur $\simeq 7\times 10^{-5}$ constat\'ee 
num\'eriquement.}
Comme $C_{\rm mc}(t)$ tend vers z\'ero plus vite que $1/t^{2+\eta}$, pour un certain $\eta>0$, nous obtenons le r\'esultat important suivant~:
la variance du d\'ephasage du condensat $\mbox{Var}_{\rm mc}[\hat{\theta}(t)-\hat{\theta}(0)]$ 
pr\'esente aux temps longs une croissance affine typique d'un r\'egime diffusif avec retard~:
\be
\mbox{Var}_{\rm mc}[\hat{\theta}(t)-\hat{\theta}(0)] \underset{t\gg t_{\rm coll}}{=}  2 D (t-t_0) + o(1)
\label{eq:varasympt}
\ee
en tiret\'e sur la figure \ref{fig:VaretC},
le retard $t_0$ \'etant d\^u \`a la largeur non nulle de la fonction de corr\'elation $C_{\rm mc}(\tau)$~:
\bea
\label{eq:Dint}
D &=& \int_0^{+\infty} \dd\tau\, C_{\rm mc}(\tau) \\
t_0 &=& \frac{\int_0^{+\infty} \dd\tau\, \tau C_{\rm mc}(\tau)}{\int_0^{+\infty}\dd\tau\, C_{\rm mc}(\tau)}
\label{eq:t0rapint}
\eea

\begin{figure}[t]
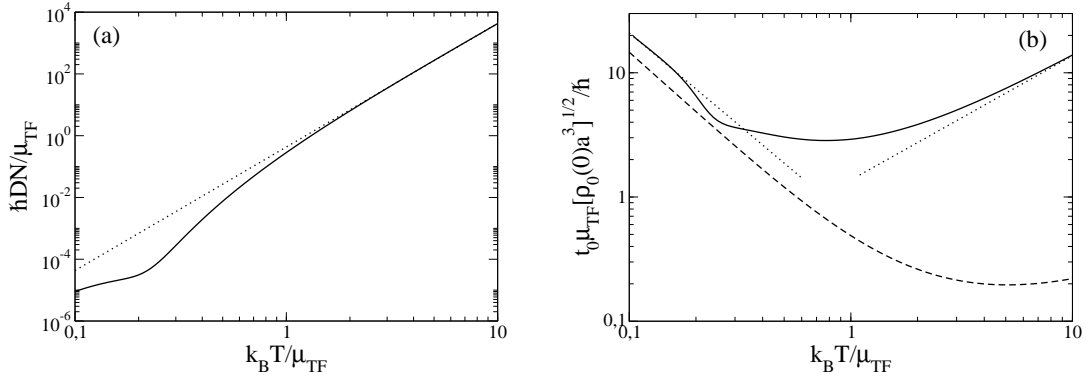

\centerline{\includegraphics[width=0.4\textwidth,clip=]{kergo3.eps}\quad \quad \includegraphics[width=0.417\textwidth,clip=]{kergo5.eps}}
\caption{Dans les conditions des figures \ref{fig:Gamma} et \ref{fig:VaretC} {\yvan pr\'ec\'edentes}, (a) coefficient de diffusion $D$ de la phase 
du condensat harmoniquement pi\'eg\'e et (b) temps de retard $t_0$ \`a la diffusion de phase en fonction de la temp\'erature  (en trait plein noir)
d\'eduits des \'equations (\ref{eq:Dint},\ref{eq:t0rapint}) et (\ref{eq:Cmcergo}). Ces quantit\'es sont ind\'ependantes de 
l'ensemble statistique.  Avec les adimensionnements choisis, les courbes sont universelles. 
On a repr\'esent\'e en pointill\'e en (a) un ajustement par une loi en $T^4$ \`a haute temp\'erature 
et en (b) un ajustement par une loi en $T$ \`a haute temp\'erature et par une loi en $T^{-3/2}$ \`a basse temp\'erature.
Les lois de puissance repr\'esent\'ees \`a haute temp\'erature sont seulement indicatives car elles omettent certainement
des facteurs logarithmiques en $k_B T/\mu_{\rm TF}$.
En (b) le tiret\'e donne l'estimation $1/{\Gamma}(\epsilon=k_B T)$ du temps de collision $t_{\rm coll}$ typique entre les quasi-particules. 
}
\label{fig:Diff_et_t0}
\end{figure}

Nous repr\'esentons le coefficient de diffusion $D$ de la phase du condensat en fonction de la temp\'erature sur la figure \ref{fig:Diff_et_t0}a.
Il pr\'esente une croissance \`a haute temp\'erature ($k_B T> \mu_{\rm TF}$) bien plus rapide que dans le cas spatialement homog\`ene~:
elle n'\'etait que lin\'eaire (\`a des facteurs logarithmiques pr\`es), elle semble ici \^etre en $T^4$ (pointill\'e sur la figure). 
Le temps
de retard \`a la diffusion $t_0$ est port\'e en fonction de la temp\'erature sur la figure \ref{fig:Diff_et_t0}b. Nous le comparons \`a l'estimation
$t_{\rm coll}\simeq 1/{\Gamma}(\epsilon=k_B T)$ du temps de collision entre quasi-particules, en tiret\'e~: celle-ci rend bien compte de la
remont\'ee brutale de $t_0$ \`a basse temp\'erature mais reproduit avec beaucoup de retard et en la sous-estimant grandement celle
\`a haute temp\'erature.
La remont\'ee de $t_0$ est bien repr\'esent\'ee par une loi en $T^{-3/2}$ \`a basse temp\'erature, et semble \^etre lin\'eaire en $T$
\`a haute temp\'erature (voir les pointill\'es).

Cherchons \`a retrouver par un raisonnement simple les lois de puissance constat\'ees. Si loi d'\'echelle il y a, elle doit survivre \`a l'approximation
de taux sur les \'equations cin\'etiques lin\'earis\'ees~; nous pouvons donc prendre l'expression approch\'ee (\ref{eq:approxtauxCmc}) de $C_{\rm mc}(t)$ comme point de d\'epart et la reporter
dans les expressions (\ref{eq:Dint}) et (\ref{eq:t0rapint}) de $D$ et $t_0$. 

\`A haute temp\'erature, les int\'egrales sur $\epsilon$ donnant $D$ et $t_0$ dans l'approximation de taux sont domin\'ees par
les \'energies d'ordre $k_B T$~; nous posons donc $\epsilon = k_B T \bar{\epsilon}$ et faisons tendre $T$ vers $+\infty$ \`a $\bar{\epsilon}$ 
fix\'e sous le signe int\'egral. Les comportements de $\rho(\epsilon)$ et $B(\epsilon)$ \`a haute \'energie sont connus.
Seul celui de $\Gamma(k_B T \bar{\epsilon})$ manque~; pour l'obtenir, nous remarquons sur
(\ref{eq:Lexpli}) que $L(k_B T\bar{\epsilon}, k_B T\bar{\epsilon}')$ tend vers une constante lorsque $T\to +\infty$. L'approximation
$L(\epsilon,\epsilon')\simeq L(\epsilon-\epsilon',\epsilon')\simeq\mbox{cte}$ d\'eclenche cependant une divergence infrarouge logarithmique 
dans les int\'egrales sur $\epsilon'$ dans (\ref{eq:Gamma_eps}), qui s'interrompt \`a $\epsilon'\lesssim \mu_{\rm TF}$, si bien que
\footnote{Un calcul plus pr\'ecis conduit \`a remplacer dans (\ref{eq:equivgam}) le symbole $\sim$ par $=$ et le facteur $\ln \frac{k_B T}{\mu_{\rm TF}}$
par $[\ln \frac{k_B T}{\mu_{\rm TF}}+\frac{\bar{\epsilon}}{4}+\ln(1-\eee^{-\bar{\epsilon}})+\frac{31}{15}-3\ln 2+ O(\mu_{\rm TF}/k_B T)]$.}
\be
\frac{\hbar\Gamma(k_BT \bar{\epsilon})}{\mu_{\rm TF} [\rho_0(\mathbf{0})a^3]^{1/2}} \underset{k_BT/\mu_{\rm TF}\to +\infty}{\sim}
\frac{512\sqrt{2}}{15\pi^{1/2}} \frac{1}{\bar{\epsilon}^{2}} \frac{\mu_{\rm TF}}{k_B T} \ln \frac{k_B T}{\mu_{\rm TF}}
\label{eq:equivgam}
\ee
Tout ceci conduit bien aux lois d'\'echelle $D\approx T^4$ et $t_0\approx T$ \`a haute temp\'erature, \`a des facteurs logarithmiques pr\`es.

\`A basse temp\'erature, nous proc\'edons de m\^eme. Le comportement de $\Gamma(k_B T \bar{\epsilon})$ est en $T^{3/2}$ lorsque $T\to 0$ \`a $\bar{\epsilon}$
fix\'e, comme l'\'equivalent (\ref{eq:gambassener}) le laissait entrevoir et comme un calcul le confirme. 
Le seul pi\`ege \`a \'eviter est que $B(k_B T\bar{\epsilon})-\Lambda k_BT\bar{\epsilon}$
est d'ordre $T^{3/2}$ lorsque $T\to 0$, et non pas $T$ comme on pourrait le croire, car les termes dominants de $B(k_B T \bar{\epsilon})$ et de 
$\Lambda k_BT \bar{\epsilon}$, tous deux lin\'eaires
en $k_B T \bar{\epsilon}$, se compensent exactement, voir l'\'equation (\ref{eq:Lambda}). Ceci conduit
aux lois de puissance exactes (sans facteur correctif logarithmique) $D\propto T^4$ et $t_0\propto T^{-3/2}$ \`a basse temp\'erature~; 
seule la seconde est accessible
sur l'intervalle en temp\'erature de la figure \ref{fig:Diff_et_t0}, mais nous avons v\'erifi\'e la premi\`ere num\'eriquement. 

{\yvan
Afin d'encourager une \'etude exp\'erimentale avec des atomes froids, terminons par une petite \'etude des limites fondamentales \`a l'observabilit\'e 
de la diffusion de phase d'un condensat pi\'eg\'e. Il existe bien entendu plusieurs difficult\'es pratiques \`a surmonter, comme $(i)$ la r\'eduction significative
des fluctuations de l'\'energie et du nombre de particules dans le gaz pour att\'enuer le brouillage balistique de la phase, concurrent dangereux de la diffusion,
$(ii)$ la mise en place d'un syst\`eme de d\'etection sensible et non biais\'e du d\'ephasage du condensat ou de la fonction de coh\'erence $g_1(t)$
de type Ramsey comme il est propos\'e dans les r\'ef\'erences \cite{chapitre,CohFer}, $(iii)$ la r\'eduction des bruits techniques du dispositif exp\'erimental,
$(iv)$ le pi\'egeage des atomes dans une cellule sous un vide suffisamment pouss\'e pour rendre n\'egligeables les pertes d'atomes froids par 
collision avec le gaz chaud r\'esiduel (pertes \`a un corps)~: des dur\'ees de vie de l'ordre de l'heure sont envisageables sous environnement
cryog\'enique \cite{Libbrecht,antiH}. Ces aspects pratiques varient suivant les \'equipes et sortent du cadre de cet article. En revanche, les pertes de particules
dues aux collisions \`a trois corps, avec formation d'un dim\`ere et d'un atome rapide, sont intrins\`eques aux gaz d'alcalins et 
constituent une limite fondamentale. Chaque perte d'un atome change, \`a un instant al\'eatoire, la vitesse de variation de la phase 
$\frac{\dd}{\dd t}\hat{\theta}$, puisque celle-ci est fonction de $N$, ce qui ajoute une composante stochastique \`a son \'evolution \cite{PRAlong,Sinatra98}. 
Pour calculer la variance du d\'ephasage du condensat induit par les pertes \`a trois corps, nous nous pla\c{c}ons
\`a l'ordre z\'ero en la fraction non condens\'ee, c'est-\`a-dire dans le cas d'un condensat pur \`a temp\'erature nulle pr\'epar\'e \`a l'instant $0$
avec un nombre initialement bien d\'efini $N$ de particules, comme dans la r\'ef\'erence \cite{PRAlong} dont nous pouvons r\'ecup\'erer 
(en les adaptant au cas pi\'eg\'e et aux pertes \`a trois corps) les expressions (G7) et (64)~:
\be
\mbox{Var}_{\rm pertes} [\hat{\theta}(t)-\hat{\theta}(0)]= \left(\frac{\dd\mu_{\rm TF}}{\hbar\dd N}\right)^2 \int_0^t\dd\tau \int_0^t \dd\tau' 
\left[\langle \hat{N}(\tau)\hat{N}(\tau')\rangle-\langle\hat{N}(\tau)\rangle \langle\hat{N}(\tau')\rangle\right]
\underset{\Gamma_3 t\to 0}{\sim} \left(\frac{\dd\mu_{\rm TF}}{\hbar\dd N}\right)^2 N \Gamma_3 t^3
\label{eq:varpertes1}
\ee
Nous avons introduit le taux de d\'ecroissance $\Gamma_3$ du nombre de particules, reli\'e comme suit \`a la constante $K_3$ des pertes \`a trois corps et
au profil de densit\'e Thomas-Fermi $\rho_0(\rr)$ du condensat~:
\be
\frac{\dd}{\dd t}N \equiv -\Gamma_3 N = -K_3 \int\dd^3r \,[\rho_0(\rr)]^3
\ee
Nous obtenons une \'ecriture plus parlante, directement comparable \`a nos r\'esultats sans pertes, en r\'ecrivant (\ref{eq:varpertes1}) 
sous forme adimensionn\'ee~:
\be
\overline{\mbox{Var}}_{\rm pertes} [\hat{\theta}(t)-\hat{\theta}(0)] \underset{\Gamma_3 t\to 0}{\sim} \frac{8}{525\pi} \bar{K}_3 {\bar{t}}\,^3
\ee
o\`u $\overline{\mbox{Var}}$ et $\bar{t}$ sont la variance du d\'ephasage et le temps \'ecoul\'e dans les unit\'es de la figure \ref{fig:VaretC}, 
et $\bar{K}_3=m K_3/(\hbar a^4)$.
La constante r\'eduite $\bar{K}_3$ est une propri\'et\'e intrins\`eque de l'esp\`ece atomique utilis\'ee dans l'exp\'erience 
(m\^eme s'il est possible de la faire varier \`a l'aide d'une r\'esonance de Feshbach magn\'etique \cite{Lev}).  
Pour estimer l'ordre de grandeur de $\bar{K}_3$ dans un gaz d'atomes froids, prenons l'exemple du rubidium 87 dans le sous-niveau hyperfin fondamental $|F=1,m_F=-1\rangle$ en champ magn\'etique quasi nul~:
les mesures donnent $K_3=6\times 10^{-42}$ m${}^6$/s et $a=5,31$ nm \cite{Egorov}
donc $\bar{K}_3\simeq 10$. Sur la figure \ref{fig:VaretC}a ($k_B T=\mu_{\rm TF}$), au temps r\'eduit $\bar{t}=5$ d'entr\'ee dans le r\'egime asymptotique
de la diffusion de phase, on voit que la variance parasite induite par les pertes pour cette valeur de $\bar{K}_3$
est environ le triple de la variance utile~; leurs d\'ependances en temps
tr\`es diff\'erentes devraient cependant permettre de les s\'eparer. La situation est beaucoup plus favorable
\`a plus haute temp\'erature, $k_B T\gg \mu_{\rm TF}$, l'effet des pertes sur la variance du d\'ephasage \'etant par exemple 
encore n\'egligeable au temps r\'eduit $\bar{t}=100$ sur la figure \ref{fig:VaretC}b ($k_B T=10 \mu_{\rm TF}$).
}

\subsection{Discussion de l'hypothèse ergodique}
\label{subsec:discuss_ergo}

\begin{figure}[t]
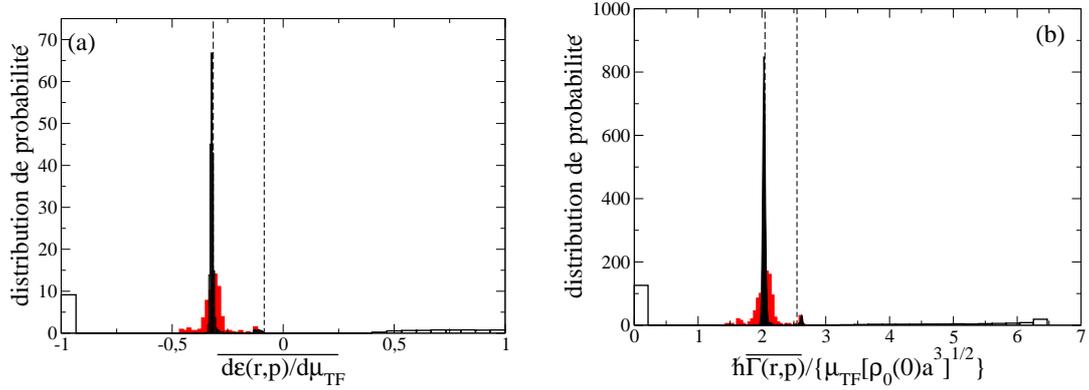

\centerline{\includegraphics[width=0.4\textwidth,clip=]{histoergo2_dedmu.eps}\quad \quad \includegraphics[width=0.418\textwidth,clip=]{histoergo2_taux.eps}}
\caption{Histogramme des moyennes temporelles des grandeurs physiques (a) ${\dd\epsilon(\rr,\pp)}/{\dd\mu_{\rm TF}}$ et (b)
$\Gamma(\rr,\pp)$  pour $10^4$ valeurs initiales ind\'ependantes $(\rr(0),\pp(0))$ tir\'ees uniform\'ement sur la couche d'\'energie 
$\epsilon=\mu_{\rm TF}$ et une \'evolution hamiltonienne (\ref{eq:drdt},\ref{eq:dpdt})
des quasi-particules de Bogolioubov pendant une dur\'ee $t$ variable~: $t=0$ (barres creuses, trait noir), 
$t=5\times 10^3/\bar{\omega}$ (barres pleines rouges), $t=2,5\times 10^5/\bar{\omega}$ (barres pleines noires). 
Le potentiel harmonique est compl\`etement anisotrope,
avec des pulsations de pi\'egeage incommensurables (rapports $\omega_x : \omega_y : \omega_z = 1 : \sqrt{3} : \sqrt{5}-1 $).
 La temp\'erature, qui intervient dans le taux d'amortissement
$\Gamma(\rr,\pp)$, vaut $T=\mu_{\rm TF}/k_B$. Tiret\'es verticaux noirs~: \`a gauche, la valeur ergodique (moyenne de la grandeur
sur la couche d'\'energie $\epsilon$)~; \`a droite, la moyenne temporelle de la grandeur sur une p\'eriode de la trajectoire lin\'eaire d'\'energie $\epsilon$ 
selon la direction $Oy$ (axe du pi\`ege le plus confinant), obtenue analytiquement pour (a) (voir la note \ref{note:trajlin}) et
num\'eriquement pour (b).}
\label{fig:histoergo}
\end{figure}

\begin{figure}[ht]
\begin{center}
\includegraphics[width=0.4\textwidth,clip=]{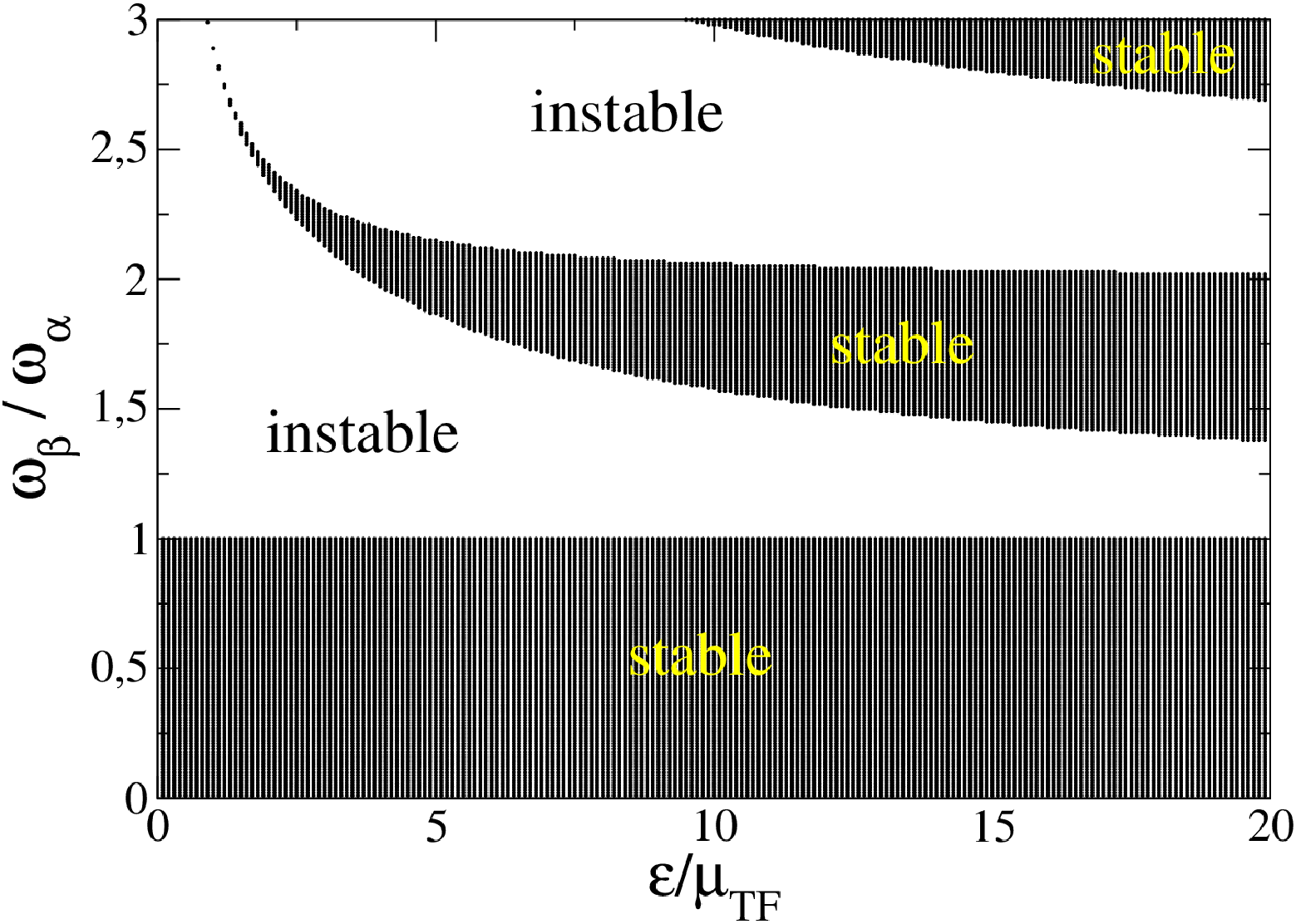} 
\end{center}
\caption{Pour la dynamique hamiltonienne classique des quasi-particules de Bogolioubov dans un potentiel harmonique, 
stabilit\'e du mouvement lin\'eaire selon un axe propre $O\alpha$ du pi\`ege vis-\`a-vis d'une perturbation initiale infinit\'esimale (un
d\'eplacement) selon un autre axe propre $O\beta$, en fonction de l'\'energie $\epsilon$ de la trajectoire et du rapport $\omega_\beta/\omega_\alpha$
des pulsations de pi\'egeage (les zones hachur\'ees sont stables).}
\label{fig:stabil}
\end{figure}

\begin{figure}[t]
\begin{center}
\includegraphics[width=\textwidth,clip=]{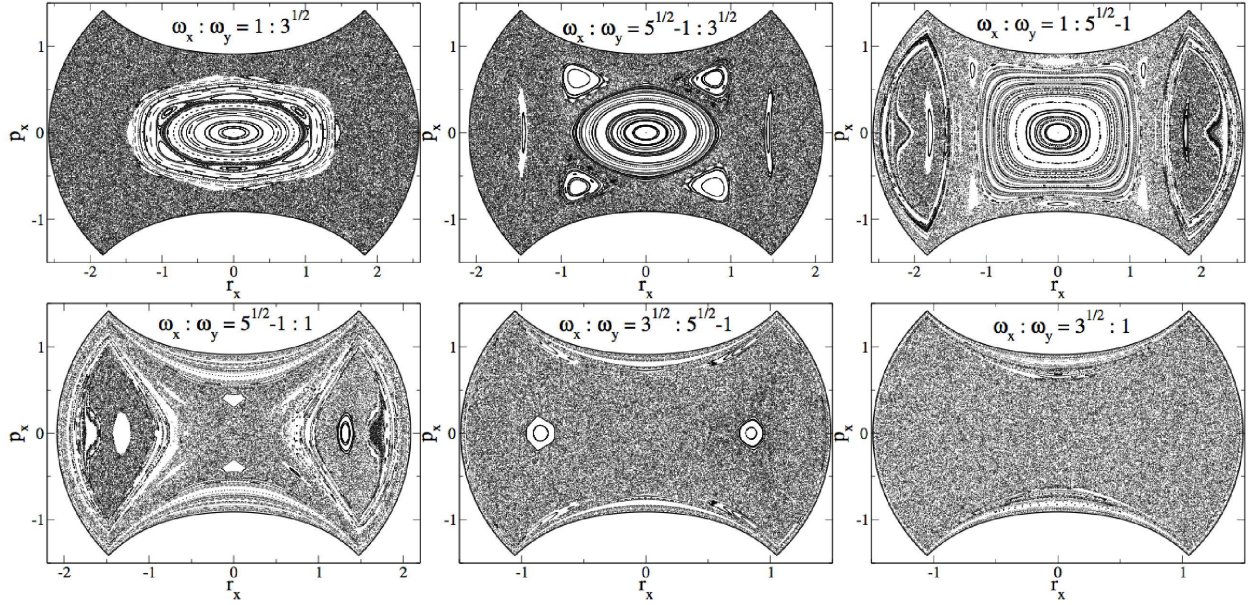}
\end{center}
\caption{Pour la dynamique hamiltonienne classique des quasi-particules de Bogolioubov dans un potentiel harmonique, 
sections de Poincar\'e dans le plan de coupe $(r_y=0,p_y(\epsilon)>0)$ de trajectoires planes dans $xOy$
(200 trajectoires ind\'ependantes, temps d'\'evolution $5000/\bar{\omega}$), 
avec un rapport $\omega_x : \omega_y$ prenant toutes les valeurs
possibles dans le pi\`ege de la figure \ref{fig:histoergo}~: $1:\sqrt{3}$, $\sqrt{3}:1$, $1:\sqrt{5}-1$, $\sqrt{5}-1:1$,
$\sqrt{3}:\sqrt{5}-1$ et $\sqrt{5}-1:\sqrt{3}$. 
{\yvan Les sections ont \'et\'e ordonn\'ees dans le sens d'un rapport $\omega_x/\omega_y$ croissant de gauche \`a droite et de haut en bas
(en effet, $1/\sqrt{3} < (\sqrt{5}-1)/\sqrt{3} < 1/(\sqrt{5}-1) < 1$). Ceci r\'ev\`ele que}
la section de Poincar\'e est d'autant plus chaotique que le rapport $\omega_x/\omega_y$ est plus grand. $r_x$ est en unit\'es 
de $(\mu_{\rm TF}/m\bar{\omega}^2)^{1/2}$ et $p_x$ en unit\'es de $(m\mu_{\rm TF})^{1/2}$.}
\label{fig:poinc}
\end{figure}

\begin{figure}[t]
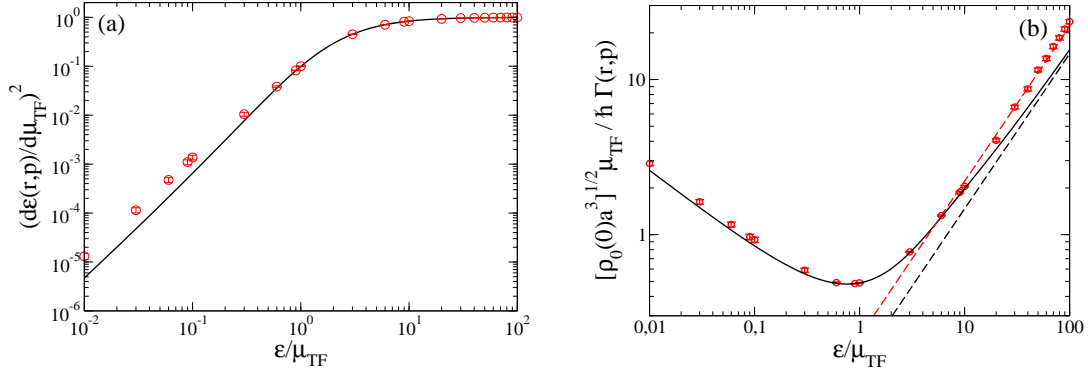

\centerline{\includegraphics[width=0.418\textwidth,clip=]{dedm2_ener_log.eps}\quad \quad \includegraphics[width=0.4\textwidth,clip=]{invtaux_ener_log.eps}}
\caption{Visualisation de l'erreur introduite par l'hypoth\`ese ergodique $\bar{O}=\langle O\rangle_\epsilon$
sur deux grandeurs physiques $O(\rr,\pp)$ intervenant dans la diffusion de la phase du condensat, en fonction de l'\'energie $\epsilon$~: 
(a) pour $O(\rr,\pp)={\dd\epsilon(\rr,\pp)}/{\dd\mu_{\rm TF}}$,
on compare comme dans l'\'equation (\ref{eq:entre1}) $\langle\bar{O}^2\rangle_\epsilon$ (cercles rouges)
\`a son approximation ergodique $\langle O\rangle^2_\epsilon$ tir\'ee de (\ref{eq:Bergoexpli}) (trait plein noir)~;
(b) pour $O(\rr,\pp)=\Gamma(\rr,\pp)$, on compare comme dans l'\'equation (\ref{eq:entre2}) $\langle 1/\bar{O}\rangle_\epsilon$ (cercles rouges) 
\`a son approximation ergodique $1/\langle O\rangle_\epsilon$ (trait plein noir). 
La moyenne temporelle $\bar{O}$ est calcul\'ee sur une dur\'ee d'\'evolution $t=5\times 10^4/\bar{\omega}$ {\yvan des quasi-particules dans le pi\`ege de la figure~\ref{fig:histoergo}}~;
la moyenne sur la couche d'\'energie $\epsilon$ est prise sur 200 trajectoires ind\'ependantes,
avec des conditions initiales $(\rr,\pp)$ tir\'ees selon la loi uniforme $\delta(\epsilon-\epsilon(\rr,\pp))/\rho(\epsilon)$, 
ce qui conduit \`a une incertitude statistique repr\'esent\'ee par les barres d'erreur sur la figure.
Droites tiret\'ees dans (b)~: \'equivalents asymptotiques (\ref{eq:invGamasympt}) de $\langle 1/\bar{\Gamma}\rangle_\epsilon$ (en rouge)  
et (\ref{eq:asymptGammaeps}) de $1/\langle \Gamma\rangle_\epsilon$ (en noir).}
\label{fig:compergo}
\end{figure}

Le mouvement classique des quasi-particules de Bogolioubov, comme l'ont montr\'e les r\'ef\'erences \cite{Graham,GrahamD} 
dans le cas d'un pi\`ege
harmonique \`a sym\'etrie de r\'evolution, est fortement chaotique aux \'energies $\epsilon\simeq \mu_{\rm TF}$ mais m\^eme \`a cette
\'energie, les sections de Poincar\'e r\'ev\`elent des \^\i lots de stabilit\'e dans l'espace des phases, qui ne sont pas travers\'es
par les trajectoires de la mer chaotique~: il n'y a donc pas ergodicit\'e au sens strict.

Qu'en est-il dans le cas d'un pi\`ege compl\`etement anisotrope~? Nous souhaitons tester l'hypoth\`ese ergodique pour deux grandeurs physiques.
La premi\`ere appara\^\i t dans nos \'equations cin\'etiques lin\'earis\'ees, il s'agit du taux d'amortissement $\Gamma(\rr,\pp)$.
La seconde intervient dans les conditions initiales de la fonction de corr\'elation $C_{\rm mc}(\tau)$ de ${\dd\hat{\theta}}/{\dd t}$, 
il s'agit de ${\dd\epsilon(\rr,\pp)}/{\dd\mu_{\rm TF}}$. Pour un \'echantillonnage uniforme de la surface d'\'energie $\epsilon$,
c'est-\`a-dire avec la distribution de probabilit\'e $\delta(\epsilon-\epsilon(\rr,\pp))/\rho(\epsilon)$ dans l'espace
des phases, nous montrons sur la figure \ref{fig:histoergo} les histogrammes de ces grandeurs apr\`es moyennage temporel 
sur chaque trajectoire {\yvan pendant} des temps respectivement de $t=0$, $t=5000/\bar{\omega}$ et $t=250\, 000/\bar{\omega}$,
\`a {\yvan l'\'energie} $\epsilon=k_B T=\mu_{\rm TF}$ pour des pulsations de pi\'egeage incommensurables.\footnote{Les \'equations du mouvement (\ref{eq:drdt},\ref{eq:dpdt}),
mises dans leur ensemble sous la forme $\frac{\dd}{\dd t}\mathbf{X}=\mathbf{f}(\mathbf{X})$, sont int\'egr\'ees
num\'eriquement avec un sch\'ema semi-implicite du second ordre, $\mathbf{X}(t+\dd t)=\mathbf{X}(t)+\dd t[1-\frac{\dd t}{2} M]^{-1} \mathbf{f}(\mathbf{X}(t))$
o\`u $M$ est la diff\'erentielle premi\`ere de $\mathbf{f}(\mathbf{X})$ en $\mathbf{X}(t)$ \cite{Recipes}. Si la trajectoire traverse la surface
du condensat entre $t$ et $t+\dd t$, il faut d\'eterminer l'instant $t_s$ de la travers\'ee avec une erreur $O(\dd t)^3$, puis
appliquer le sch\'ema semi-implicite successivement sur $[t,t_s]$ et $[t_s, t+\dd t]$, pour pallier la discontinuit\'e de $\mathbf{f}(\mathbf{X})$
et de ses d\'eriv\'ees.} Le moyennage temporel 
conduit bien \`a un r\'etr\'ecissement spectaculaire de la distribution de probabilit\'e, qui se pique autour de la moyenne
ergodique (ligne tiret\'ee \`a gauche), ce qui va dans le sens de l'hypoth\`ese ergodique. Cette dynamique de r\'etr\'ecissement se poursuit
sur des temps tr\`es longs, mais ne vient jamais \`a bout d'un petit pic lat\'eral \'eloign\'e de la moyenne ergodique. 

Un examen des trajectoires contribuant au pic lat\'eral montre qu'il s'agit de perturbations de trajectoires lin\'eaires stables le long de l'axe propre du pi\`ege
de raideur maximale. La valeur moyenne temporelle des deux grandeurs consid\'er\'ees  sur ces trajectoires lin\'eaires
est repr\'esent\'ee par le tiret\'e \`a droite sur la figure \ref{fig:histoergo}, elle est effectivement proche du pic en question.
Le diagramme de stabilit\'e d'une trajectoire lin\'eaire le long d'un axe propre $\alpha$ quelconque du pi\`ege vis-\`a-vis
d'une perturbation le long d'un autre axe propre $\beta$ est repr\'esent\'e sur la figure \ref{fig:stabil}, dans le plan (\'energie, rapport
$\omega_\beta/\omega_\alpha$). Il montre que la trajectoire lin\'eaire selon l'axe le plus confinant est bien stable, \`a toute \'energie.
\footnote{La trajectoire lin\'eaire d'une quasi-particule d'\'energie $\epsilon$ selon l'axe propre $O\alpha$ du pi\`ege s'\'ecrit
$m^{1/2}\omega_\alpha r_\alpha(t)=|\mu_{\rm TF}+\ii\epsilon|\sin(\sqrt{2}\omega_\alpha t)/\sqrt{G(t)}$ et 
$p_\alpha(t)/(2m)^{1/2}=\epsilon/\sqrt{G(t)}$ avec $G(t)=\mu_{\rm TF}+|\mu_{\rm TF}+\ii\epsilon|\cos(\sqrt{2}\omega_\alpha t)$. Ceci
correspond au choix $r_\alpha(0)=0,p_\alpha(0)\geq 0$ et vaut pour $-t_s \leq t\leq t_s$, o\`u le temps d'atteinte de la surface du condensat
est donn\'e par $\sqrt{2}\omega_\alpha t_s=\acos\frac{\epsilon-\mu_{\rm TF}}{|\mu_{\rm TF}+\ii\epsilon|}$. \`A l'ext\'erieur du condensat,
la quasi-particule oscille harmoniquement comme une particule libre pendant un temps $2\omega_\alpha^{-1}\atan[(\epsilon/\mu_{\rm TF})^{1/2}]$
avant de regagner le condensat, pour le traverser en un temps $2 t_s$, et ainsi de suite. La connaissance de la trajectoire
rend imm\'ediate l'analyse lin\'eaire de stabilit\'e num\'erique. Elle permet aussi de calculer analytiquement la moyenne temporelle
de $\dd\epsilon(\rr,\pp)/\dd\mu_{\rm TF}$ sur la trajectoire lin\'eaire~; si l'on pose $\epsc=\epsilon/\mu_{\rm TF}$, le r\'esultat s'\'ecrit
\[
\overline{\frac{\dd\epsilon(\rr,\pp)}{\dd\mu_{\rm TF}}} = \frac{\ln\frac{1+\epsc+\sqrt{2\epsc}}{(1+\epsc^2)^{1/2}}-\sqrt{2}\atan\sqrt{\epsc}}
{\acos\frac{\epsc-1}{(1+\epsc^2)^{1/2}} +\sqrt{2}\atan\sqrt{\epsc}}
\]
\label{note:trajlin}
}
Les sections de Poincar\'e des trajectoires planes dans les plans $\alpha O\beta$ sur la figure \ref{fig:poinc}
pr\'ecisent la largeur de l'\^\i lot de stabilit\'e et r\'ev\`elent
l'existence d'\^\i lots secondaires, etc.  Il n'y a donc pas pleine ergodicit\'e de notre dynamique classique, 
m\^eme aux \'energies $\epsilon\approx \mu_{\rm TF}$, m\^eme dans le cas compl\`etement anisotrope. 

Pour mesurer quantitativement l'erreur commise par l'hypoth\`ese ergodique dans le calcul de $C_{\rm mc}(0)$ et de $C_{\rm mc}(\tau>0)$, nous 
consid\'erons les \'ecarts entre
\bea
\label{eq:entre1}
\left\langle\overline{\frac{\dd\epsilon(\rr,\pp)}{\dd\mu_{\rm TF}}}^2\right\rangle_\epsilon &\mbox{et}& 
\left\langle\frac{\dd\epsilon(\rr,\pp)}{\dd\mu_{\rm TF}}\right\rangle_\epsilon^2 \\
\left\langle \frac{1}{\overline{\Gamma(\rr,\pp)}} \right\rangle_\epsilon &\mbox{et}& \frac{1}{\left\langle{\Gamma(\rr,\pp)}\right\rangle}_\epsilon
= \frac{1}{\Gamma(\epsilon)}
\label{eq:entre2}
\eea
o\`u, rappelons-le, la barre horizontale $\overline{O(\rr,\pp)}$ au-dessus d'une grandeur physique repr\'esente la moyenne temporelle sur la
trajectoire passant par $(\rr,\pp)$  dans l'espace des phases {\yvan comme dans l'\'equation (\ref{eq:approx_secu})}, 
et les crochets $\langle O(\rr,\pp)\rangle_\epsilon$ repr\'esentent
la moyenne uniforme sur la couche d'\'energie $\epsilon$ comme dans l'\'equation (\ref{eq:defGamergo}). Dans le tableau 
d'\'equations (\ref{eq:entre1},\ref{eq:entre2}), la colonne de gauche contient les grandeurs apparaissant dans $C_{\rm mc}(0)$ 
ou dans les \'equations cin\'etiques s\'eculaires avant l'approximation ergodique, et la colonne de droite ce qu'elles deviennent
apr\`es approximation ergodique. Point important, nous consid\'erons dans l'\'equation (\ref{eq:entre2})
$1/\bar{\Gamma}$ plut\^ot que $\bar{\Gamma}$ car ce sont les inverses $M^{-1}$ et $M^{-2}$ qui apparaissent dans l'expression 
(\ref{eq:Dint},\ref{eq:t0rapint})
du coefficient de diffusion $D$ et du temps de retard $t_0$, $M$ \'etant l'op\'erateur repr\'esentant le second membre des \'equations
cin\'etiques lin\'earis\'ees (\ref{eq:evolX}).\footnote{Faut-il le rappeler, $\langle\overline{\Gamma(\rr,\pp)}\rangle_\epsilon=\Gamma(\epsilon)$, la
moyenne uniforme \'etant invariante par \'evolution temporelle. Du coup, l'in\'egalit\'e entre moyenne  harmonique et moyenne arithm\'etique
impose $\langle 1/\overline{\Gamma(\rr,\pp)}\rangle_\epsilon \ge 1/\Gamma(\epsilon)$.}
Les quantit\'es \`a comparer (\ref{eq:entre1},\ref{eq:entre2}) sont repr\'esent\'ees en fonction de l'\'energie $\epsilon$ sur la figure 
\ref{fig:compergo} \`a la temp\'erature $T=\mu_{\rm TF}/k_B$. Il y a un accord remarquable sur une large plage d'\'energie 
autour de $\epsilon=\mu_{\rm TF}$. 
Les \'ecarts \`a {\yvan l'approximation} ergodique \`a tr\`es basse \'energie et \`a tr\`es haute \'energie \'etaient attendus~: dans ces limites, 
la dynamique classique devient int\'egrable \cite{Graham}. \`A haute \'energie, nous obtenons pour la grandeur $\Gamma(\rr,\pp)$ la 
pr\'ediction analytique suivante~:\footnote{
\`A l'ordre dominant en $\epsilon$, on obtient $\overline{\Gamma(\rr,\pp)}$ en moyennant l'\'equivalent (\ref{eq:Gamhomasympt})
(dans lequel $\mu_0=g\rho_0(\rr)$) sur une trajectoire harmonique non perturb\'ee par le condensat, $r_\alpha(t) = A_\alpha \cos (\omega_\alpha t + \phi_\alpha)$,
$\forall \alpha\in\{x,y,z\}$.  Consid\'erons astucieusement la quantit\'e $g\rho_0(\rr)$ \`a moyenner comme une fonction $f(\boldsymbol{\theta})$ des
angles $\theta_\alpha=\omega_\alpha t+\phi_\alpha$. C'est une fonction p\'eriodique de p\'eriode $2\pi$ selon chaque direction, d\'ecomposable
en s\'erie de Fourier, $f(\boldsymbol{\theta})=\sum_{\mathbf{n}\in\mathbb{Z}^3} c_{\mathbf{n}} \eee^{\ii\mathbf{n}\cdot\boldsymbol{\theta}}$. 
Dans le cas incommensurable, $\mathbf{n}\cdot\boldsymbol{\omega}\neq 0$ et la moyenne temporelle de $\eee^{\ii\mathbf{n}\cdot\boldsymbol{\theta}}$ est 
nulle $\forall \mathbf{n}\in \mathbb{Z}^{3*}$, si bien que $\overline{f(\boldsymbol{\theta})}=c_{\mathbf{0}}$.
Dans l'habituelle expression int\'egrale de $c_{\mathbf{0}}$, on effectue le changement de variable $x_\alpha=X_\alpha \cos\theta_\alpha$,
o\`u $X_\alpha=(\epsilon_\alpha/\mu_{\rm TF})^{1/2}$ et $\epsilon_\alpha$ est l'\'energie
du mouvement selon $O\alpha$. Il reste \`a faire tendre les $X_\alpha$ vers $+\infty$ sous le signe int\'egral
pour obtenir 
\[
\frac{\hbar\overline{\Gamma(\rr,\pp)}}{\mu_{\rm TF}[\rho_0(\mathbf{0})a^3]^{1/2}} \underset{\epsilon\to +\infty}{\sim} 
\frac{32\sqrt{2}}{15\pi^{3/2}} \left(\frac{\epsilon}{\mu_{\rm TF}}\right)^{1/2} 
\prod_\alpha \left(\frac{\mu_{\rm TF}}{\epsilon_\alpha}\right)^{1/2}
\]
En moyennant {\yvan l'inverse de} cet \'equivalent sur la distribution de probabilit\'e $2\epsilon^{-2}\delta(\epsilon-\sum_\alpha \epsilon_\alpha)$ des \'energies
par direction pour un oscillateur harmonique d'\'energie totale $\epsilon$, on tombe sur (\ref{eq:invGamasympt}).}
\be
\left\langle \frac{1}{\hbar\overline{\Gamma(\rr,\pp)}}\right\rangle_\epsilon 
\underset{\epsilon\to+\infty}{\sim}\frac{\pi^{5/2}}{56\sqrt{2}}\frac{\epsilon}{\mu_{\rm TF}^2} \frac{1}{[\rho_0(\mathbf{0})a^3]^{1/2}}
\label{eq:invGamasympt}
\ee
Elle diff\`ere de la pr\'ediction  ergodique (\ref{eq:asymptGammaeps}) par un coefficient num\'erique, et reproduit bien les r\'esultats des 
simulations num\'eriques (voir le tiret\'e rouge sur la figure \ref{fig:compergo}b). Ceci nous interdit de calculer le coefficient de diffusion
$D$ et le retard \`a la diffusion $t_0$ dans 
l'approximation s\'ecularo-ergodique \`a trop haute temp\'erature.
En ce qui concerne la grandeur ${\dd\epsilon(\rr,\pp)}/{\dd\mu_{\rm TF}}$, qui tend vers $-1$ \`a haute \'energie,
l'\'ecart ne peut \^etre significatif qu'\`a basse \'energie~; il ne le devient en fait qu'\`a tr\`es basse \'energie, et ne poserait
probl\`eme \`a notre calcul ergodique de $D$ et $t_0$ qu'\`a des temp\'eratures $k_B T\ll \mu_{\rm TF}$ rarement atteintes
dans les exp\'eriences sur les atomes froids.

\section{Conclusion}
\label{sec:conclusion}

Motivés par des progrès expérimentaux récents dans la manipulation des gaz d'atomes froids piégés 
\cite{Treutlein,Oberthaler,Schmiedmayer},
nous avons étudié théoriquement le temps de cohérence et la dynamique de phase d'un condensat de Bose-Einstein dans un gaz de bosons isolé et piégé harmoniquement, un problème fondamental important pour les applications interférométriques.
La variance du déphasage subi par le condensat au bout d'un temps $t$ croît ind\'efiniment avec $t$, ce qui limite le temps de cohérence intrinsèque du gaz. Pour $t \gg t_{\rm coll}$, où $t_{\rm coll}$ est le temps de collision typique entre les quasi-particules de Bogolioubov, elle devient une fonction quadratique du temps,
\begin{equation}
\mbox{Var}[\hat{\theta}(t)-\hat{\theta}(0)] = At^2 + 2D(t-t_0)+o(1)
\end{equation}
où $\hat{\theta}$ est l'opérateur phase du condensat. Cette loi asymptotique a la même forme que dans le cas spatialement homogène précédemment étudié \cite{PRAlong}, ce qui n'était pas garanti, mais les coefficients diffèrent bien entendu. Pour les calculer, nous considérons la limite thermodynamique dans le piège, dans laquelle le nombre de particules tend vers l'infini, 
$N\to+\infty$, à temp\'erature $T$ et potentiel chimique de Gross-Pitaevskii $\mu_{\rm GP}$ fixés. Ceci impose que les pulsations de piégeage réduites tendent vers zéro,
$\hbar \omega_\alpha/\mu_{\rm GP} \to 0$, ce que nous réinterprétons avantageusement comme une limite classique $\hbar\to0$.

Le terme dominant $At^2$ est dû aux fluctuations dans l'état initial des quantités conservées par évolution temporelle, $N$ et $E$, où $E$ est l'énergie totale du gaz. Nous donnons une expression explicite (\ref{eq:grandA})-(\ref{eq:acalc1})-(\ref{eq:acalc2}) 
du coefficient $A$ dans un ensemble généralisé, mélange statistique quelconque d'ensembles microcanoniques avec des fluctuations au plus normales de $N$ et de $E$. 
Dans ce cas, $A=O(1/N)$. Nous obtenons une forme plus simple (\ref{eq:Apart}) dans le cas d'un mélange statistique d'ensembles canoniques de même température mais de nombre de particules variable. Aux températures habituelles, plus grandes que $\mu_{\rm GP}/k_B$, et pour des fluctuations poissoniennes du nombre de particules, la contribution à $A$ des fluctuations thermiques de $E$ est rendue négligeable par un facteur d'ordre la fraction non condensée $\propto(T/T_c)^{3}$. Il faut réduire d'autant la variance de $N$ pour voir l'effet des fluctuations thermiques sur l'\'etalement balistique de la phase
du condensat.

Le terme sous-dominant $2D(t-t_0)$ ne dépend pas de l'ensemble dans lequel le système est préparé, du moins au premier ordre non nul $1/N$ à la limite thermodynamique, et il est le seul qui subsiste dans l'ensemble microcanonique.
Le calcul de ses deux ingrédients, le coefficient de diffusion $D$ de la phase et le retard à la diffusion $t_0$,
nécessite la connaissance à tout temps de la fonction de corrélation de $\dd{\hat{\theta}}/\dd t$ dans l'ensemble microcanonique,
et donc la résolution d'équations cinétiques linéarisées sur les nombres d'occupation des quasi-particules de Bogolioubov. 
Ce sont en effet les fluctuations temporelles de ces nombres d'occupation pour une réalisation donnée du système qui stochastisent l'évolution de la phase du condensat.
Nous adoptons pour ce faire une description semi-classique, dans laquelle le mouvement des quasi-particules dans le gaz piégé est
traité classiquement dans l'espace des phases $(\rr,\pp)$, mais le champ bosonique des quasi-particules est traité quantiquement,
au travers des opérateurs nombres d'occupation $\hat{n}(\rr,\pp)$. Dans les observables quantiques de la forme 
$\hat{A}=\sum_k a_k \hat{n}_k$, comme $\dd{\hat{\theta}}/\dd t$, la moyenne $a_k$ et la somme sur les modes quantiques $k$
de Bogolioubov sont alors remplacées, selon un principe de correspondance, par une moyenne temporelle et une intégrale sur les trajectoires classiques (voir les équations (\ref{eq:A})-(\ref{eq:Asc})). Les équations cinétiques linéarisées sur les fluctuations 
$\delta \hat{n}(\rr,\pp)$ comportent une partie de transport, selon le mouvement hamiltonien classique des quasi-particules,
et une intégrale de collision, locale en position, qui décrit les processus d'interaction Beliaev-Landau à trois quasi-particules. {\yvan
Elles prennent la m\^eme forme que les \'equations de Boltzmann quantiques lin\'earis\'ees sur la fonction de distribution semi-classique 
$n(\rr,\pp,t)$ des quasi-particules dans l'espace des phases.} On les simplifie dans la limite séculaire $\omega_\alpha t_{\rm coll} \gg1$ et sous l'hypoth\`ese d'un mouvement classique ergodique des quasi-particules. Cette hypoth\`ese, selon laquelle les fluctuations $\delta \hat{n}(\rr,\pp)$ moyennées sur une trajectoire ne dépendent plus que de l'énergie de cette dernière, ne tient que si le piège est complètement anisotrope~; nous lui apportons dans ce cas une justification num\'erique soigneuse.

Les quantités cherchées $D$ et $t_0$, correctement adimensionnées, sont des fonctions universelles
de $k_BT/\mu_{\rm TF}$ où $\mu_{\rm TF}$ est la limite Thomas-Fermi de $\mu_{\rm GP}$, 
et sont en particulier indépendantes des rapports $\omega_\alpha/\omega_{\beta}$ des pulsations de piégeage. 
Elles sont représentées sur la figure \ref{fig:Diff_et_t0}. 
Un sous-produit intéressant et plus directement mesurable de notre étude est le taux d'amortissement $\Gamma(\epsilon)$
des modes de Bogolioubov d'énergie $\epsilon$ dans le piège. \`A la température adimensionn\'ee $k_BT/\mu_{\rm TF}$ fixée, il est lui 
aussi décrit par une fonction universelle de $\epsilon/\mu_{\rm TF}$ indépendante des pulsations de piégeage,  
voir la figure \ref{fig:Gamma}. 
Ces résultats participent d'une nouvelle classe d'universalité, celle des pièges harmoniques complètement anisotropes, 
fort différente de celle théoriquement mieux défrichée des systèmes spatialement homogènes, et recevront, espérons-le, une confirmation expérimentale prochaine.

\section*{Remerciements}
Nous remercions les membres de l'\'equipe \og fermions froids\fg\ et \og puces \`a atomes\fg\ du LKB,  en particulier  Christophe Salomon,
pour d'utiles discussions.

\section*{Annexe A. Comportement de $\Gamma(\epsilon)$ \`a basse et \`a haute \'energie}

Pour obtenir les comportements aux limites (\ref{eq:gambassener}) et (\ref{eq:asymptGammaeps}) du taux d'amortissement $\Gamma(\epsilon)$ des
modes de Bogolioubov dans un pi\`ege \`a l'approximation s\'ecularo-ergodique, 
nous r\'ecrivons l'int\'egrale dans l'espace des phases (\ref{eq:defGamergo}) comme une moyenne sur le potentiel
chimique de Gross-Pitaevskii local $\mu_0=g\rho_0(\rr)$ du taux d'amortissement $\Gamma_h(\epsilon,\mu_0,k_B T)$ des
modes d'\'energie $\epsilon$ dans un syst\`eme homog\`ene de densit\'e $\mu_0/g$ et de temp\'erature $T$~:
\be
{\Gamma}(\epsilon) = \int_0^{\mu_{\rm TF}} \dd\mu_0 P_\epsilon(\mu_0) \Gamma_h(\epsilon,\mu_0,k_B T)
\label{eq:GamepsP0}
\ee
avec
\be
P_\epsilon(\mu_0) \equiv \frac{1}{\rho(\epsilon)} \int \frac{\dd^3r\,\dd^3p}{(2\pi\hbar)^3} \delta(\epsilon-\epsilon(\rr,\pp))
\delta(\mu_0-g\rho_0(\rr)) = \frac{4}{\pi\rho(\epsilon)} \frac{1}{(\hbar\bar{\omega})^3} 
\frac{\epsilon^2(\mu_{\rm TF}-\mu_0)^{1/2}}
{(\mu_0^2+\epsilon^2)^{1/2} [(\mu_0^2+\epsilon^2)^{1/2}+\mu_0]^{1/2}}
\ee

Dans la limite $\epsilon\to 0$, nous rempla\c{c}ons d'abord heuristiquement l'int\'egrande dans l'\'equation (\ref{eq:GamepsP0}) par un \'equivalent
\`a basse \'energie, en utilisant~:
\bea
\label{eq:equivPeps}
P_\epsilon(\mu_0) &\underset{\epsilon\to 0}{\sim}& \frac{3}{8\sqrt{2}} \frac{\epsilon^{1/2}(\mu_{\rm TF}-\mu_0)^{1/2}} {\mu_{\rm TF}^{1/2}\mu_0^{3/2}} \\
\frac{\hbar\Gamma_h(\epsilon,\mu_0,k_B T)}{2} &\underset{\epsilon\to 0}{\sim}& \epsilon \left(\frac{\mu_0 a^3}{g}\right)^{1/2} F(k_B T/\mu_0)
\label{eq:equivGammaeps}
\eea
Dans l'\'equation (\ref{eq:equivPeps}), nous avons utilis\'e l'\'equation (\ref{eq:fbasse})~; le r\'esultat (\ref{eq:equivGammaeps}) se trouve
dans la r\'ef\'erence \cite{Giorgini}, o\`u la fonction $F$ est calcul\'ee et \'etudi\'ee. Comme $F(\theta)\underset{\theta\to+\infty}{\sim} 
\frac{3\pi^{3/2}}{4}\theta$, ceci fait appara\^\i tre dans l'\'equation (\ref{eq:GamepsP0}) 
l'int\'egrale divergente ${\epsilon^{3/2}}\int_0^{\mu_{\rm TF}} \dd\mu_0/{\mu_0^2}$. On voit bien cependant qu'il faut
couper cette int\'egrale \`a $\mu_0>\epsilon$ pour que l'\'equivalent 
(\ref{eq:equivGammaeps}) reste utilisable, d'o\`u la loi d'\'echelle ${\Gamma}(\epsilon) \approx \epsilon^{1/2}$, 
domin\'ee par le bord du condensat pi\'eg\'e et tr\`es diff\'erente de la loi d'annulation lin\'eaire du cas homog\`ene.
Pour trouver le pr\'efacteur dans la loi, il suffit de faire le changement d'\'echelle $\mu_0=\epsilon \nu_0$ dans l'int\'egrale et d'utiliser 
l'approximation de \og haute temp\'erature\fg\, de la r\'ef\'erence \cite{PRAGora} sur $\Gamma_h$, 
uniform\'ement valable pr\`es du bord du condensat pi\'eg\'e,
\be
\frac{\hbar \Gamma_h(\epsilon,\mu_0,k_B T)}{2} \underset{k_B T\gg \epsilon,\mu_0}{\sim} k_B T \left(\frac{\mu_0 a^3}{g}\right)^{1/2} \phi(\epsilon/\mu_0)
\ee
avant de passer \`a la limite $\epsilon\to 0$ sous le signe int\'egral,
ce qui conduit \`a  l'\'equation cherch\'ee (\ref{eq:gambassener})
avec \footnote{En pratique, la fonction $\phi$ se d\'eduit de l'\'equation (\ref{eq:Gamrp}) en effectuant l'approximation de champ
classique $1+\bar{n}(\rr,\qq)\simeq \bar{n}(\rr,\qq) \simeq k_B T/\epsilon(\rr,\qq)$. Dans le calcul num\'erique de $\mathcal{I}$, effectu\'e en prenant $\check{\epsilon}
=1/\nu_0$ comme variable d'int\'egration, on r\'eduit l'effet de la troncature num\'erique \`a l'aide du d\'eveloppement asymptotique 
$\phi(\epsc) \underset{\epsc\to+\infty}{=} 4\left(\frac{2\pi}{\epsc}\right)^{1/2} \left[2\ln \frac{\epsc}{2} + \frac{1-\ln(\epsc/2)}{\epsc}
+\frac{23+6\ln(\epsc/2)}{8\epsc^2} + O(\frac{\ln\epsc}{\epsc^3})\right]$, qui corrige et am\'eliore celui de l'\'equation (35) 
de la r\'ef\'erence \cite{PRAGora}.}
\be
\mathcal{I}=\int_0^{+\infty} \dd\nu_0 \frac{\nu_0^{1/2} \phi(1/\nu_0)}{(1+\nu_0^2)^{1/2} [\nu_0+(1+\nu_0^2)^{1/2}]^{1/2}}
=4, 921\, 208\, \ldots
\ee

Dans la limite $\epsilon\to +\infty$, nous utilisons le fait que, dans le cas homog\`ene, le taux d'amortissement des quasi-particules
se r\'eduit au taux de collision $\rho_0 \sigma v$ d'une particule de vitesse $v=(2\epsilon/m)^{1/2}$ avec les particules du condensat,
de vitesse nulle et de densit\'e $\rho_0$, avec la section efficace $\sigma=8\pi a^2$ des bosons indiscernables (c'est un processus
de Beliaev)~:
\be
\frac{\hbar \Gamma_h(\epsilon,\mu_0, k_B T)}{2} \underset{\epsilon\to +\infty}{\sim} \mu_0 a \frac{(2m\epsilon)^{1/2}}{\hbar}
\label{eq:Gamhomasympt}
\ee
En utilisant de m\^eme le d\'eveloppement \`a haute \'energie (\ref{eq:fhaute}) de $\rho(\epsilon)$, nous trouvons que
$P_\epsilon(\mu_0) \sim (8/\pi) (\mu_{\rm TF}-\mu_0)^{1/2} \epsilon^{-3/2}$. Le report de ces \'equivalents dans l'\'equation
(\ref{eq:GamepsP0}) donne bien (\ref{eq:asymptGammaeps}).
